\documentclass[journal]{new-aiaa}
\usepackage{textcomp}

\usepackage{abstract,pstricks,booktabs,amsfonts,amsmath,fancyhdr}
\usepackage{graphicx}
\usepackage{subcaption}
\usepackage{titlesec}
\usepackage{lastpage}
\usepackage{dblfloatfix}
\usepackage{float}
\usepackage{anyfontsize}
\usepackage{makecell}
\usepackage{bm}
\usepackage{url}
\usepackage{natbib}
\usepackage{textcase}

\DeclareMathOperator{\sn}{sn}
\DeclareMathOperator{\cn}{cn}
\DeclareMathOperator{\dn}{dn}

\title{Averaged Solar Torque Rotational Dynamics for Defunct Satellites}

\author{Conor J. Benson \footnote{PhD Candidate, Aerospace Engineering Sciences, 429 UCB, 3775 Discovery Drive, Boulder, CO, 80303} and Daniel J. Scheeres \footnote{Distinguished Professor and A. Richard Seebass Chair, Aerospace Engineering Sciences, 429 UCB, 3775 Discovery Drive, Boulder, CO, 80303}}
\affil{University of Colorado Boulder, Boulder, CO, 80303}

\begin{document}

\maketitle

\begin{abstract}
 Spin state predictions for defunct satellites in geosynchronous earth orbit (GEO) are valuable for active debris removal and servicing missions as well as material shedding studies and attitude-dependent solar radiation pressure (SRP) modeling. Previous studies have shown that solar radiation torques can explain the observed spin state evolution of some GEO objects via the Yarkovsky-O'Keefe-Radzievskii-Paddack (YORP) effect. These studies have focused primarily on uniform rotation. Nevertheless, many objects are in non-principal axis rotation (i.e. tumbling). Recent exploration of the tumbling regime for the family of retired GOES 8-12 satellites has shown intriguing YORP-driven behavior including  spin-orbit coupling, tumbling cycles, and tumbling period resonances. To better explore and understand the tumbling regime, we develop a semi-analytical tumbling-averaged rotational dynamics model. The derivation requires analytically averaging over the satellite's torque-free rotation, defined by Jacobi elliptic functions. Averaging is facilitated by a second order Fourier series approximation of the facet illumination function. The averaged model is found to capture and explain the general long-term behavior of the full dynamics while reducing computation time by roughly three orders of magnitude. This improved computation efficiency promises to enable rapid exploration of general long-term rotational dynamics for defunct satellites and rocket bodies. 
\end{abstract}

\section{Introduction}

The long-term orbital evolution of debris in geosynchronous earth orbit (GEO) has been studied extensively over the past 50 years \cite{allan1964, schildknecht2007, rosengren2019}.  Lacking atmospheric drag and other natural de-orbit mechanisms, GEO debris will remain on orbit indefinitely \citep{rosengren2019}. On the other hand, less work has been done to understand the attitude dynamics of this debris. Many GEO debris objects are retired and otherwise defunct satellites and rocket bodies. The spin rates of these large debris objects are diverse and evolve over time \citep{papushev,cognion,earl,benson2018a}. Understanding their attitude evolution will benefit orbit prediction since attitude-dependent solar radiation pressure (SRP) is the largest non-gravitational perturbation at GEO. Also, spin rate knowledge for these large objects will help predict debris shedding. Most high area-to-mass ratio GEO debris is thought to be multi-layer insulation (MLI) from defunct satellites and rocket bodies \citep{liou}. Finally, given the growing debris population and the large cost to construct, launch, and operate GEO satellites, many organizations are developing active debris removal (ADR) and satellite servicing missions. To grapple and de-spin large, potentially non-cooperative satellites, spin state predictions are vital. With variable spin rates, forecasting windows of slow rotation will reduce collision risk as well as time and energy needed to de-spin. Also, understanding how end of life satellite configurations (e.g. solar array orientation) affect long-term spin state evolution is important. Improved knowledge could help inform decommission procedures to minimize post-disposal spin rates and variability, further facilitating ADR and servicing missions.  

Leveraging studies of asteroid dynamics, Albuja et al.~\cite{albuja2015,albuja2018} investigated the influence of solar radiation and thermal re-emission torques on defunct satellite spin states. The combined influence of these torques on a body's spin state is called the Yarkovsky-O'Keefe-Radzievskii-Paddack (YORP) effect \citep{rubincam}. Albuja et al.~\cite{albuja2018} found that the YORP effect could explain the observed spin rate evolution of the defunct GOES 8 and 10 weather satellites. The authors closely predicted the rapid observed spin down of GOES 8 in 2014 and its subsequent transition from uniform rotation to non-principal axis tumbling \cite{albuja2018}. Benson et al. \cite{benson2020b} found that solar array orientation greatly impacts YORP-driven uniform spin state evolution, consistent with the dramatically different observed evolution of GOES 8 and 10. This demonstrated the potential to dictate post-disposal spin state evolution with proper end of life configurations. Propagating the GOES dynamics into the tumbling regime, Benson et al.~\cite{benson2020b} found that the satellite's rotational angular momentum vector tends to track the time-varying sun direction. Further exploration has uncovered cyclic behavior where the satellite transitions repeatedly between uniform rotation and tumbling as well as tumbling period resonances. Additional work is needed to understand these behaviors. All study of tumbling YORP for defunct satellites has considered the full dynamics (i.e. Euler's equations of motion) \cite{albuja2018,benson2020b}. These equations are not amenable to long-term numerical propagation as they require short integration time steps to maintain solution accuracy. Furthermore, Euler's equations are expressed in terms of fast variables (i.e. attitude and angular velocity). Since we are interested in studying changes over long periods of time, slowly varying osculating elements (e.g. the rotational angular momentum vector and kinetic energy) are more appropriate. This is directly comparable to orbital dynamics, where the averaged Lagrange and Gauss planetary equations, written in terms of osculating orbital elements, have been used extensively to study long-term orbital evolution \cite{vallado}. This success motivates development of analogous tumbling-averaged dynamical equations for osculating rotational elements, namely the rotational angular momentum vector and kinetic energy. 

A number of authors have investigated spin-averaged attitude dynamics models. Albuja et al. \cite{albuja2015} extended the uniform spin-averaged asteroidal YORP work of Scheeres \cite{scheeres2007} to defunct satellites. These models are not applicable to tumbling satellites as the motion is driven by two generally incommensurate periods rather than one for the uniform case \citep{sa1991}. Several tumbling-averaged YORP models have been developed for asteroids \citep{cicalo,breiter2011}. These asteroidal models average over the spin state and heliocentric orbit given the slow spin state evolution. Orbit averaging is not appropriate for defunct satellites due to the possibility for angular momentum sun-tracking. Also, these models only account for diffuse reflections which is insufficient for defunct satellites since many surfaces are dominated by specular reflections. 

In this paper we develop a fast, semi-analytical tumbling-averaged attitude dynamics model that accounts for specular and diffuse reflections as well as absorption and instantaneous thermal re-emission of solar radiation. To allow for analytical averaging, we approximate the facet illumination function with its second order Fourier series expansion. For the time-being, we neglect all other perturbations including gravitational/magnetic torques and internal energy dissipation. First we describe relevant frames, dynamics, and the radiation torque equations in Section II. In Section III, we illustrate the YORP-driven tumbling behavior of the full model. Motivated by these results, we then derive the semi-analytical averaged dynamics in Section IV, leaving details for the appendices. Here, we also validate and explore the averaged model. We finish by discussing implications of the findings and providing conclusions.

\section{Preliminaries}
\subsection{Frames}

\begin{figure}[h]
	\centering
	\subcaptionbox{$\mathcal{O}$ and $\mathcal{H}$ frames}{\includegraphics[width=3in]{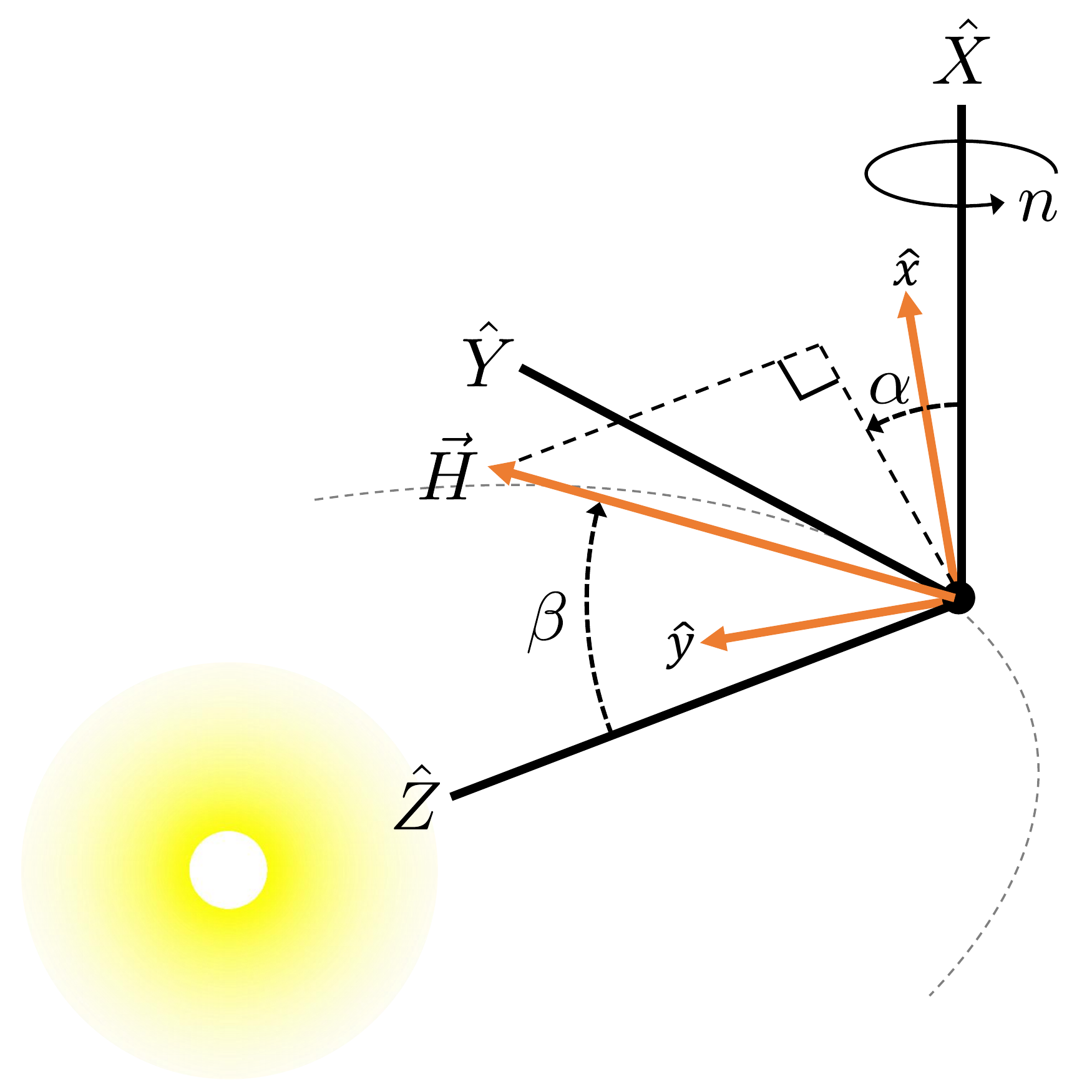}}
	\subcaptionbox{$\mathcal{H}$ and $\mathcal{B}$ frames}{\includegraphics[width=2in]{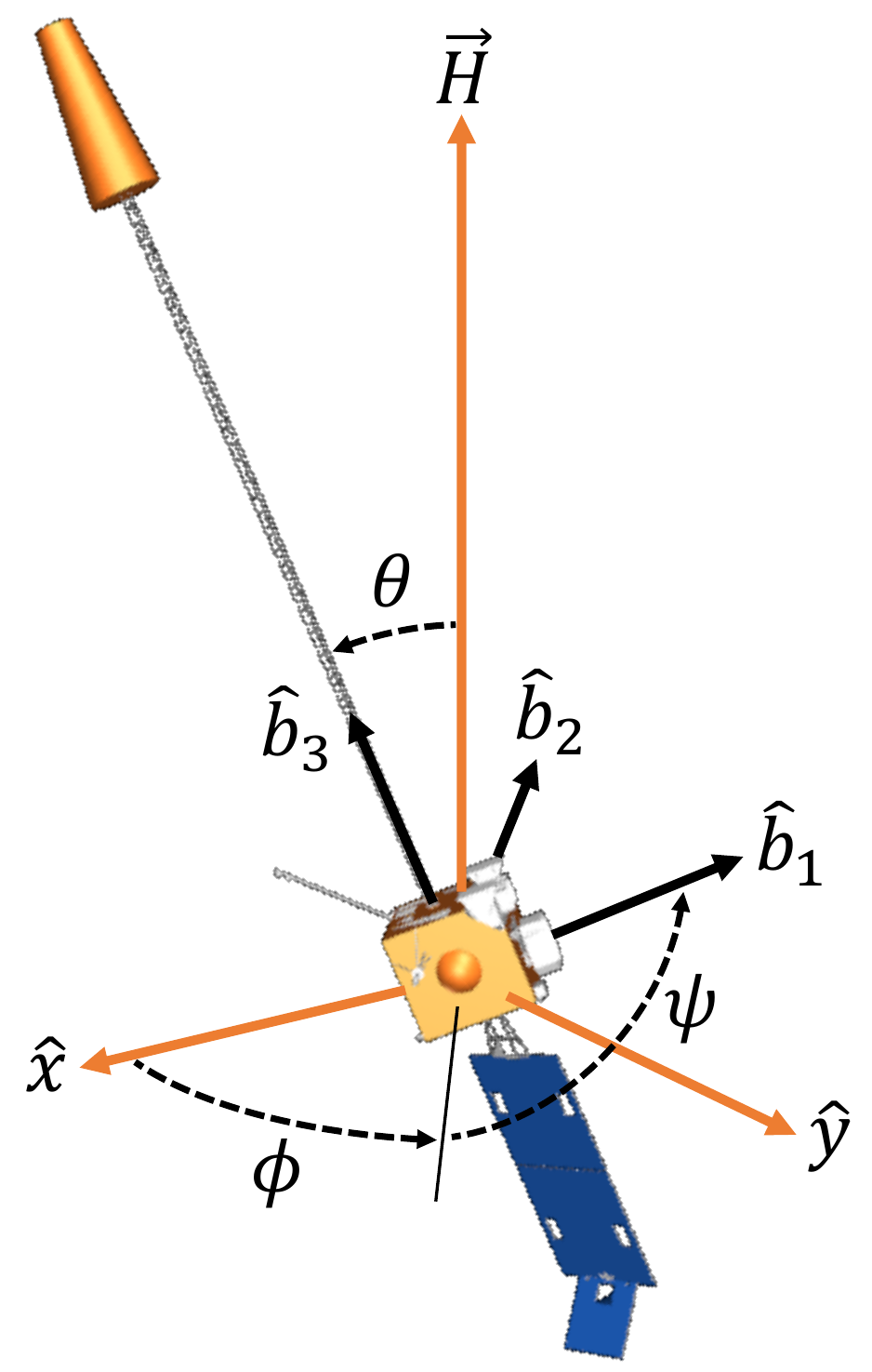}}
	\caption{Relevant frames and rotations.}
\label{fig:frames}
\end{figure}

For this paper we will assume the satellite is in a circular heliocentric orbit at 1 astronomical unit (AU), neglecting its much smaller earth orbit. This approximation was validated by Albuja et al. \cite{albuja2018} for the GOES 8 and 10 satellites. The rotating orbit frame is denoted by  $\mathcal{O}$:$\{\bm{\hat{X}}$,$\bm{\hat{Y}}$,$\bm{\hat{Z}}\}$. This frame is centered at the satellite with $\bm{\hat{X}}$ along the orbit angular momentum direction, $\bm{\hat{Z}}$ pointed towards the sun, and $\bm{\hat{Y}}$ in the orbital velocity direction (see Figure~\ref{fig:frames}a). The angular velocity of $\mathcal{O}$ with respect to the inertial frame $\mathcal{N}$ is $\boldsymbol{\omega}_{\mathcal{O}/\mathcal{N}}=n\bm{\hat{X}}$ where $n$ is the heliocentric mean motion. The next frame is the angular momentum frame $\mathcal{H}$:$\{\bm{\hat{x}}$,$\bm{\hat{y}}$,$\bm{\hat{z}}\}$. Here $\bm{\hat{z}}$ is along the satellite's rotational angular momentum vector $\bm{H}$. Rotation from $\mathcal{O}$ to $\mathcal{H}$ is given by the rotation matrix $HO=R_2(\beta)R_3(\alpha)$. $R_i$ denotes a principal rotation about the $i$th axis \citep{schaub}. Consulting Figure~\ref{fig:frames}a, the "clocking" angle $\alpha$ and "coning" angle $\beta$ are the spherical coordinates of $\bm{\hat{H}}$ in the $\mathcal{O}$ frame. 

The final relevant frame is the satellite body frame $\mathcal{B}$:$\{\bm{\hat{b}}_1$,$\bm{\hat{b}}_2$,$\bm{\hat{b}}_3\}$. Rotation from $\mathcal{H}$ to $\mathcal{B}$, shown in Figure~\ref{fig:frames}b, is given by (3-1-3) ($\phi$-$\theta$-$\psi$) Euler angles with the rotation matrix $BH$ \cite{schaub},
\small
\begin{singlespace}
\begin{equation}
BH =
\begin{bmatrix}
\cos\phi\cos\psi - \cos\theta\sin\phi\sin\psi & \cos\psi\sin\phi + \cos\phi\cos\theta\sin\psi & \sin\psi\sin\theta \\
- \cos\phi\sin\psi - \cos\psi\cos\theta\sin\phi & \cos\phi\cos\psi\cos\theta - \sin\phi\sin\psi & \cos\psi\sin\theta\\
\sin\phi\sin\theta & -\cos\phi\sin\theta & \cos\theta
\end{bmatrix}
=
\begin{bmatrix}
a_{x1} & a_{y1} & a_{z1} \\
a_{x2} & a_{y2} & a_{z2} \\
a_{x3} & a_{y3} & a_{z3} \\
\end{bmatrix}
\label{eq:BH}
\end{equation}
\end{singlespace}
\normalsize
So an arbitrary vector $\bm{f}$ in the $\mathcal{H}$ frame is given by,
equivalently in matrix form,
\begin{singlespace}
\begin{equation}
\begin{bmatrix}
f_x \\
f_y \\ 
f_z \\
\end{bmatrix}
=
\begin{bmatrix}
a_{x1} & a_{x2} & a_{x3} \\
a_{y1} & a_{y2} & a_{y3} \\
a_{z1} & a_{z2} & a_{z3} \\
\end{bmatrix}
\begin{bmatrix}
f_1 \\
f_2 \\ 
f_3 \\
\end{bmatrix}
\label{eq:Hvec}
\end{equation}
\end{singlespace}
where $f_1$, $f_2$, and $f_3$ are the $\mathcal{B}$ frame components.

\subsection{Osculating Elements}

Given the sun-tracking behavior observed in the full dynamical simulations, we are interested in developing our equations in the rotating $\mathcal{O}$ frame. Using the transport theorem, a method to calculate time derivatives in rotating frame \citep{schaub}, we find the time derivative of $\bm{H}$ with respect to the $\mathcal{O}$ frame, 
\begin{equation}
\frac{^\mathcal{O}d}{dt}(\bm{H})=-\boldsymbol{\omega}_{\mathcal{O}/\mathcal{N}}\times\bm{H}+\bm{M}
\label{eq:Horbdot}
\end{equation}
where $\bm{M}=\dot{\bm{H}}$ is the net external torque. Then, expressing $\bm{H}$ in the $\mathcal{O}$ frame, we have
\begin{singlespace}
\begin{equation}
  \begin{bmatrix}
    H_X\\
    H_Y\\
    H_Z\\
   \end{bmatrix}=
  \begin{bmatrix}
    H\cos{\alpha}\sin{\beta}\\
    H\sin{\alpha}\sin{\beta}\\
    H\cos{\beta}\\
   \end{bmatrix}
\label{eq:Horb}
\end{equation}
\end{singlespace}
\noindent where ($H_X$, $H_Y$, $H_Z$) are the $\mathcal{O}$ frame components and $H=|\bm{H}|$. Taking the time derivative of the Eq.~\ref{eq:Horb}, solving for $\dot{\alpha}$, $\dot{\beta}$, and $\dot{H}$, and substituting the results from Eq.~\ref{eq:Horbdot}, we ultimately obtain,

\begin{equation}
\dot{\alpha}=\frac{M_y+Hn\cos{\alpha}\cos{\beta}}{H\sin{\beta}}
\label{eq:alphadot}
\end{equation}
\begin{equation}
\dot{\beta}=\frac{M_x+Hn\sin{\alpha}}{H}
\label{eq:betadot}
\end{equation}
\begin{equation}
\dot{H}=M_z
\label{eq:Hdot}
\end{equation}
where ($M_x$, $M_y$, $M_z$) denote the torque components in the angular momentum frame. Note that $\dot{\alpha}$ is singular for $\beta=$ 0$^{\circ}$ and 180$^{\circ}$ due to $\sin{\beta}$ in the denominator of Eq.~\ref{eq:alphadot}. While not implemented in our model, one could replace $\alpha$ and $\beta$ with the alternate coordinates $v=\sin{\alpha}\sin{\beta}$ and $w=\cos{\alpha}\sin{\beta}$ when $\bm{H}$ is very near the sun/anti-sun line. These coordinates were simply obtained by finding expressions that cancel $\sin{\beta}$ in the denominator of Eq.~\ref{eq:alphadot}. This alternate set will instead have a $\beta$ ambiguity since $\sin{\beta}$ is symmetric about $\beta=$ 90$^{\circ}$.

\begin{figure}[h]
	\centering
	\includegraphics[height=2in]{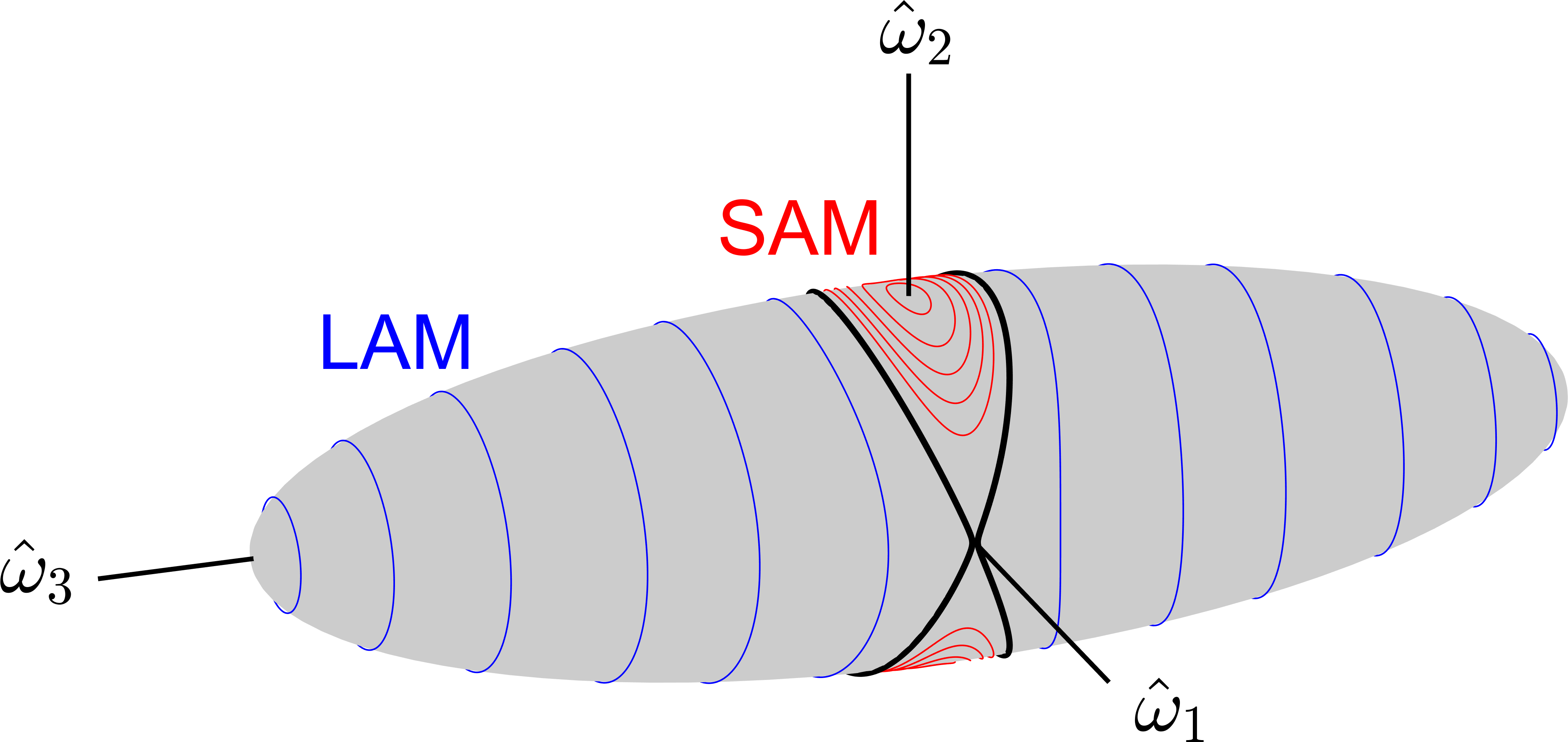}
	\caption{Angular velocity curves for long (LAM) and short (SAM) axis modes.}
\label{fig:polhode}
\end{figure}

Another quantity of interest, the dynamic moment of inertia $I_d$, is given by $I_d=H^2/2T$ where $\bm{H}=[I]\boldsymbol{\omega}$ and the rotational kinetic energy $T=\frac{1}{2}\boldsymbol{\omega}{\cdot}[I]\boldsymbol{\omega}$. $[I]$ and $\boldsymbol{\omega}$ are the body's inertia tensor and inertial angular velocity of the $\mathcal{B}$ frame respectively.  With principal inertias $I_s\;{\geq}\;I_i\;{\geq}\;I_l$, we will assume the long axis convention with $[I]=\mathrm{diag}([I_i,I_s,I_l])$ \cite{sa1991}. For torque-free rigid body rotation, $I_d$ defines the closed path that $\boldsymbol{\omega}$ takes through the body frame, known as a polhode \cite{landau}. $I_d$ is constrained to $[I_l,I_s]$ since $T$ is bounded for a given $H$. When $I_l<I_d<I_i$, the satellite is said to be in a long axis mode (LAM) because $\boldsymbol{\omega}$ circulates about the satellite's long axis ($\bm{\hat{b}}_3$) \cite{landau}. When $I_i<I_d<I_s$, the satellite is in a short axis mode (SAM) where $\boldsymbol{\omega}$ instead circulates about the short axis ($\bm{\hat{b}}_2$). $I_d=I_l$ and $I_d=I_s$ correspond to principal axis rotation about $\bm{\hat{b}}_3$ and $\bm{\hat{b}}_2$ respectively. Finally $I_d=I_i$ denotes motion along the separatrix between LAMs and SAMs or uniform rotation about the intermediate axis, both of which are unstable. Various polhodes are illustrated in Figure~\ref{fig:polhode} for the GOES 8 satellite assuming constant $H$. Here, the separatrices are shown in black.

Taking the time derivative of $I_d$, we ultimately obtain, 
\begin{equation}
\dot{I}_d=-\frac{2I_d}{H}\Bigg[\frac{I_d-I_i}{I_i}a_{z1}M_1+\frac{I_d-I_s}{I_s}a_{z2}M_2+\frac{I_d-I_l}{I_l}a_{z3}M_3\Bigg]
\label{eq:Iddot2}
\end{equation}
where ($M_1$, $M_2$, $M_3$) denote the net torque components in the body frame. Complementing $I_d$ is another fundamental quantity called the effective spin rate $\omega_e=H/I_d$, which is proportional to $\boldsymbol{\omega}$ (see Appendix A). Analogous to osculating orbital elements that define an instantaneous unperturbed two-body (Keplerian) orbit \cite{vallado}, $\alpha$, $\beta$, $I_d$, and $H$ (or $\omega_e$) define the instantaneous unperturbed rotation state, which changes slowly over time due solar radiation torques and/or other perturbations.

\subsection{Full Dynamics}

For the full dynamics, the body frame angular velocity $\boldsymbol{\omega}$ evolution is given by Euler's equations,

\begin{equation}
[I]\dot{\boldsymbol{\omega}}=-[\tilde{\boldsymbol{\omega}}][I]\boldsymbol{\omega}+\bm{M}
\label{eq:wdot}
\end{equation}
where $[\tilde{\;\;\;}]$ is the skew-symmetric cross product operator.

The body's inertial attitude is tracked using quaternions \citep{schaub},
\begin{singlespace}
\begin{equation}
	\begin{bmatrix}
    \dot{\beta}_0 \\
    \dot{\beta}_1 \\
    \dot{\beta}_2 \\
    \dot{\beta}_3 \\
   \end{bmatrix}
   =
	\frac{1}{2}\begin{bmatrix}
    -\beta_1 & -\beta_2 & -\beta_3\\
    \beta_0 & -\beta_3 & \beta_2\\
    \beta_3 & \beta_0 & -\beta_1\\
    -\beta_2 & \beta_1 & \beta_0\\
   \end{bmatrix}
   \begin{bmatrix}
    \omega_1\\
    \omega_2\\
    \omega_3\\
   \end{bmatrix}
\label{eq:quatkde}
\end{equation}
\end{singlespace}
\noindent where $\beta_0$ is the scalar component. In this paper, the full dynamics are propagated with MATLAB's ode113 numerical integrator with 1e-12 absolute and relative tolerances. 

\subsection{Solar Torque Model}
For this work, the faceted solar radiation force model provided by Scheeres \cite{scheeres2007} is used. This model accounts for absorption, specular reflection, and Lambertian diffuse reflection and re-emission. The satellite is assumed to be in thermal equilibrium, so all absorbed radiation is immediately re-emitted. The solar radiation force acting on the $i$th satellite facet is given by, 
\begin{equation}
\bm{f}_i=-P_{SRP}\Big[\{{\rho_i}s_i(2\bm{\hat{n}}_{i}\bm{\hat{n}}_{i}-U)+U\}\cdot\bm{\hat{u}}\\
+c_{di}\bm{\hat{n}}_{i}\Big]A_i\max(0,\bm{\hat{u}}\cdot\bm{\hat{n}}_i)
\label{eq:srpforce}
\end{equation}
Here, $P_{SRP}$ is the solar radiation pressure (nominally 4.56${\times}10^{-6}\;\mathrm{N/m^2}$ at 1 AU), $\rho_i$ is the total facet reflectivity, $s_i$ is the fraction of total reflectivity that is specular, $\bm{\hat{n}}_i$ is the facet unit normal vector, $U$ is the 3$\times$3 identity matrix, $\bm{\hat{u}}$ is the satellite to sun unit vector (equivalent to $\bm{\hat{Z}}$), $A_i$ is the facet area, and $c_{di}=B(1-s_i)\rho_i+B(1-\rho_i)$ where $B$ is the scattering coefficient (2/3 for Lambertian reflection). The operation $\bm{\hat{n}}_i\bm{\hat{n}}_i$ represents a matrix outer product. The illumination function $\max(0,\bm{\hat{u}}\cdot\bm{\hat{n}}_i)$ ensures that only illuminated facets contribute. Self-shadowing by other facets and multiple reflections are not currently considered.

The solar radiation torque acting on the faceted satellite model can then be calculated as,

\begin{equation}
\bm{M}={\sum_{i=1}^{n_f}}{\bm{r}_i}\times\bm{f}_i
\label{eq:M}
\end{equation}
where $\bm{r}_i$ is the center of mass to the facet centroid position vector and $n_f$ is the number of facets.

\subsection{GOES Model}

\begin{figure}[H]
	\centering
	\includegraphics[width=4in]{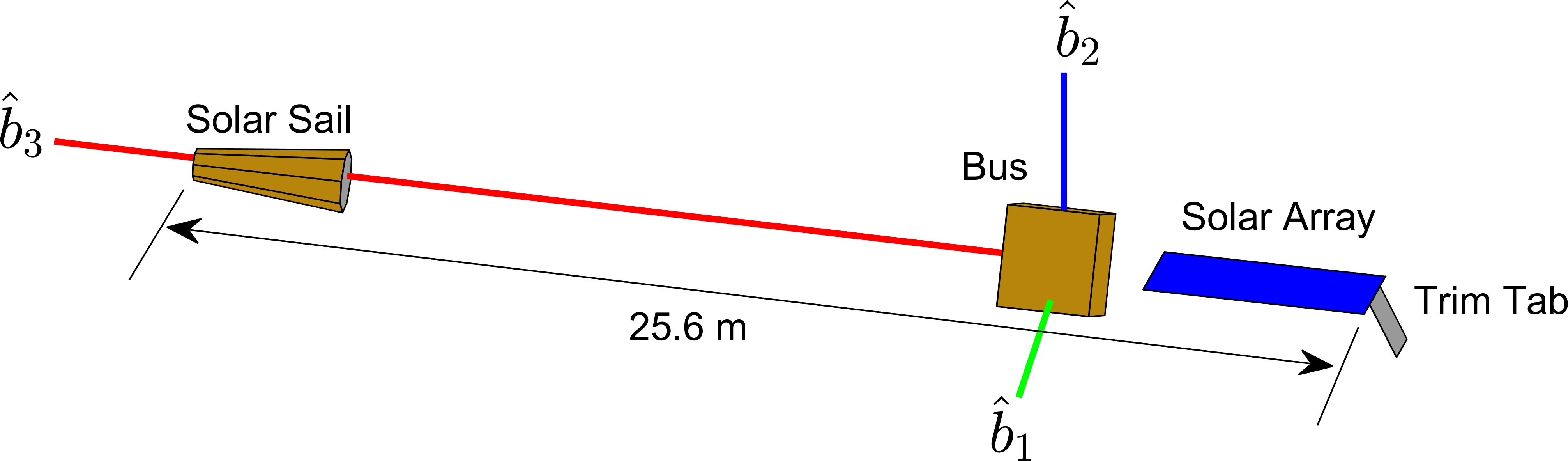}
	\caption{GOES 8 shape model with principal axes and major components labeled.}
\label{fig:goes_baxes}
\end{figure}

We will now briefly discuss the GOES model used to explore the YORP dynamics in this paper. GOES 8-12 are a family of five retired GEO weather satellites. They are notable for their asymmetry and well-documented dimensions \citep{databook}. When uncontrolled, this asymmetry provides the opportunity for large net solar torques. The 26 facet GOES shape model used for this work is provided in Figure~\ref{fig:goes_baxes} with GOES 8's approximate end of life principal axes and solar array angle  $\theta_{sa}$ of 17$^{\circ}$ \citep{benson2020b}. For GOES 8 with a dry mass of 972 kg, the end of life principal inertias are $I_l =$ 980.5, $I_i =$ 3432.1, and $I_s =$ 3570.0 kg${\cdot}$m$^2$ \citep{benson2020b}, Also, $\theta_{sa}$ is measured positive around $-\bm{\hat{b}}_3$, and $\theta_{sa}=$ 0$^{\circ}$ when the solar array sun-side and $+\bm{\hat{b}}_2$ face are parallel. See Ref. \cite{benson2020b} for how $\theta_{sa}$ impacts the model inertias. Table~\ref{tab:goesoptical} provides the optical properties assumed for the various GOES model components \citep{benson2020b}. Note that most of the materials are MLI or aluminized tape which provide almost exclusively specular reflections.

\begin{singlespace}
\begin{table}[H]
	\centering
	\caption{GOES Model Optical Properties}
	\begin{tabular}{llllll}
		\hline
		Component & Material & $\rho_i$ & $s_i$ \\ 
		\hline 
		Bus & MLI & 0.60 & 1 \\
		Solar Array front & Solar cell & 0.27 & 1 \\
		Solar Array back & Graphite & 0.07 & 0 \\
		Trim Tab front & Al tape & 0.83 & 1 \\
		Trim Tab back & Graphite & 0.07 & 0 \\
		Solar Sail sides/top & Al Kapton & 0.66 & 1 \\
		Solar Sail base & Al tape & 0.83 & 1 \\
		\hline
	\end{tabular}
	\label{tab:goesoptical}
\end{table}
\end{singlespace}

\section{Full YORP Dynamics}

\begin{figure}[H]
	\centering
	\subcaptionbox{Effective Spin Rate}{\includegraphics[width=3in]{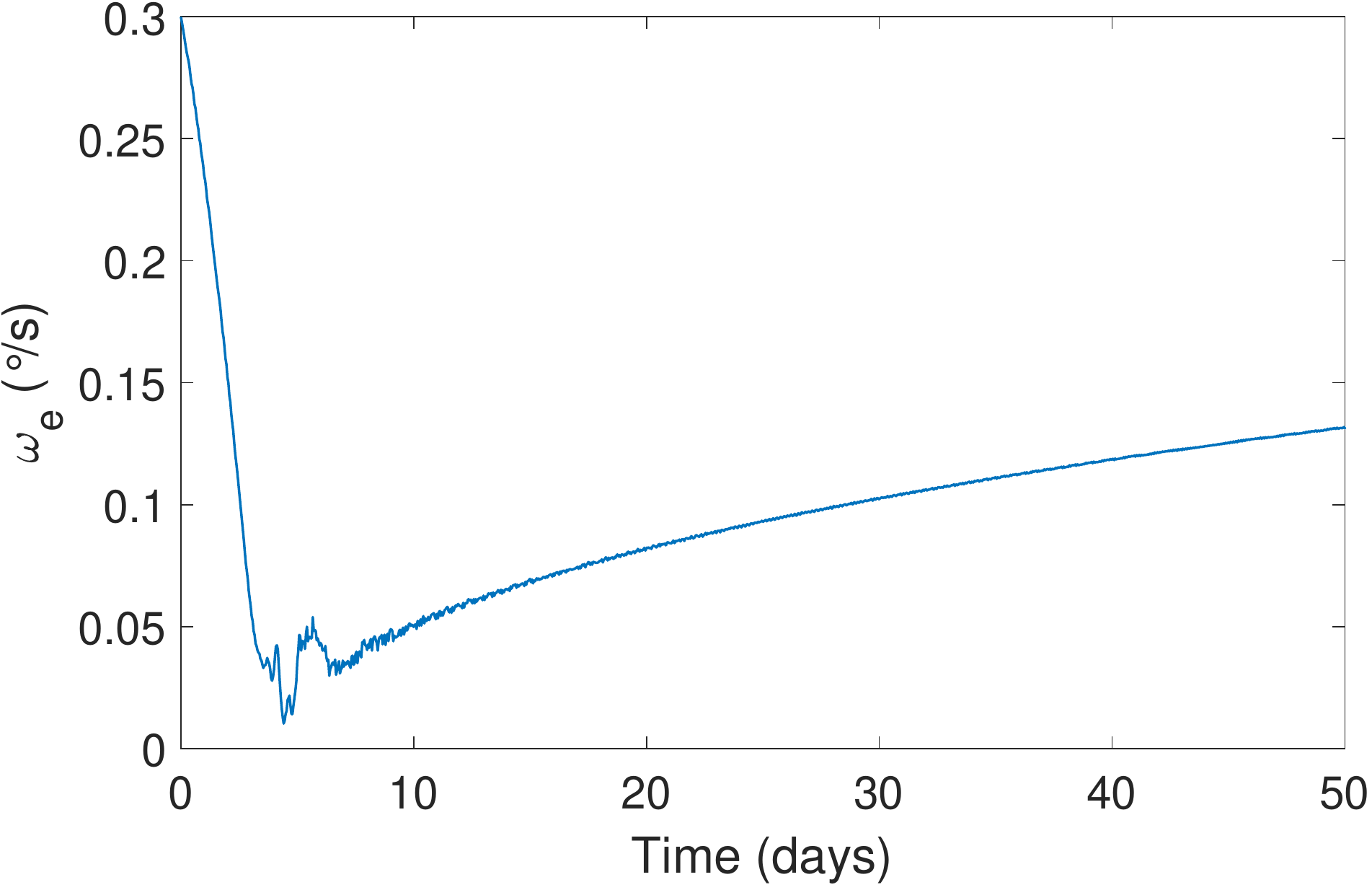}}
	\subcaptionbox{Scaled Dynamic Moment of Inertia}{\includegraphics[width=3in]{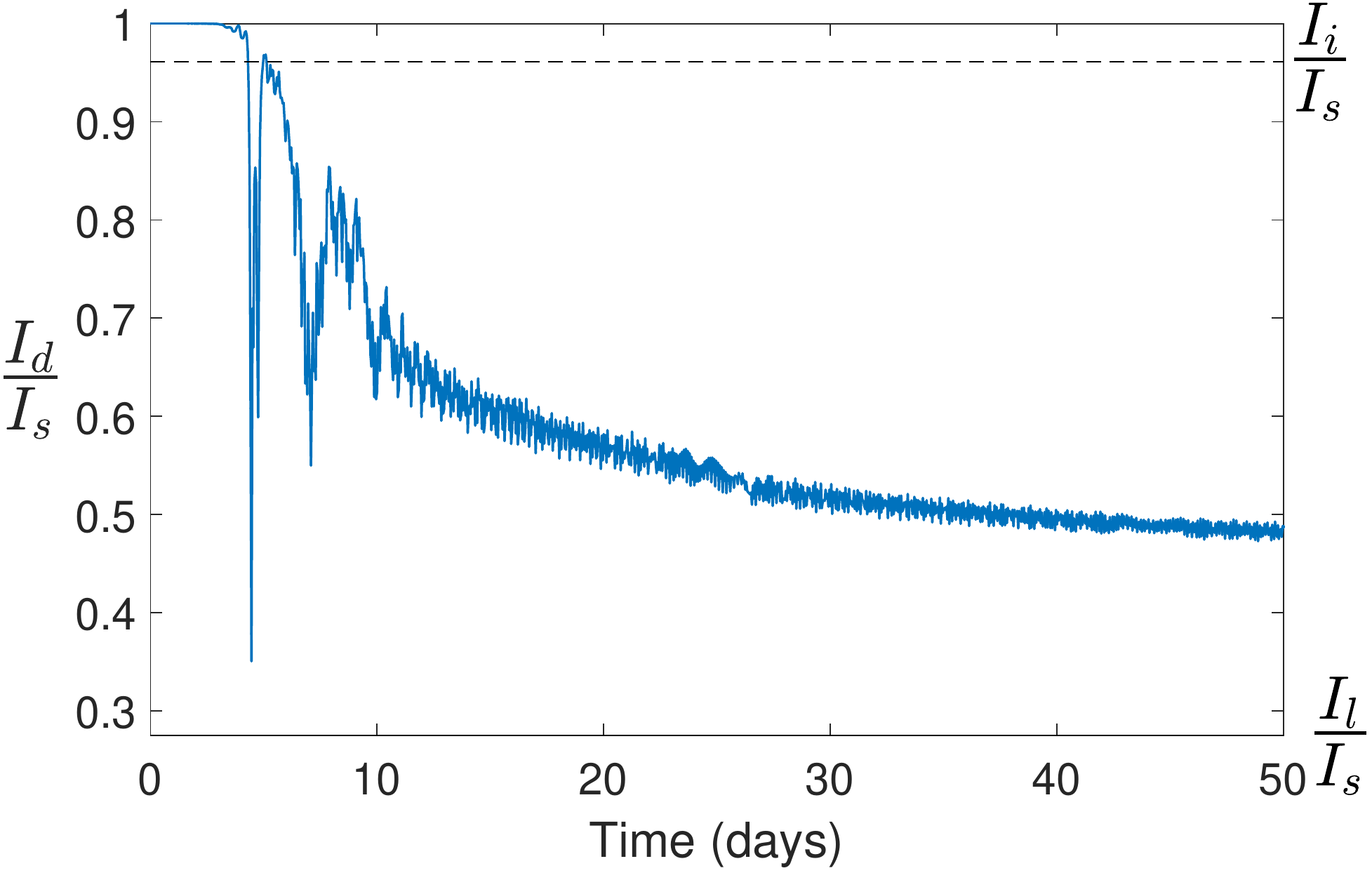}}
	\subcaptionbox{Clocking Angle}{\includegraphics[width=3in]{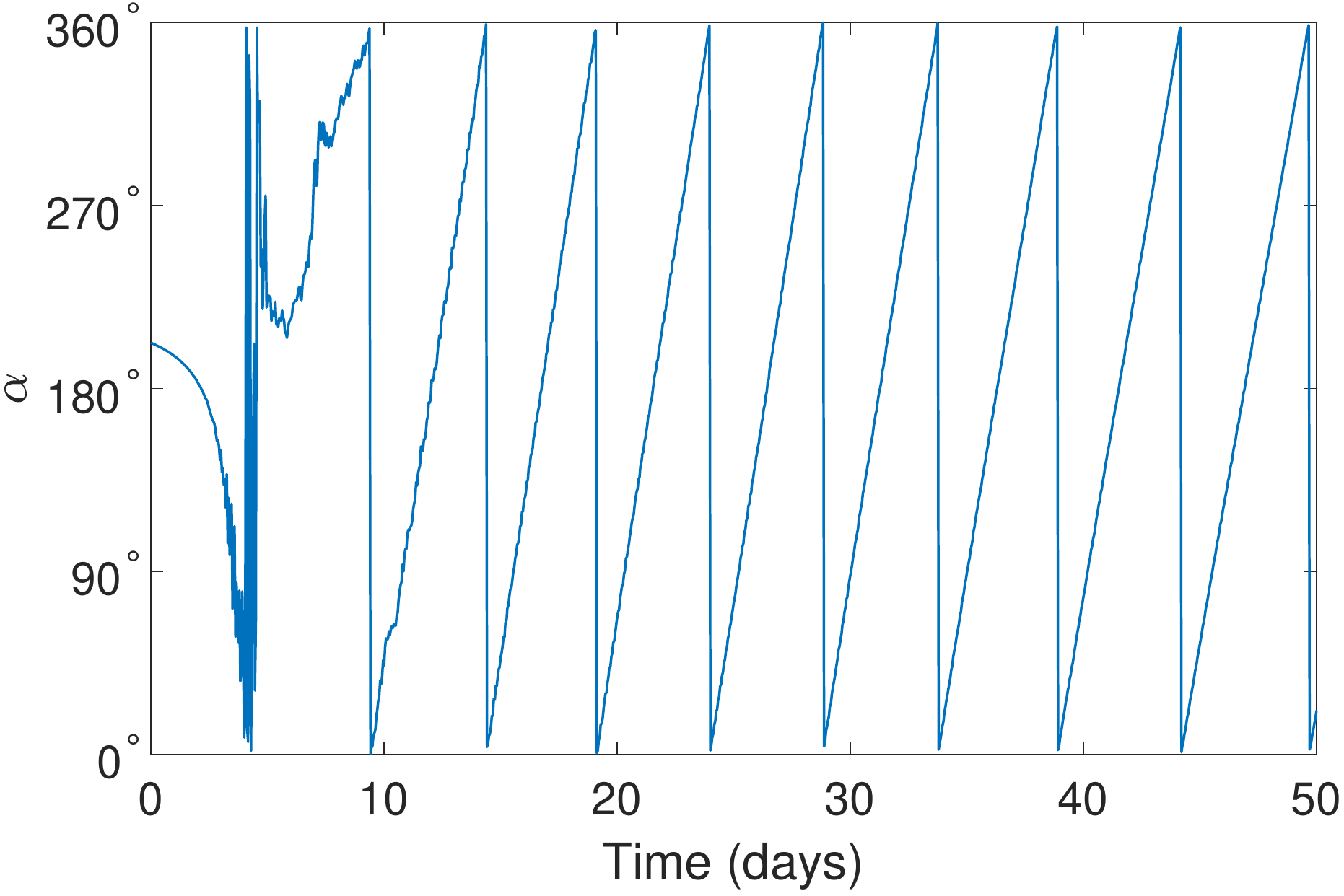}}
	\subcaptionbox{Angle between $\bm{H}$ and $\bm{\hat{u}}$}{\includegraphics[width=3in]{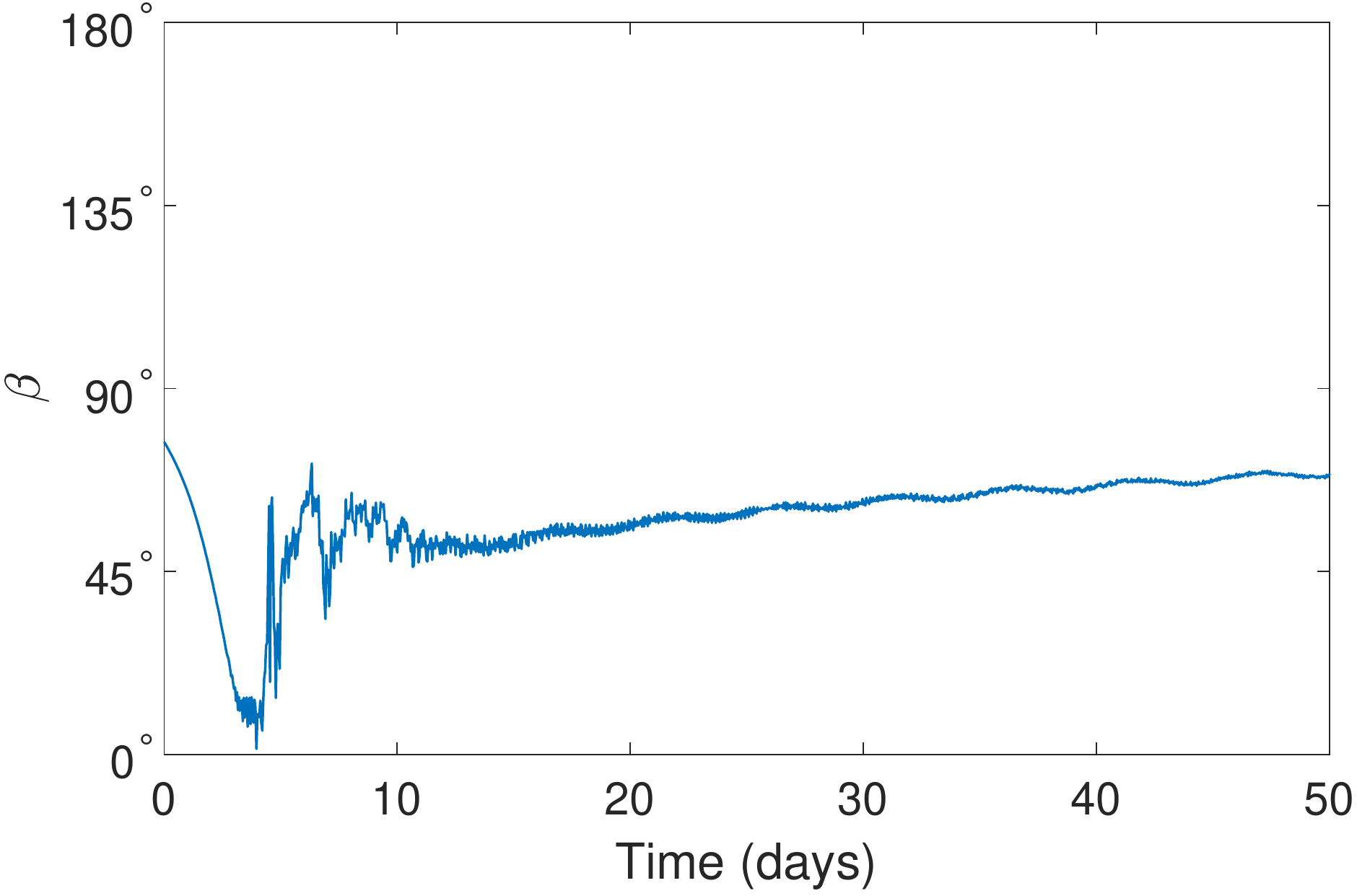}}	
	\caption{Run 1 - transition from uniform rotation to tumbling.}
\label{fig:run1_evol_zoom}
\end{figure}

We will now provide simulation results from the full dynamics model (Eqs.~\ref{eq:wdot} - \ref{eq:M}) to illustrate the complex, yet structured YORP-driven dynamical evolution. This will motivate our development of the tumbling-averaged model in Section IV. Again, we neglect the satellite's geosynchronous orbit and assume that the sun rotates in the inertial frame at earth's mean motion $n$ (${\sim}$0.986$^{\circ}$/day). The GOES 8 shape model and mass parameters given above are utilized. We will discuss two simulation runs, Run 1 and Run 2. Run 1 demonstrates uniform to tumbling transition, spin-orbit coupling, and tumbling cycles. Run 2 demonstrates these behaviors in addition to tumbling period resonances. Starting with Run 1, the satellite is placed in uniform rotation about $+\bm{\hat{b}}_2$ with $P_e=2\pi/{\omega_e}=$ 20 min and a pole direction with $\alpha_o=$ 202$^{\circ}$ and $\beta_o=$ 77$^{\circ}$. The initial evolution is provided in Figure~\ref{fig:run1_evol_zoom}. Starting in uniform rotation, Figure~\ref{fig:run1_evol_zoom}a shows that $\omega_e$ decreases rapidly over the first four days as the satellite spins down. During this initial spin down, Figure~\ref{fig:run1_evol_zoom}d shows that $\beta$ decreases as the pole moves towards the sun-line. Once $\omega_e$ reaches a sufficiently small value, the satellite transitions to non-principal axis rotation, apparent in Figure~\ref{fig:run1_evol_zoom}b. Here, $I_d$ decreases as the rotation moves from uniform rotation to SAM to LAM, crossing the separatrix denoted by the dashed line. From approximately five days onward, $\omega_e$ increases and $I_d$ decreases as the satellite spins up further about $+\bm{\hat{b}}_3$, the minimum inertia axis.  During this time, $\alpha$ and $\beta$ increase as the pole begins precessing about the sun-line with $\alpha$ taking roughly five days to complete a cycle. 

\begin{figure}[H]
	\centering
	\subcaptionbox{Effective Spin Rate}{\includegraphics[width=3in]{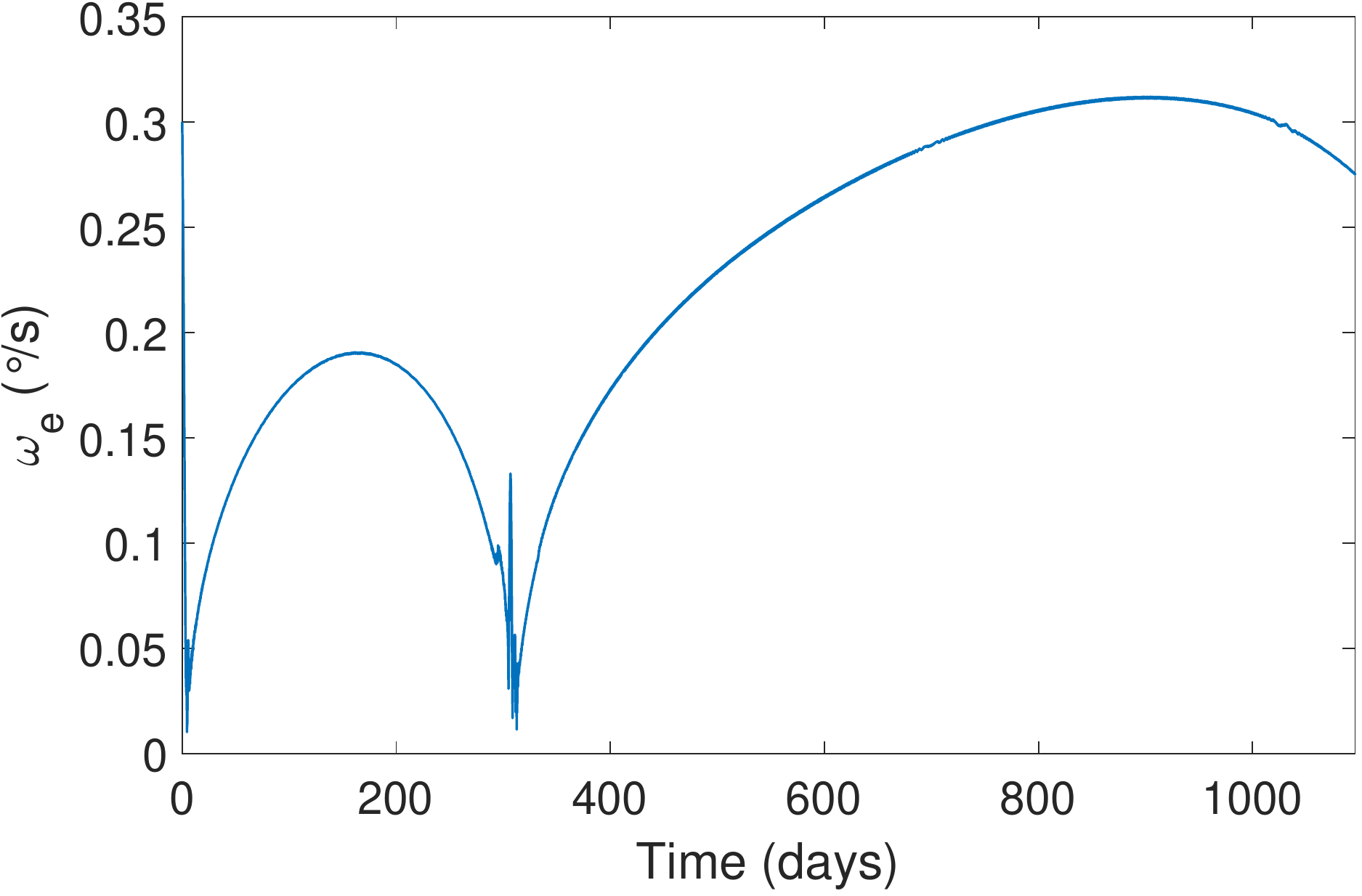}}
	\subcaptionbox{Scaled Dynamic Moment of Inertia}{\includegraphics[width=3in]{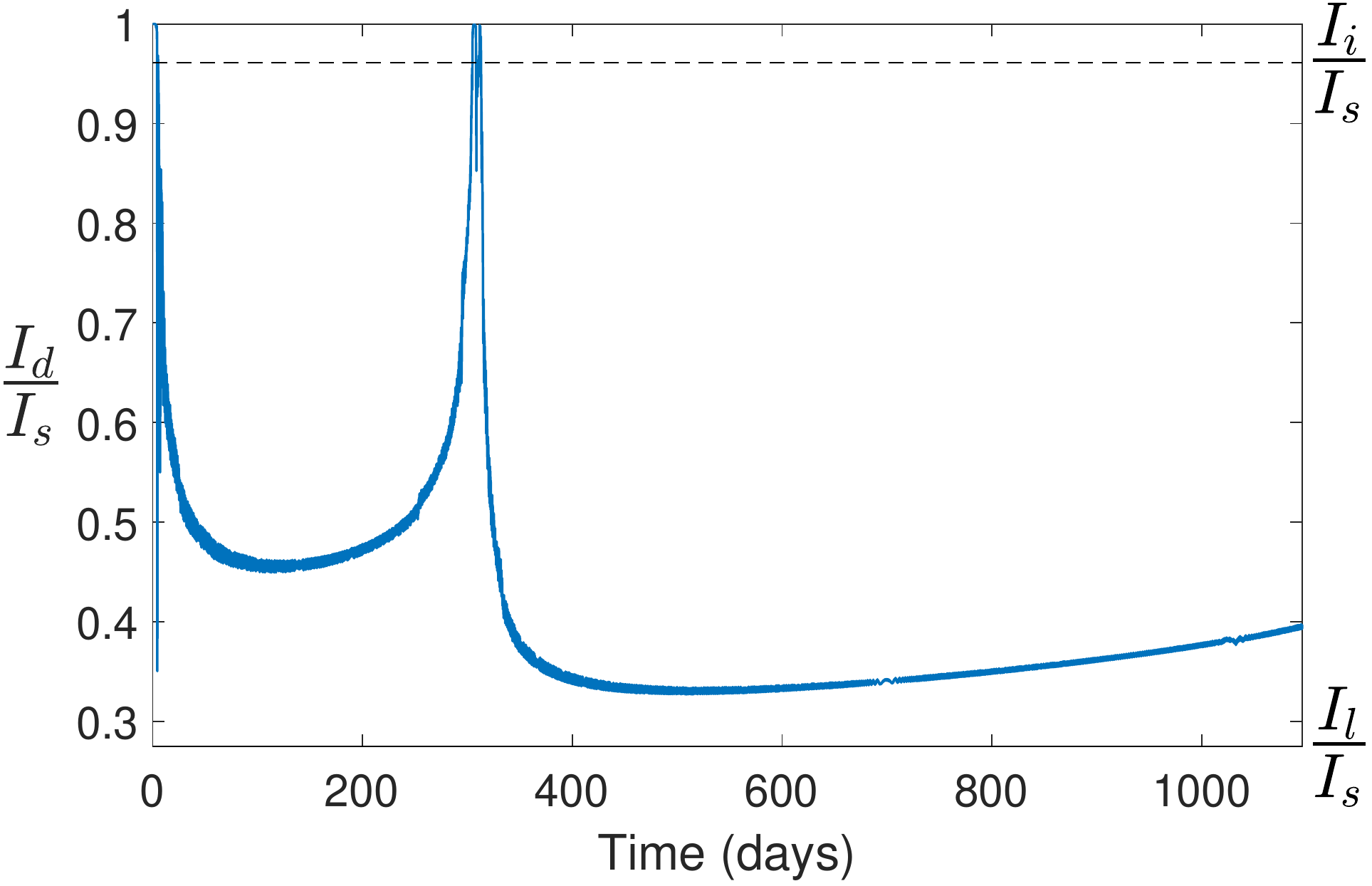}}
	\subcaptionbox{Clocking Angle}{\includegraphics[width=3in]{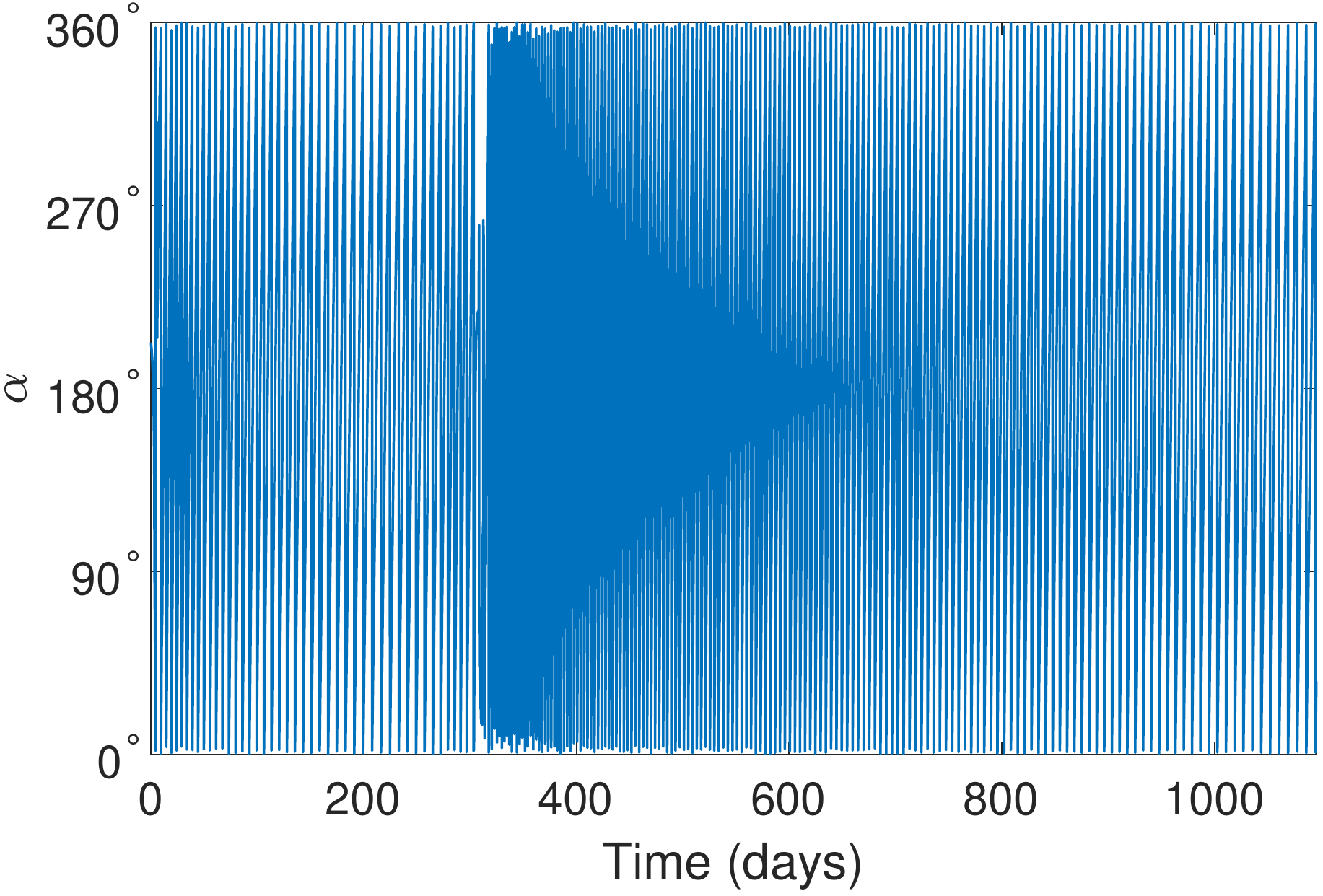}}
	\subcaptionbox{Angle between $\bm{H}$ and $\bm{\hat{u}}$}{\includegraphics[width=3in]{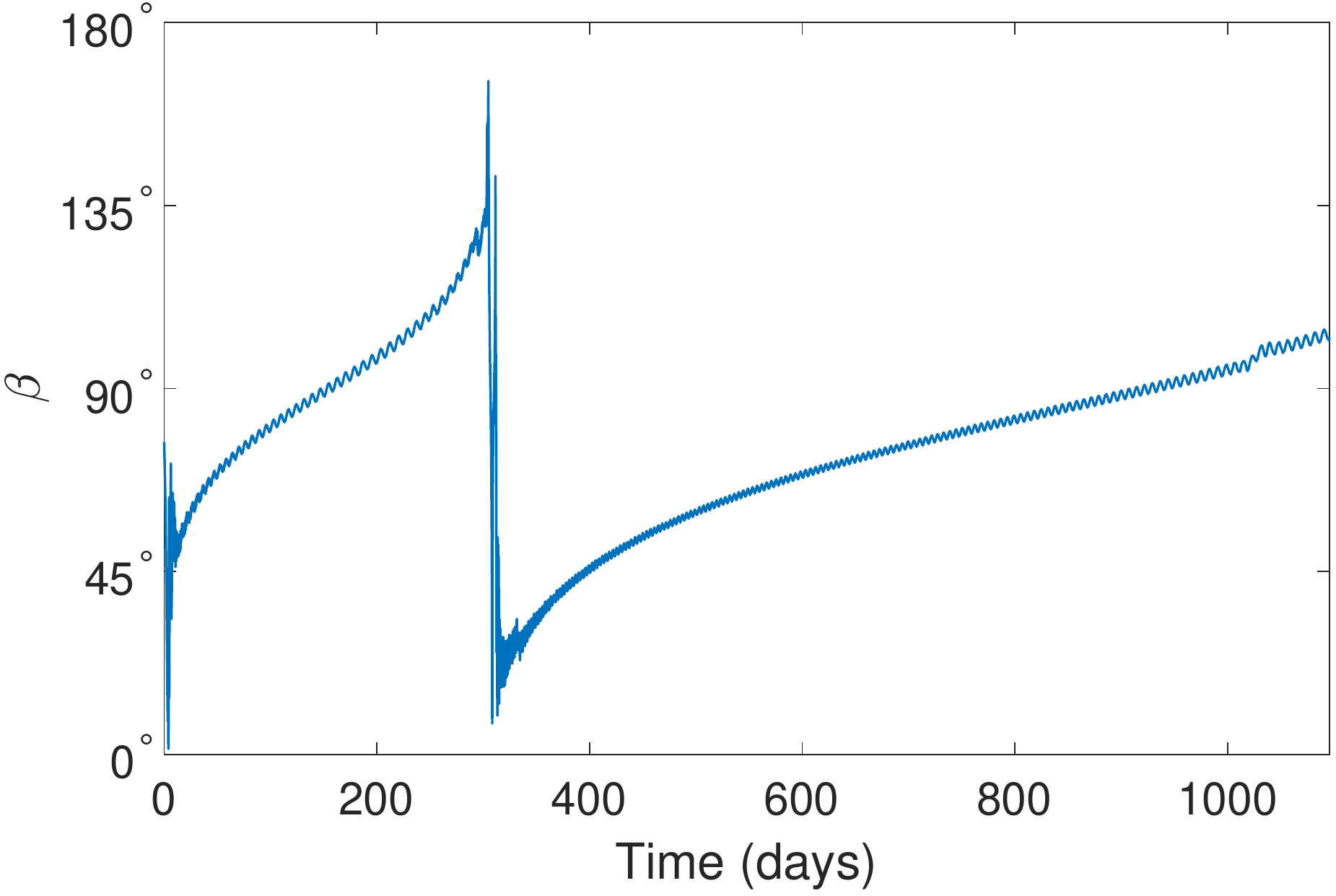}}	
	\caption{Run 1 - long-term dynamical evolution.}
\label{fig:run1_evol}
\end{figure}

Proceeding further in time, Figure~\ref{fig:run1_evol} shows evolution of the Run 1 solution over three years. On this timescale, we see that the satellite continues in this long axis spin up state until around 160 days when $\beta$ reaches 90$^{\circ}$. At this point, $\omega_e$ decreases and $I_d$ increases as the satellite moves back towards uniform rotation. This trend continues until 285 days when the satellite is finally rotating slowly in near-uniform rotation with $\beta$ approaching 180$^{\circ}$. Given the small $\omega_e$, $\beta$ decreases rapidly towards 0$^{\circ}$. During this time, $\omega_e$ briefly increases, then decreases with $I_d\;{\approx}\;I_s$. Once $\beta$ nears 0$^{\circ}$, the satellite again spins up about $+\bm{\hat{b}}_3$ and enters a second, much longer, tumbling cycle. 

\begin{figure}[H]
	\centering
	\includegraphics[width=3.5in]{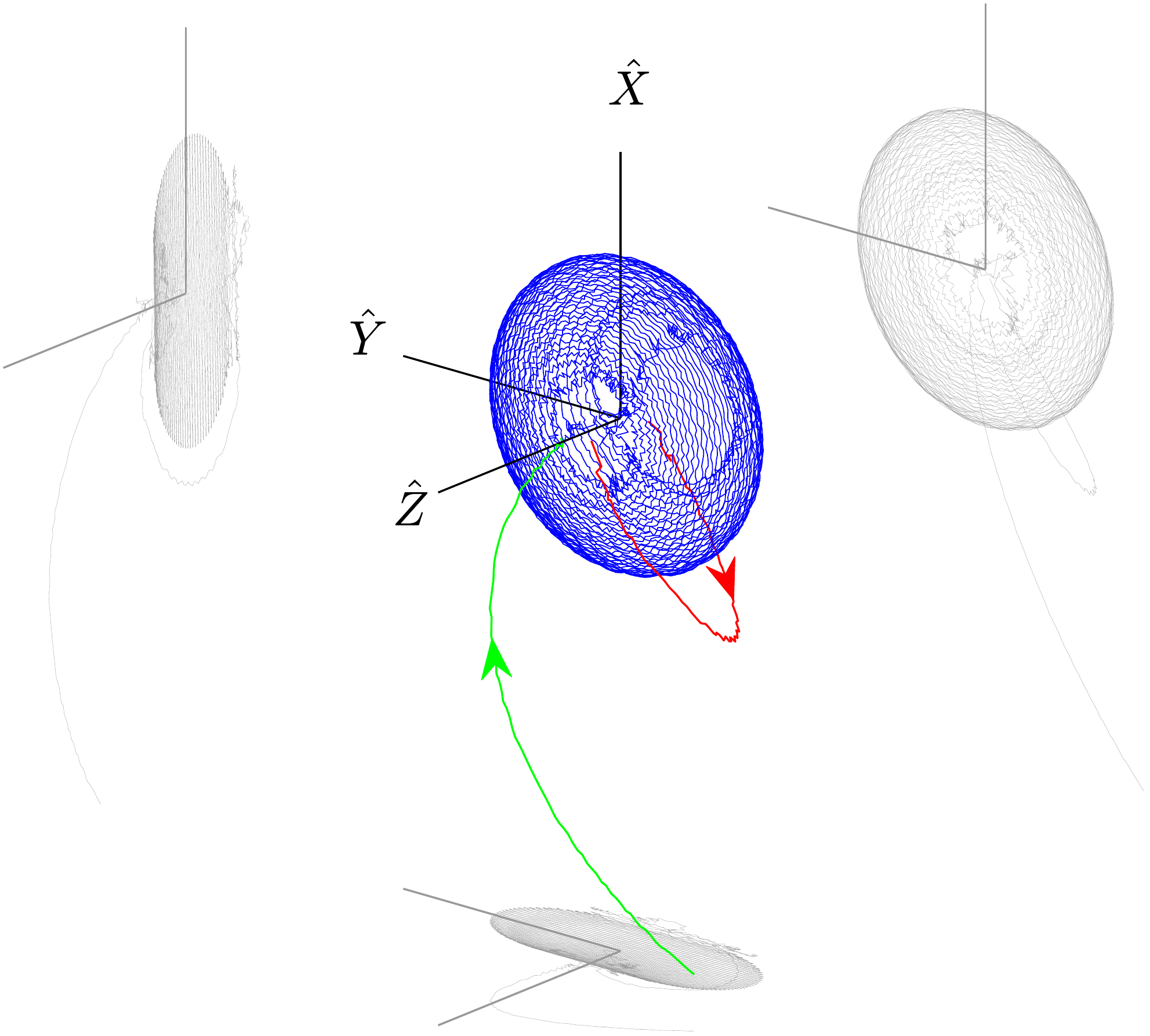}
	\caption{Run 1 - $\bm{H}$ evolution in $\mathcal{O}$ frame over the first tumbling cycle (0 to 308 days). The gray lines are projections of this evolution on the three orthogonal planes.}
\label{fig:run1_hvecorb}
\end{figure}

To better visualize the pole evolution during these tumbling cycles, Figure~\ref{fig:run1_hvecorb} shows the evolution of $\bm{H}$ in the $\mathcal{O}$ frame over the first tumbling cycle in Run 1 (from 0 to 309 days in Figure~\ref{fig:run1_evol}). The green section is the initial uniform spin down from 0 to 4 days as $\omega_e$ decreases and $\bm{H}$ moves towards the sun-line ($\bm{\hat{Z}}$). The blue tumbling segment from 4 days to 305 days, shows $\bm{H}$ precess about $\bm{\hat{Z}}$ while slowly moving in the $-\bm{\hat{Z}}$ direction. The near-uniform return from $\beta$ near 180$^{\circ}$ to 0$^{\circ}$ is shown in red. The second tumbling cycle is not shown for clarity but follows this same general behavior.

\begin{figure}[H]
	\centering
	\subcaptionbox{Effective Spin Rate}{\includegraphics[width=3in]{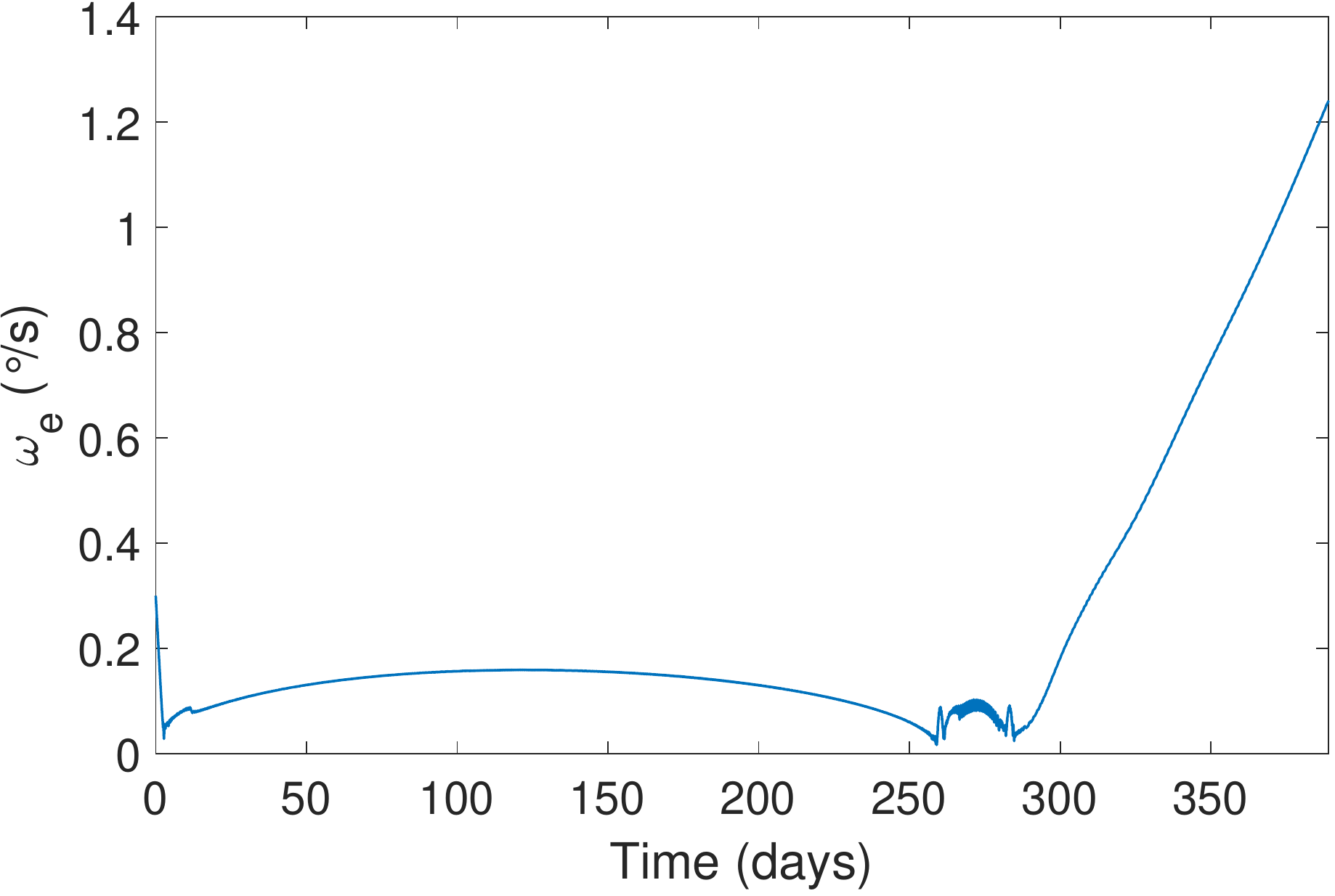}}
	\subcaptionbox{Scaled Dynamic Moment of Inertia}{\includegraphics[width=3in]{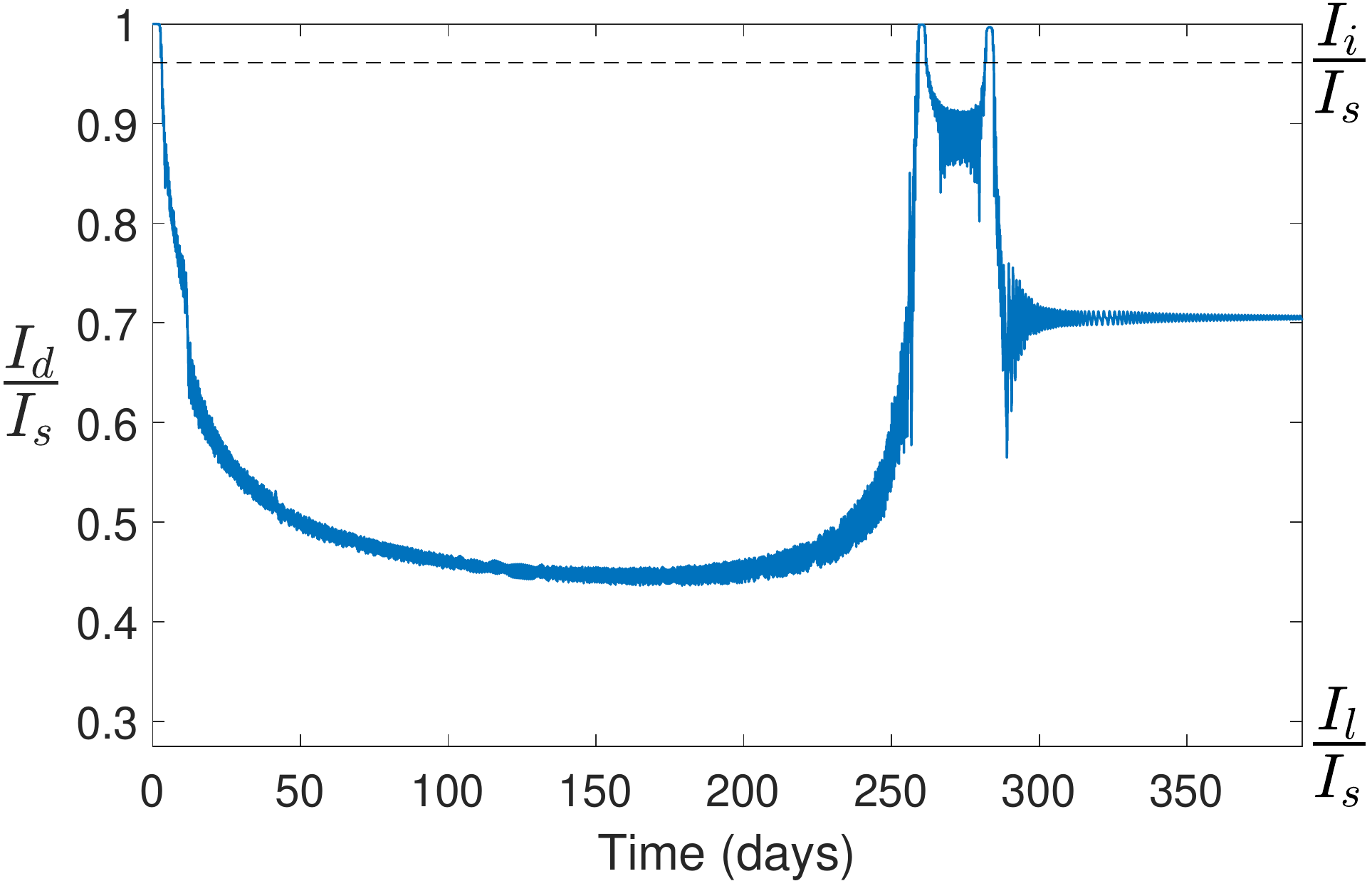}}
	\subcaptionbox{Ratio of Fundamental Tumbling Periods}{\includegraphics[width=3in]{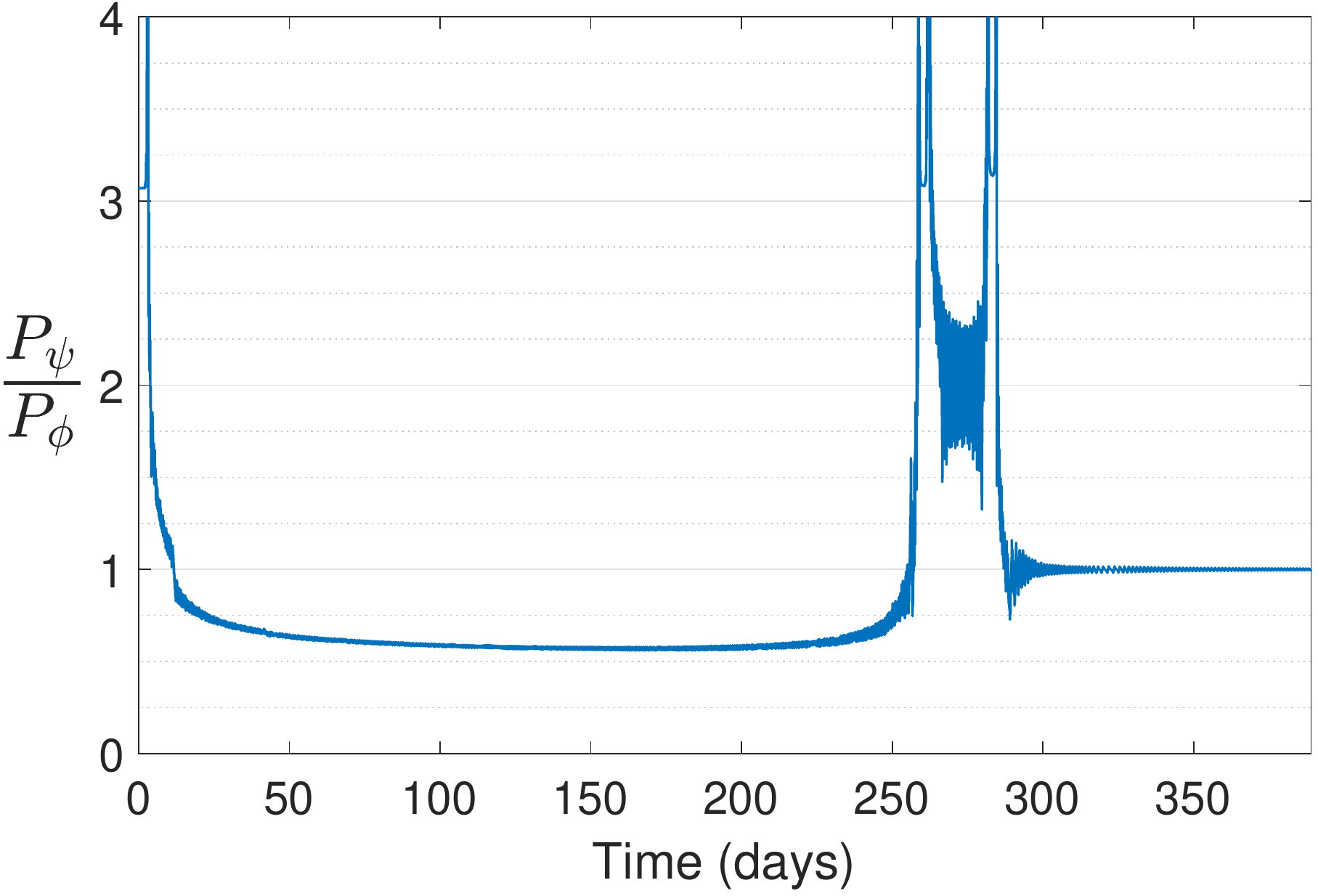}}	
	\subcaptionbox{Angle between $\bm{H}$ and $\bm{\hat{u}}$}{\includegraphics[width=3in]{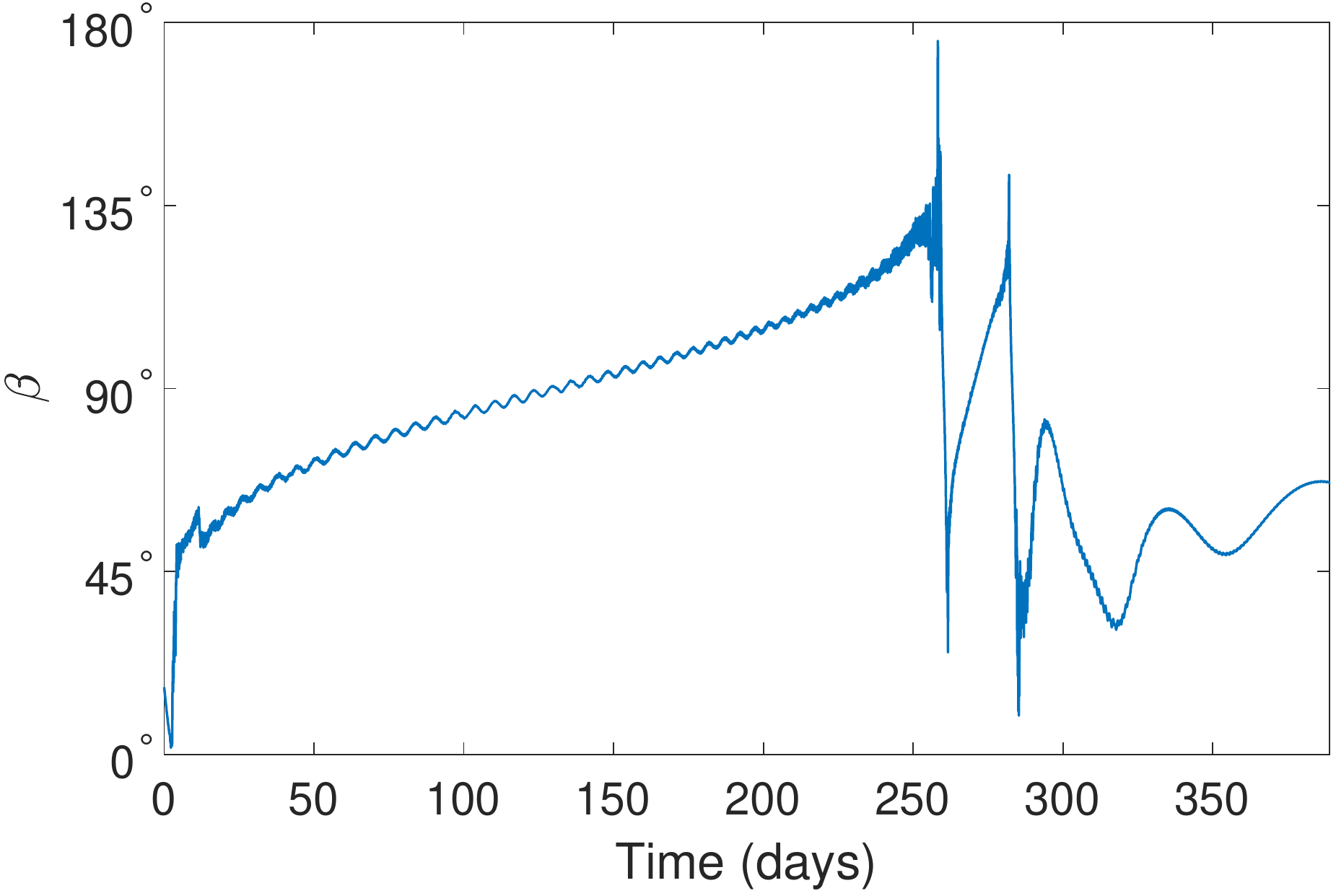}}	
	\caption{Run 2 - long-term dynamical evolution.}
\label{fig:run2_evol}
\end{figure}

For Run 2, we illustrate tumbling period resonances. The satellite is again placed in uniform rotation with $P_e=$ 20 min but now with a pole given by $\alpha_o=$ 202$^{\circ}$ and $\beta_o=$ 17$^{\circ}$. The resulting long-term evolution is provided in Figure~\ref{fig:run2_evol}. As with Run 1, $\omega_e$ decreases rapidly, the satellite transitions to tumbling, and it proceeds through a tumbling cycle. This first cycle is followed by a second, shorter cycle. After this tumbling cycle, the satellite again spins up about the minimum inertia axis but this time is captured in a $P_\psi/P_{\bar{\phi}}=$ 1 tumbling period resonance at roughly 290 days rather than entering another cycle. $P_{\bar{\phi}}$ is the average precession period of the satellite's long axis ($\bm{\hat{b}}_3$) about $\bm{H}$ and $P_\psi$ is the rotation period about $\bm{\hat{b}}_3$ itself. See Appendix A for the fundamental period expressions. Given the nearly axisymmetric mass distribution of GOES 8 ($I_s\;\approx\;I_i>I_l$), $\dot{\phi}$ is nearly constant and the average precession period $P_{\bar{\phi}}$ is essentially equal to the true precession period. So at this 1:1 resonance, the satellite returns to the same inertial attitude at multiples of $P_{\bar{\phi}}$ and $P_\psi$. Figure~\ref{fig:run2_evol} shows that $\omega_e$ increases steadily while $P_{\bar{\phi}}$ and $P_\psi$ remain in lock step with one another. Since the period ratio $P_\psi/P_{\bar{\phi}}$ is only a function of $I_l$, $I_i$, $I_s$, and $I_d$, constant $P_\psi/P_{\bar{\phi}}$ requires that $I_d$ be constant as well. While in this resonance, $\beta$ oscillates between 40$^{\circ}$ and 70$^{\circ}$ with a slight secular increase over time. Carefully examining Figure~\ref{fig:run2_evol}c, the satellite's long axis spin up is briefly perturbed when passing through the 1:1 period resonance near 11 days. Also, the period ratio over the second tumbling cycle (from 260 to 285 days) oscillates around a 2:1 ratio. Tumbling resonances were often observed in other simulation runs with 1:1 and 2:1 resonances being most common. Higher order resonances were occasionally observed. 

\section{Averaged YORP Dynamics}

To better understand the behavior illustrated by the full dynamics model, we will now develop, validate, and explore the semi-analytical tumbling-averaged model. For this paper, we will assume the tumbling periods are non-resonant to simplify the averaging. Analysis of specific tumbling resonances and their stability will feature in a follow-up paper.  

\subsection{Averaging Approach}
Following Cicalo and Scheeres \cite{cicalo}, we aim to average Eqs.~\ref{eq:alphadot} - \ref{eq:Iddot2} over the satellite's tumbling motion. For this approach, we assume that the variables $\alpha$, $\beta$, $H$ and $I_d$ vary slowly compared to the satellite's intrinsic rotation. We also assume that the solar radiation torque is a relatively small perturbation on the satellite's torque-free motion. So we average over the torque-free motion (i.e. with respect to $\phi$, $\theta$, and $\psi$) assuming constant values for the average parameters $\overline{\alpha}$, $\overline{\beta}$, $\overline{H}$ and $\overline{I}_d$. 

Torque-free rigid body rotation is defined by two fundamental tumbling periods $P_{\bar{\phi}}$ and $P_\psi$ \citep{sa1991,bensonda14}. Again, $P_{\bar{\phi}}$ is the average precession period of the satellite's minimum inertia axis ($\bm{\hat{b}}_3$) about $\bm{H}$ and $P_\psi$ is the rotation period about $\bm{\hat{b}}_3$ itself. $P_\theta$ is proportional to $P_\psi$ and is therefore not independent. The average time needed for $\phi$ to increase by $2\pi$ is generally not constant. Nevertheless, we will assume constant $\dot{\phi}$ to greatly simplify the averaging process. Fortunately, $\dot{\phi}$ is essentially constant for bodies with roughly axisymmetric inertia tensors, making this a good approximation for many GEO satellites and rocket bodies. Furthermore, assuming $P_{\bar{\phi}}$ and $P_\psi$ are non-resonant, we can separately average over the independent precession ($\phi$) and coupled nutation ($\theta$) and rotation ($\psi$) motions. Expressing this mathematically for the general variable $F$, we have,

\begin{equation}
{\langle\dot{F}\rangle}_\phi=\frac{1}{2\pi}\int_0^{2\pi}{\dot{F}(\phi,\theta,\psi)}d{\phi}
\label{eq:phiavg}
\end{equation} 
and
\begin{equation}
\dot{\overline{F}}=\frac{1}{P_\psi}\int_0^{P_\psi}{{\langle\dot{F}\rangle}_\phi}\Big(\theta(t),\psi(t)\Big)dt
\label{eq:psiavg}
\end{equation} 

To evaluate Eq.~\ref{eq:psiavg}, we leverage the complete elliptic integral of the first kind $K$ (see Appendix A) \citep{landau,numericalrecipes}. Rewriting Eq.~\ref{eq:psiavg} with the linearly scaled time variable $\tau$, noting that ${\Delta}t=P_{\psi}$ corresponds to ${\Delta}\tau=4K$, 
\begin{equation}
\dot{\overline{F}}=\frac{1}{4K}\int_0^{4K}{{\langle\dot{F}\rangle}_\phi}\Big(\theta(\tau),\psi(\tau)\Big)d\tau
\label{eq:psiavgK}
\end{equation} 
Averaging over $\tau$ involves the Jacobi elliptic functions $\cn\tau$, $\sn\tau$, and $\dn\tau$ (see the Appendices).

The tumbling-averaged equations of motion are then given by,

\begin{equation}
\dot{\overline{\alpha}}=\frac{\overline{M_y}+\overline{H}n\cos{\overline{\alpha}}\cos{\overline{\beta}}}{\overline{H}\sin{\overline{\beta}}}
\label{eq:alphadotavg}
\end{equation}
\begin{equation}
\dot{\overline{\beta}}=\frac{\overline{M_x}+\overline{H}n\sin{\overline{\alpha}}}{\overline{H}}
\label{eq:betadotavg}
\end{equation}
\begin{equation}
\dot{\overline{H}}=\overline{M_z}
\label{eq:Hdotavg}
\end{equation}

\begin{equation}
\dot{\overline{I}}_d=-\frac{2\overline{I}_d}{\overline{H}}\Bigg[\frac{\overline{I}_d-I_i}{I_i}\overline{a_{z1}M_1}+\frac{\overline{I}_d-I_s}{I_s}\overline{a_{z2}M_2}+\frac{\overline{I}_d-I_l}{I_l}\overline{a_{z3}M_3}\Bigg]
\label{eq:Iddotavg}
\end{equation}
\begin{equation}
\dot{\overline{\omega}}_e=\frac{1}{\overline{I_d}}\Bigg[\overline{M_z}-\frac{\overline{H}}{\overline{I_d}}\dot{\overline{I}}_d\Bigg]
\label{eq:wedotavg}
\end{equation}

\section{Non-Resonant Averaged YORP}

We must evaluate the six averaged torque components $\overline{M_x}$, $\overline{M_y}$, $\overline{M_z}$, $\overline{a_{z1}M_1}$, $\overline{a_{z2}M_2}$, and $\overline{a_{z3}M_3}$. To facilitate the analytical averaging, we follow Ref. \cite{cicalo} and approximate $\max(0,\bm{\hat{u}}\cdot\bm{\hat{n}}_i)$ using its second order Fourier series expansion,
\begin{equation}
\max(0,\bm{\hat{u}}\cdot\bm{\hat{n}}_i)\;{\approx}\; g_i = \frac{1}{3\pi}+\frac{1}{2}(\bm{\hat{u}}\cdot\bm{\hat{n}}_i)+\frac{4}{3\pi}(\bm{\hat{u}}\cdot\bm{\hat{n}}_i)^2
\label{eq:illuminationfunction}
\end{equation}
where, given our frame definitions, $u_x = -\sin{\beta}$, $u_y = 0$, and $u_z=\cos{\beta}$. So $\bm{\hat{u}}\cdot\bm{\hat{n}} = {u_x}n_x+{u_z}n_z$. 

With this approximation,
\begin{singlespace}
\begin{equation}
\renewcommand{\arraystretch}{1.5}
\begin{bmatrix}
\overline{M_x} \\ \overline{M_y} \\ \overline{M_z} 
\end{bmatrix} = -P_{SRP}{\sum_{i=1}^{n_f}}\Bigg[c_{si}\overline{(\bm{\hat{u}}\cdot\bm{\hat{n}}_i)g_i\bm{d}_i}+c_{ai}\overline{g_i\bm{r}_i\times\bm{\hat{u}}}+c_{di}\overline{g_i\bm{d}_i}\Bigg]A_i
\label{eq:HM}
\end{equation}
\end{singlespace}
\noindent where $\bm{d}_i = \bm{r}_i\times\bm{\hat{n}}_i$ and the constants $c_{si}=2{\rho_i}s_i$ and $c_{ai} = (1-{\rho_i}s_i)$.

From Eqs.~\ref{eq:BH} and \ref{eq:Hvec}, we see that all $\mathcal{H}$ frame $x$ and $y$ vector components will contain either $\cos{\phi}$ or $\sin{\phi}$. So products with odd combined powers of $x$ and $y$ will average to zero over $\phi$. Expanding Eq.~\ref{eq:HM}, including only non-zero terms, and dropping the $i$th facet indices from the averaged products for brevity, $\overline{M_x}$, $\overline{M_y}$, and $\overline{M_z}$ are then given by, 

\begin{equation}
\begin{split}
\overline{M_x}= -P_{SRP}{\sum_{i=1}^{n_f}}\Bigg[&
u_x\Big(\frac{1}{2}c_{di} + \frac{1}{3\pi}c_{si}\Big)\overline{d_xn_x} 
+u_xu_z\Big(c_{si} + \frac{8}{3\pi}c_{di}\Big)\overline{d_xn_xn_z} 
+ \frac{4}{3\pi}c_{si}u_x^3\overline{d_xn_x^3}   \\ & 
+ \frac{4}{\pi}c_{si}u_xu_z^2\overline{d_xn_xn_z^2} 
+ \frac{1}{2}c_{ai}u_xu_z\overline{r_yn_x} 
+ \frac{8}{3\pi}c_{ai}u_xu_z^2\overline{r_yn_xn_z}\Bigg]A_i
\end{split}
\label{eq:Mx}
\end{equation}
\begin{equation}
\begin{split}
\overline{M_y}=-P_{SRP}{\sum_{i=1}^{n_f}}\Bigg[&
u_x\Big(\frac{1}{2}c_{di} + \frac{1}{3\pi}c_{si}\Big)\overline{d_yn_x}
+u_xu_z\Big(c_{si} +\frac{8}{3\pi}c_{di}\Big)\overline{d_yn_xn_z}
+ \frac{4}{3\pi}c_{si}u_x^3\overline{d_yn_x^3} \\ &  
+ \frac{4}{\pi}c_{si}u_xu_z^2\overline{d_yn_xn_z^2} 
+\frac{1}{3\pi}c_{ai}u_x\overline{r_z}
-\frac{1}{2}c_{ai}u_xu_z\overline{r_xn_x}
+ \frac{1}{2}c_{ai}u_xu_z\overline{r_zn_z} \\ & 
- \frac{8}{3\pi}c_{ai}u_xu_z^2\overline{r_xn_xn_z} 
+ \frac{4}{3\pi}c_{ai}u_x^3\overline{r_zn_x^2} 
+ \frac{4}{3\pi}c_{ai}u_xu_z^2\overline{r_zn_z^2}\Bigg]A_i
\end{split}
\label{eq:My}
\end{equation}
\begin{equation}
\begin{split}
\overline{M_z}=-P_{SRP}{\sum_{i=1}^{n_f}}\Bigg[&
\frac{1}{3\pi}c_{di}\overline{d_z} 
+ u_z\Big(\frac{1}{2}c_{di} + \frac{1}{3\pi}c_{si}\Big)\overline{d_zn_z}
+ \Big(\frac{1}{2}c_{si} + \frac{4}{3\pi}c_{di}\Big)\Big(u_x^2\overline{d_zn_x^2}  
+ u_z^2\overline{d_zn_z^2}\Big) \\ & 
+ \frac{4}{\pi}c_{si}u_x^2u_z\overline{d_zn_x^2n_z} 
+ \frac{4}{3\pi}c_{si}u_z^3\overline{d_zn_z^3} 
-\frac{1}{2}c_{ai}u_x^2\overline{r_yn_x} 
- \frac{8}{3\pi}c_{ai}u_x^2u_z\overline{r_yn_xn_z}\Bigg]A_i
\end{split}
\label{eq:Mz}
\end{equation}
Solutions for the various averaged quantities in Eqs.~\ref{eq:Mx}, \ref{eq:My}, and \ref{eq:Mz} are provided in Appendix B. Note that these quantities are implicitly dependent on $\overline{I}_d$.

The terms $\overline{a_{z1}M_1}$, $\overline{a_{z2}M_2}$, and $\overline{a_{z3}M_3}$ 
are given by,
\begin{equation}
\begin{split}
\overline{a_{z*}M_*}=-P_{SRP}{\sum_{i=1}^{n_f}}\Bigg[&
\frac{1}{3\pi}c_{di}\overline{a_{z*}d_*}
+ u_z\Big(\frac{1}{2}c_{di} + \frac{1}{3\pi}c_{si}\Big)\overline{a_{z*}d_*n_z} \\ &
+ \Big(\frac{1}{2}c_{si} + \frac{4}{3\pi}c_{di}\Big)\Big(u_x^2\overline{a_{z*}d_*n_x^2}  
+ u_z^2\overline{a_{z*}d_*n_z^2}\Big) \\ & 
+ \frac{4}{\pi}c_{si}u_x^2u_z\overline{a_{z*}d_*n_x^2n_z} 
+ \frac{4}{3\pi}c_{si}u_z^3\overline{a_{z*}d_*n_z^3} 
+c_{ai}\overline{ga_{z*}\delta_*}\Bigg]A_i
\end{split}
\label{eq:az*M*}
\end{equation}
where $*=1,2,3$. Also, $\delta_1=(r_2u_3-r_3u_2)$, $\delta_2=(r_3u_1-r_1u_3)$, and $\delta_3=(r_1u_2-r_2u_1)$. To calculate $\overline{a_{z*}d_*}$, $\overline{a_{z*}{d_*}n_z}$, etc., we note that $d_z = a_{z1}d_1+a_{z2}d_2+a_{z3}d_3$. From the averaged Appendix B equations that include $d_z$ (Eqs.~\ref{eq:l_fz}, \ref{eq:l_fznz}, \ref{eq:l_fznx2}, \ref{eq:l_fznz2}, \ref{eq:l_fznx2nz}, \ref{eq:l_fznz3} for LAMs and Eqs.~\ref{eq:s_fz}, \ref{eq:s_fznz}, \ref{eq:s_fznx2}, \ref{eq:s_fznz2}, \ref{eq:s_fznx2nz}, \ref{eq:s_fznz3} for SAMs), we retain just the terms containing $d_*$. Solutions for $\overline{g{a_{z*}}\delta_*}$ are provided separately in Appendix B. Overall, Eqs.~\ref{eq:Mx} - \ref{eq:az*M*} depend on $\overline{I_d}$ and $\overline{\beta}$ but are independent of $\overline{\alpha}$ and $\overline{H}$. 

\subsection{Averaged Model Validation}

To gain insight about the YORP-driven behavior of the full dynamics model, we now investigate the tumbling-averaged model. First, we will validate the analytically averaged torques using the full torque-free dynamics model (Eqs.~\ref{eq:wdot} - \ref{eq:M}). For the full model, we numerically average Eq.~\ref{eq:M} over time using trapezoidal integration and use $\max(0,\bm{\hat{u}}\cdot\bm{\hat{n}}_i)$ rather than its 2nd order Fourier series approximation. The full model is averaged for ${\Delta}t=200P_e$ where again $P_e=2\pi/\omega_e$. This span is more than sufficient for the time-averaged torques to converge. 

\begin{figure}[H]
	\centering
\subcaptionbox{SAM+ $I_d=3500$ $\mathrm{kg{\cdot}m^2}$}{\includegraphics[width=3in]{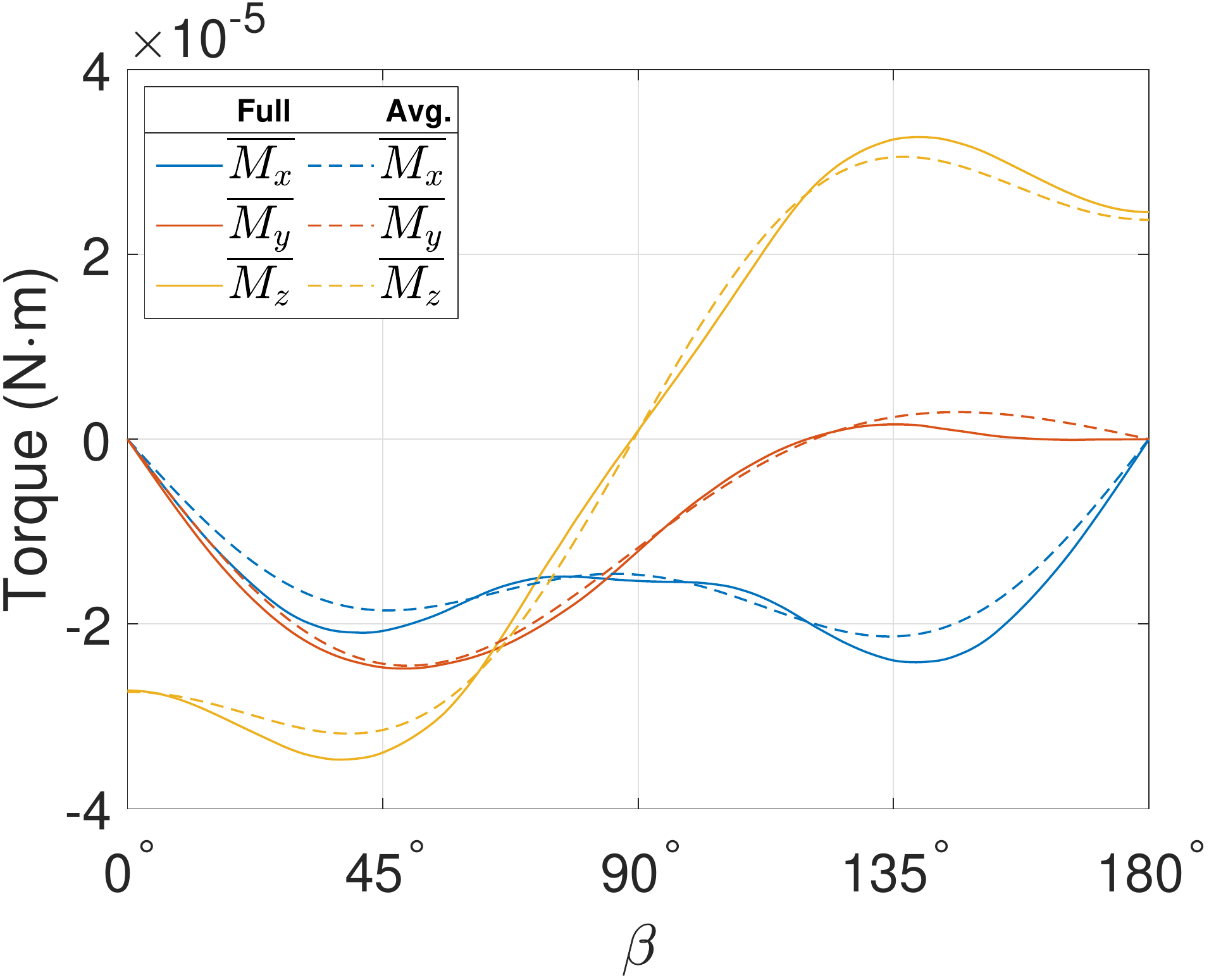}}
\subcaptionbox{LAM+ $I_d=3000$ $\mathrm{kg{\cdot}m^2}$}{\includegraphics[width=3in]{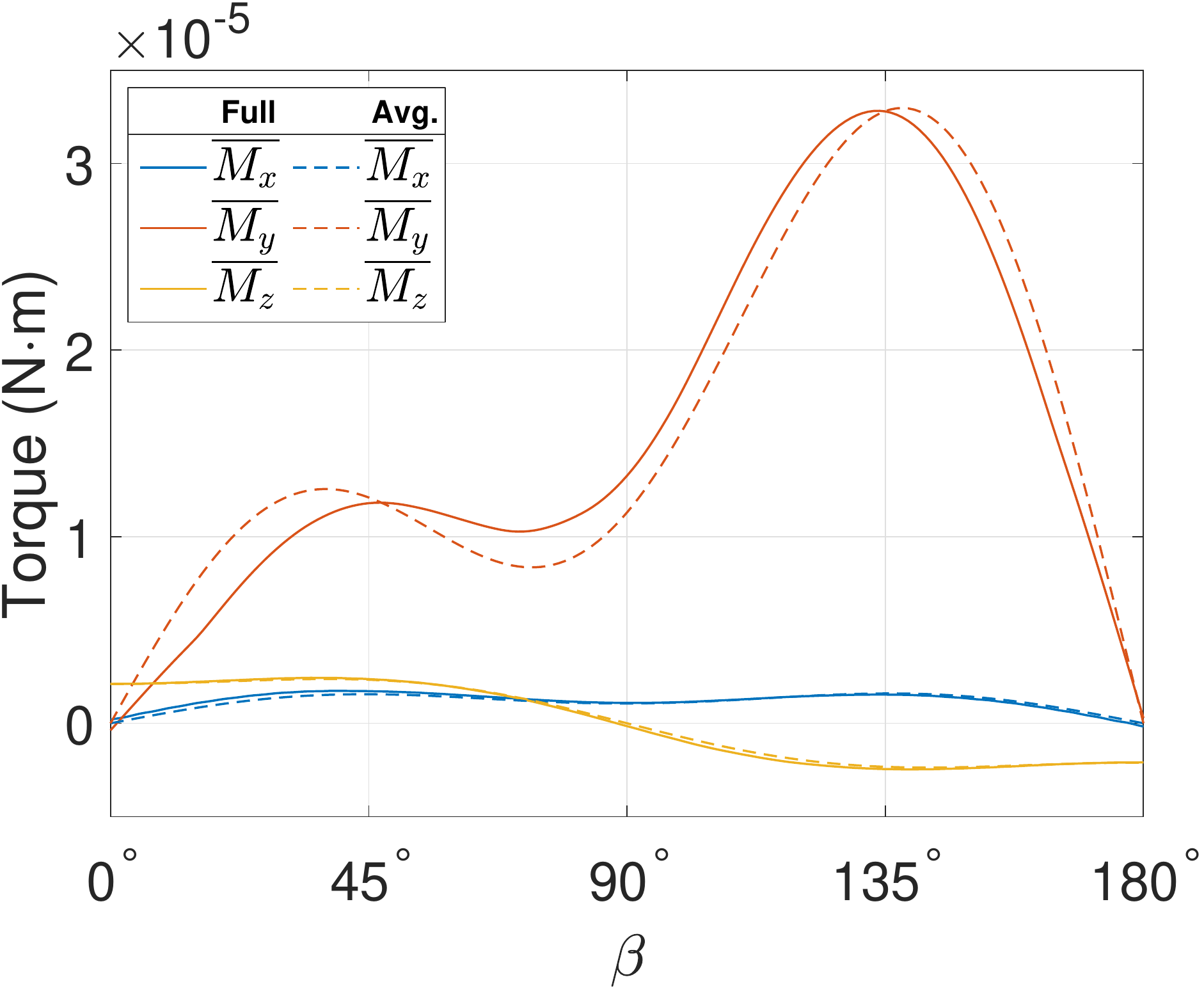}}
\subcaptionbox{SAM- $I_d=3500$ $\mathrm{kg{\cdot}m^2}$}{\includegraphics[width=3in]{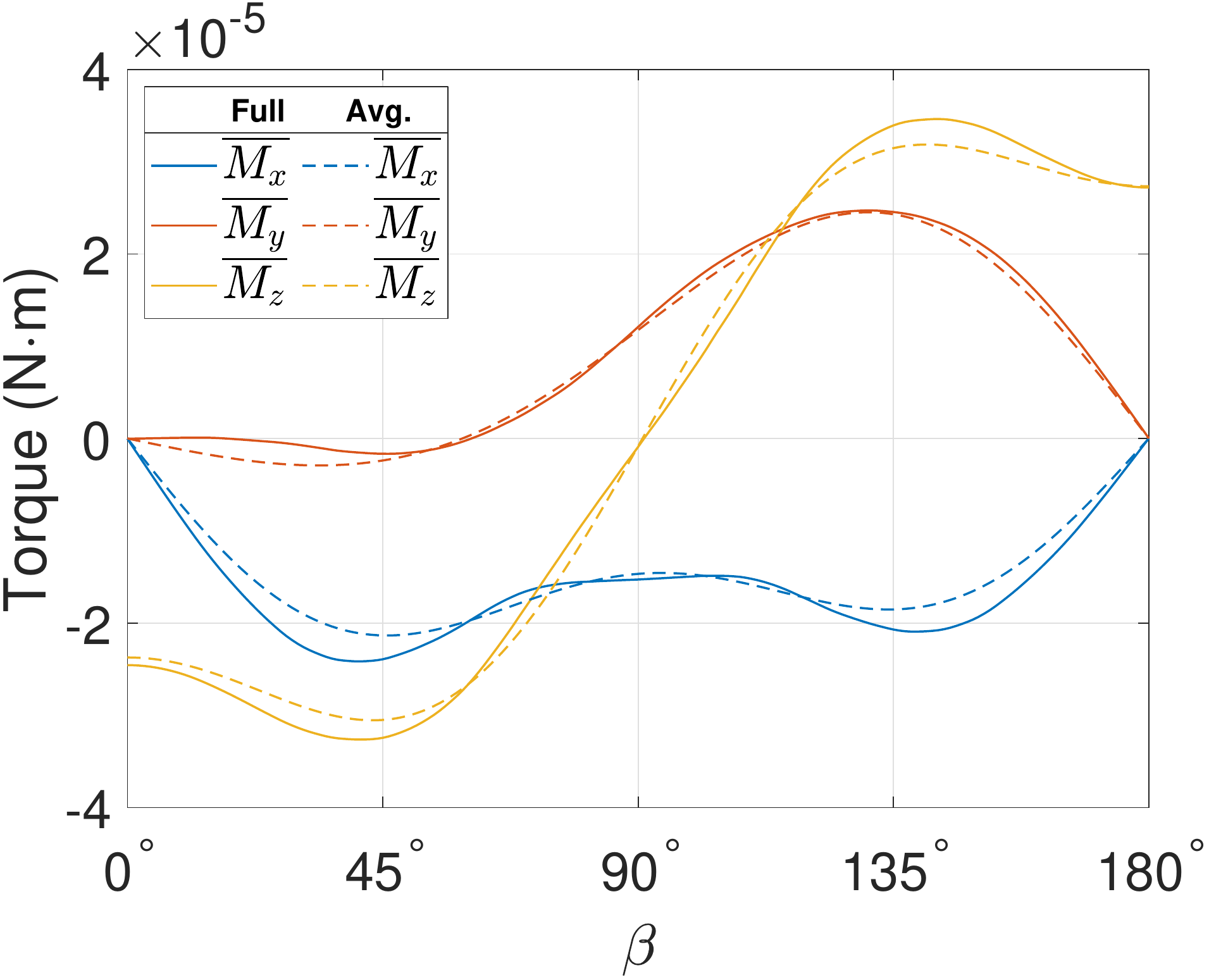}}
\subcaptionbox{LAM- $I_d=3000$ $\mathrm{kg{\cdot}m^2}$}{\includegraphics[width=3in]{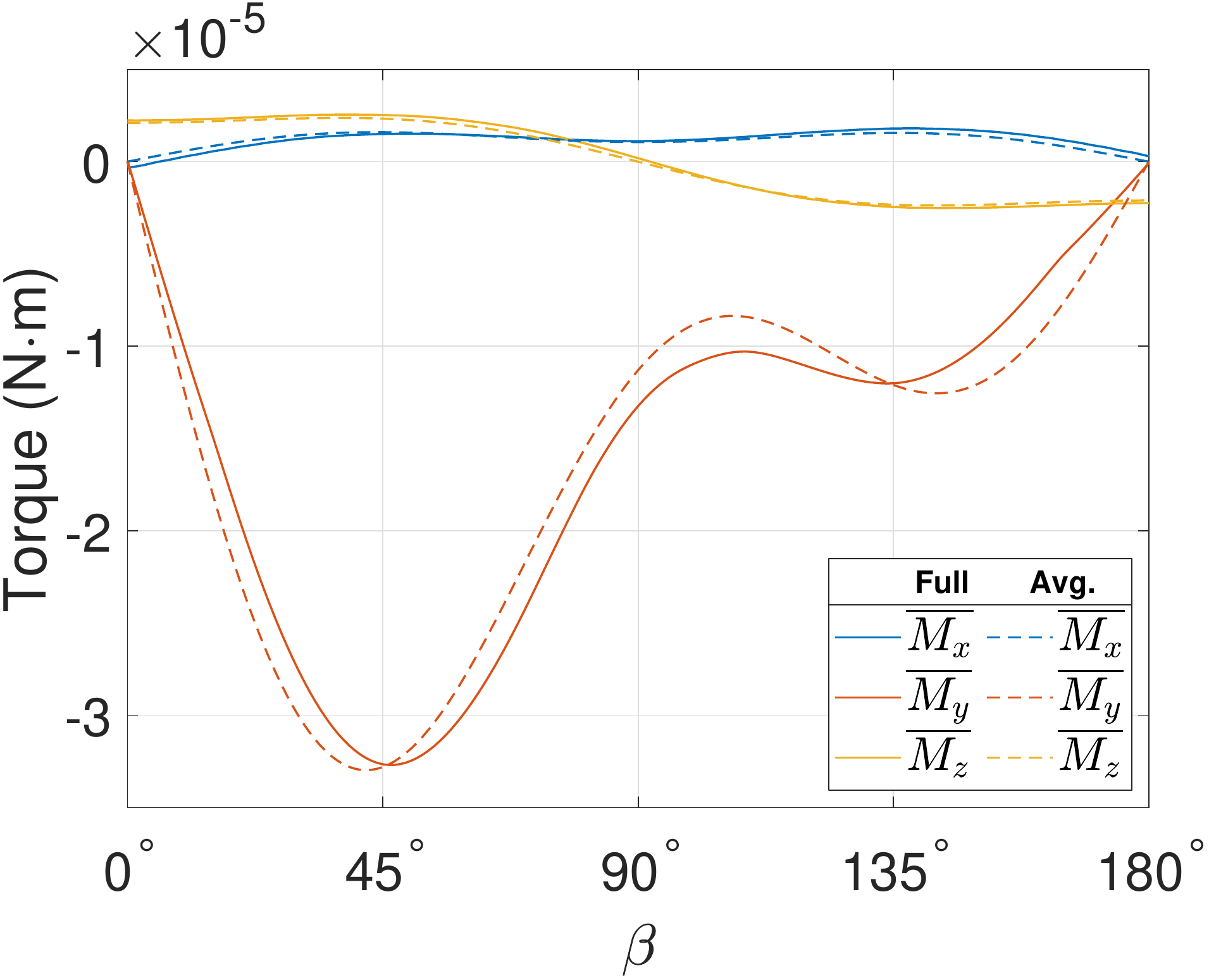}}
	\caption{Comparison of full and analytically averaged torques for GOES 8 in the $\mathcal{H}$ frame. The full model is solid and the analytically averaged model is dashed.}
\label{fig:avg_torques}
\end{figure}

\begin{figure}[h]
	\centering
\subcaptionbox{SAM+ $I_d=3500$ $\mathrm{kg{\cdot}m^2}$}{\includegraphics[width=3in]{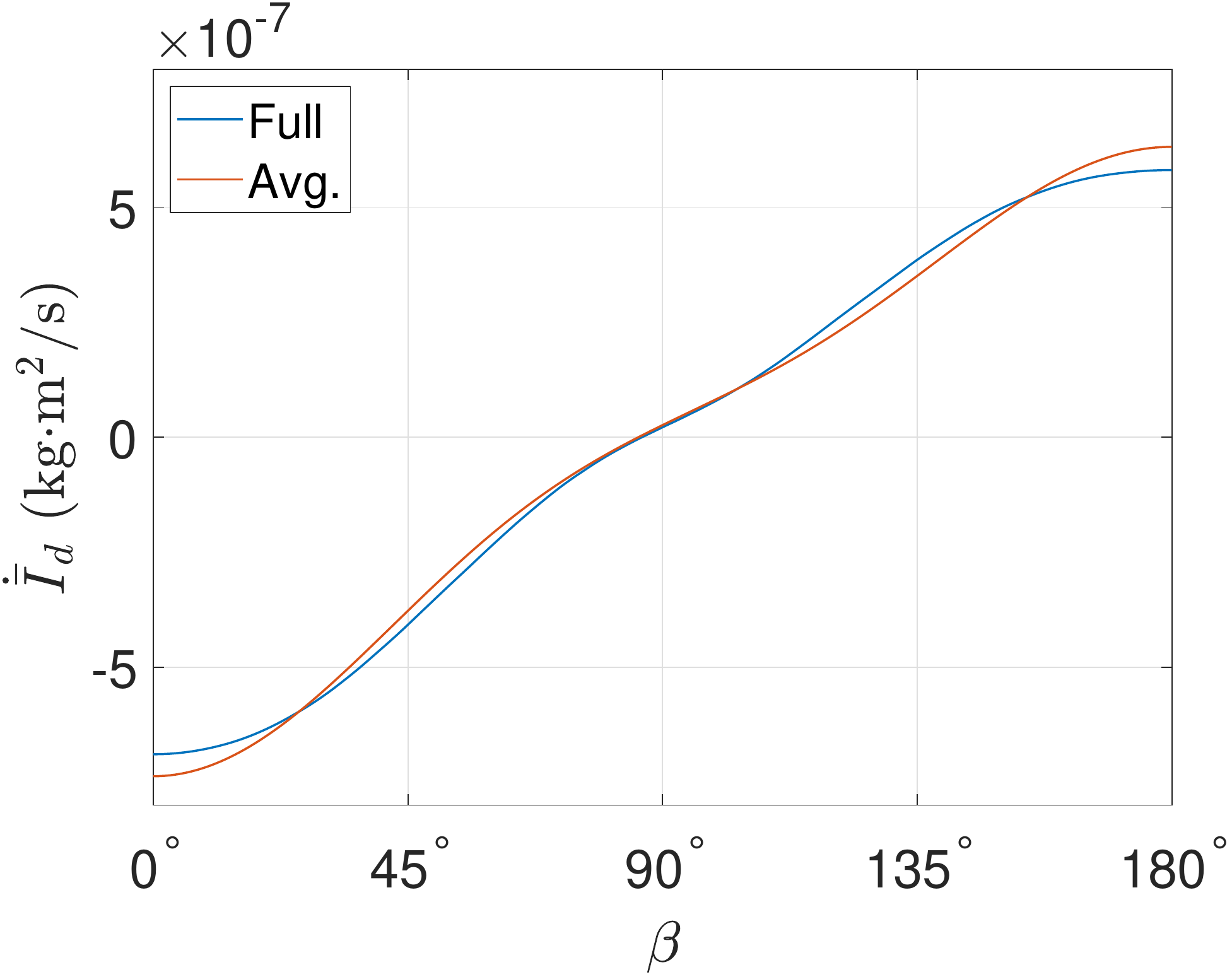}}
\subcaptionbox{LAM+ $I_d=3000$ $\mathrm{kg{\cdot}m^2}$}{\includegraphics[width=3in]{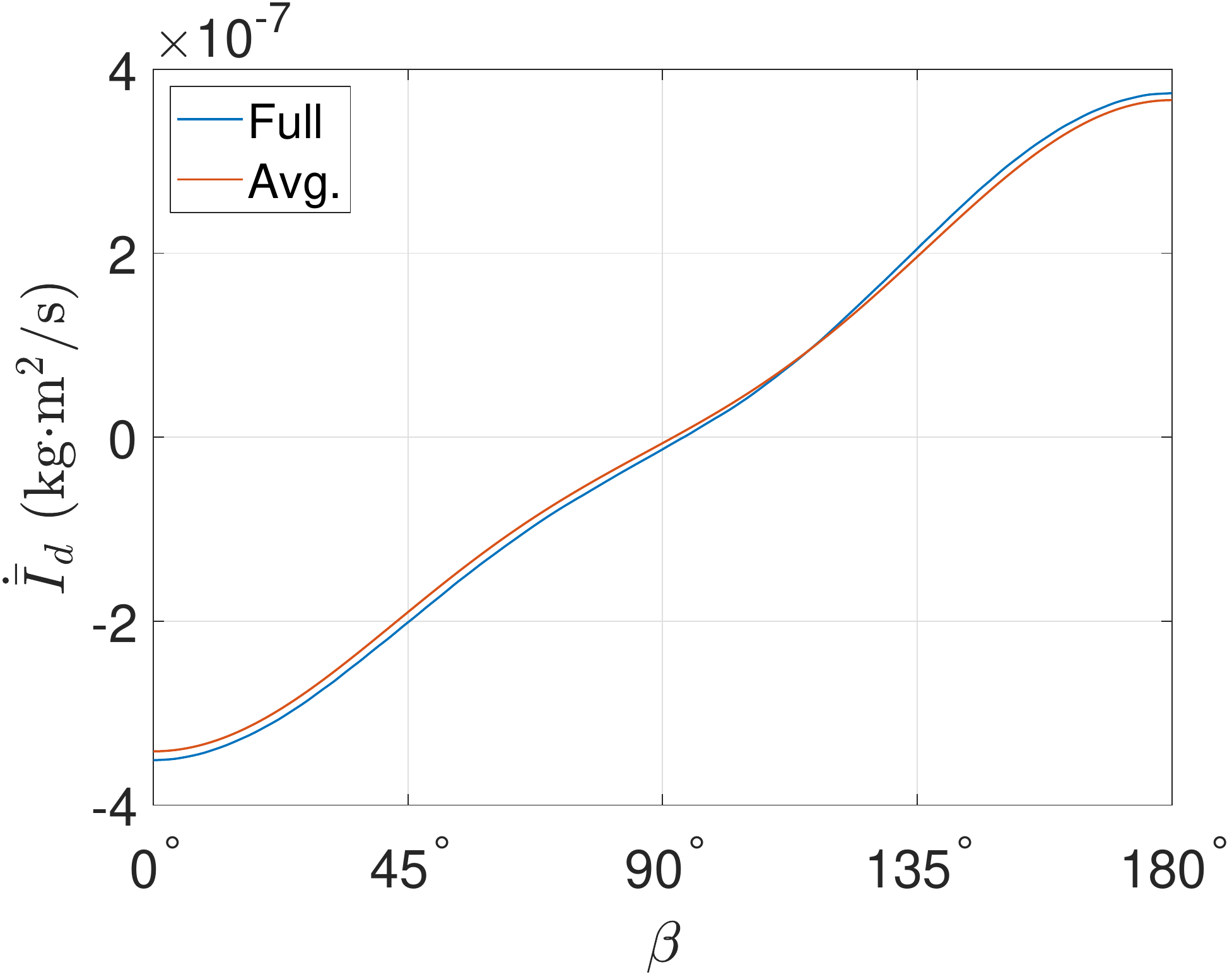}}
	\caption{GOES 8 $\dot{\overline{I}}_d$ vs. $\beta$.}
\label{fig:avg_Iddot}
\end{figure}

Figure~\ref{fig:avg_torques} shows the average torques in the $\mathcal{H}$ frame for the full and analytically averaged models. Both SAM and LAM states are tested. We see that in all cases, the models only differ quantitatively, sharing the same general structure. For the SAM cases, we see that $\overline{M_z}$ is negative for $\beta<$ 90$^{\circ}$ and positive for $\beta>$ 90$^{\circ}$. So the satellite will spin down when $\beta<$  90$^{\circ}$. Also, $\overline{M_x}\;{\leq}\;0$ across all $\beta$, so $\bm{H}$ will tend to be pushed towards the sun line. For the LAM cases in Figure~\ref{fig:avg_torques}, $\overline{M_y}$ has the largest magnitude of the three torque components. $\overline{M_y}$ drives $\dot{\overline{\alpha}}$ and therefore precession of $\bm{H}$ around the sun line. The precession rate $\dot{\overline{\alpha}}$ varies significantly with $\beta$. Also, $\overline{M_x}\;{\geq}\;0$ for all $\beta$, pushing $\bm{H}$ away from the sun line. $\overline{M_z}$ changes sign at $\beta=$ 90$^{\circ}$, so the satellite will first spin up and then down as $\beta$ increases. Continuing the comparison, Figure~\ref{fig:avg_Iddot} shows $\dot{\overline{I}}_d$ for the full and analytically averaged models assuming an arbitrary $\omega_e=$ 2$\pi$ rad/s. Again, they differ only quantitatively. We see that for both the SAM and LAM states the satellite will be pushed towards more excited tumbling (smaller $I_d$) for $\beta<$ 90$^{\circ}$ and towards uniform rotation (larger $I_d$) for $\beta>$ 90$^{\circ}$. $\dot{\overline{I}}_d$ solutions for LAM/SAM $-$ were virtually indistinguishable from the $+$ solutions and have been excluded from Figure~\ref{fig:avg_Iddot} for brevity. Overall, the $+/-$ solutions for both LAMs and SAMs differ insignificantly for all components except $\overline{M_y}$, where the solution is mirrored around $\beta=90^{\circ}$ and has an opposite sign.  So for the $+$ and $-$ solutions,  $\dot{\overline{\alpha}}$ will have opposite signs and $\bm{H}$ will precess about the sun line in opposite directions. This symmetric structure is due to the particular satellite geometry. For a fixed $\bm{H}$, the $+/-$ LAM/SAM spin states essentially flip the satellite model 180$^{\circ}$ while maintaining the same inertial precession direction. As a result, some averaged torque contributions from the GOES solar array will change for $+/-$ LAM/SAM due to different reflective properties for the front and back array faces. On the other hand, contributions from the axisymmetric solar sail will not change.


\begin{figure}[H]
	\centering
	\subcaptionbox{Effective Spin Rate}{\includegraphics[width=2.8in]{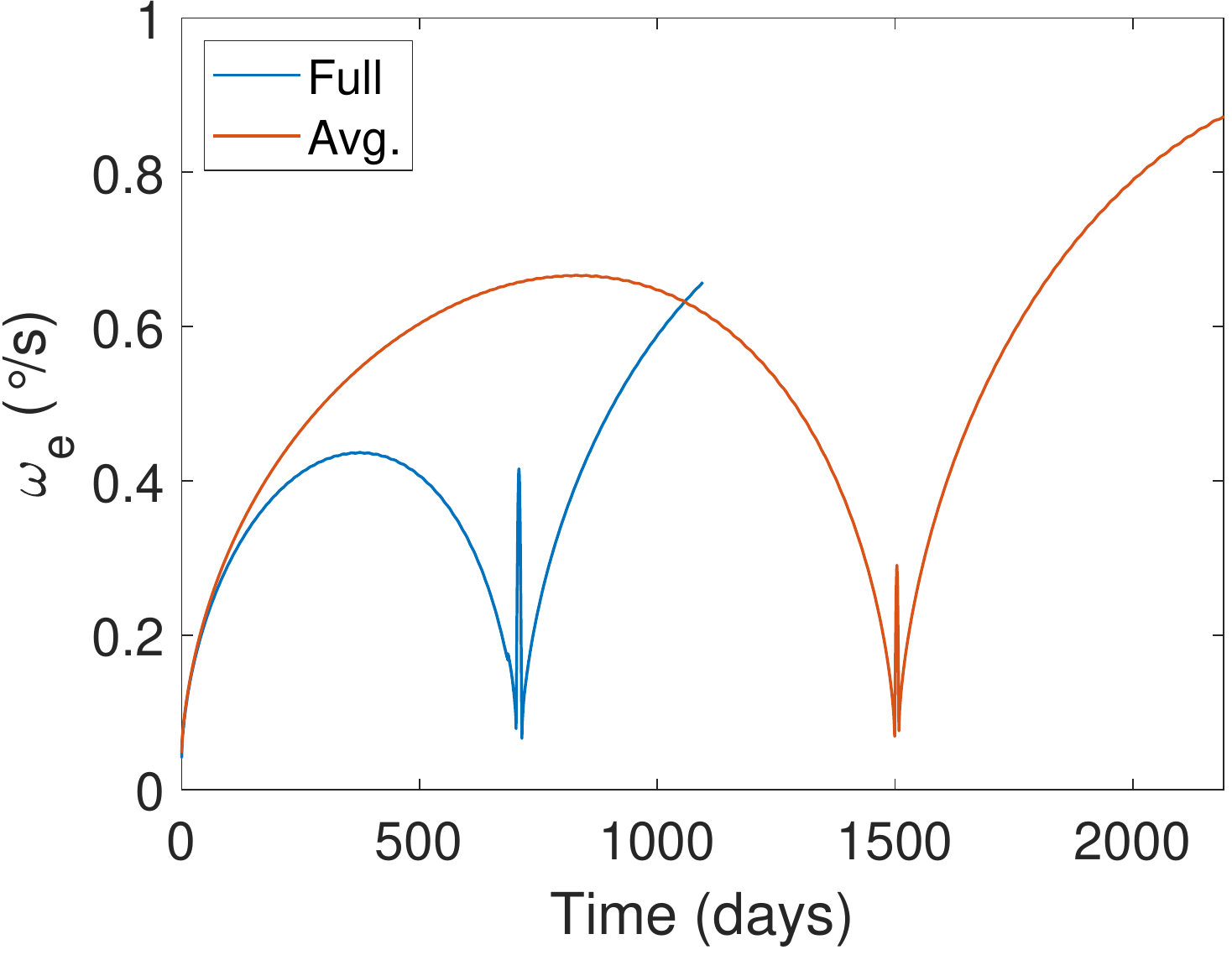}}
	\subcaptionbox{Scaled Dynamic Moment of Inertia}{\includegraphics[width=2.8in]{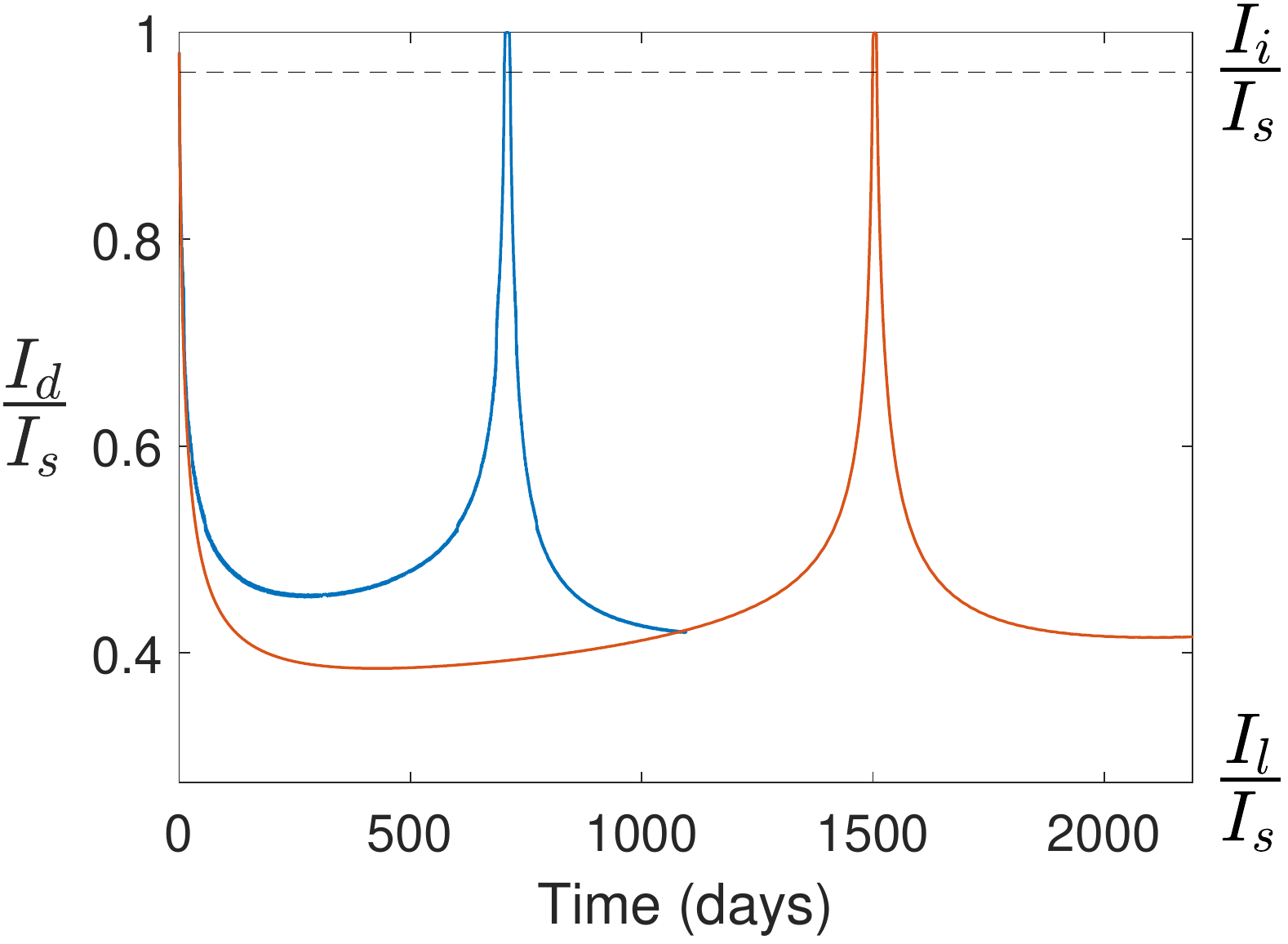}}
	\subcaptionbox{Clocking Angle}{\includegraphics[width=2.8in]{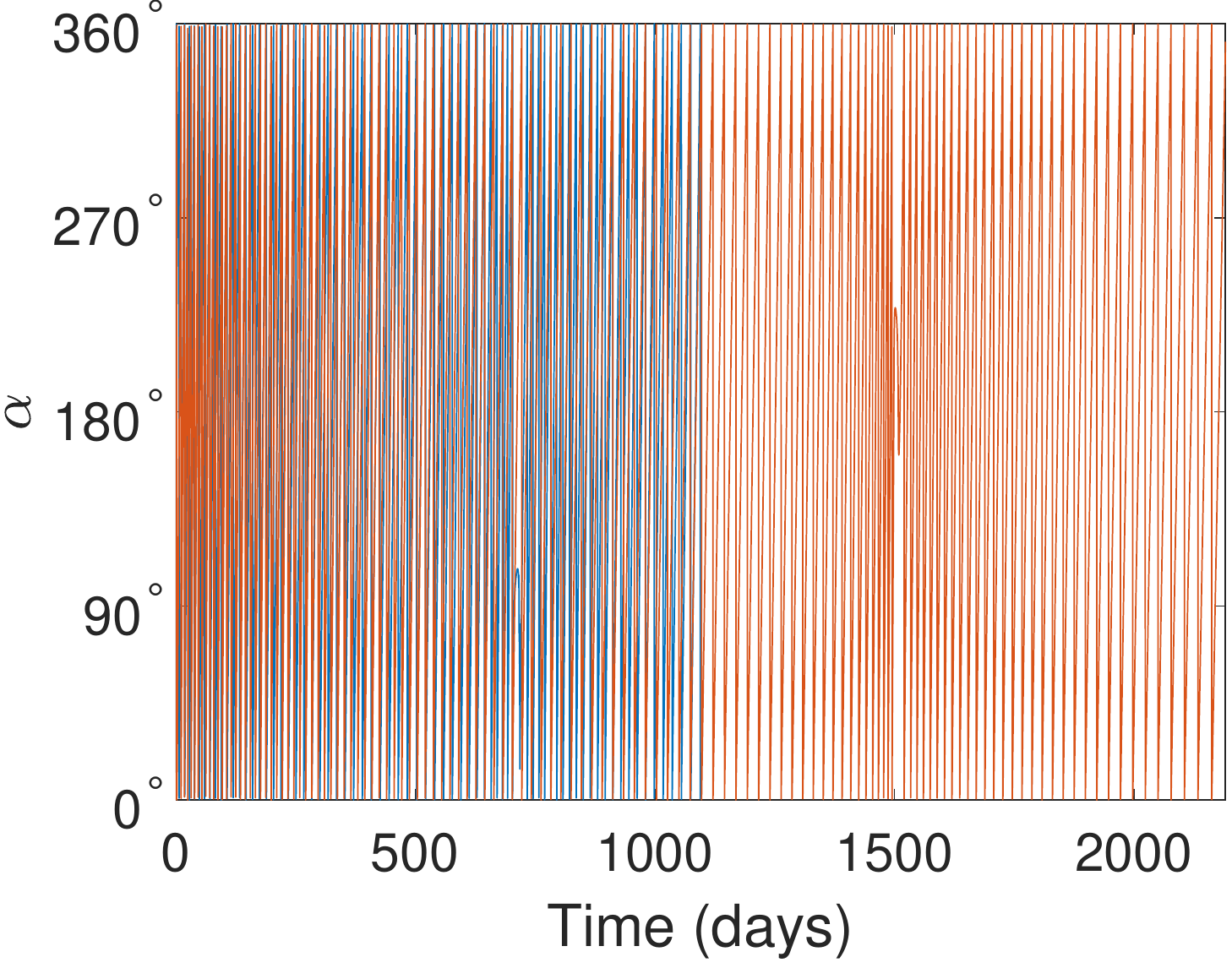}}	
	\subcaptionbox{Angle between $\bm{H}$ and $\bm{\hat{u}}$}{\includegraphics[width=2.8in]{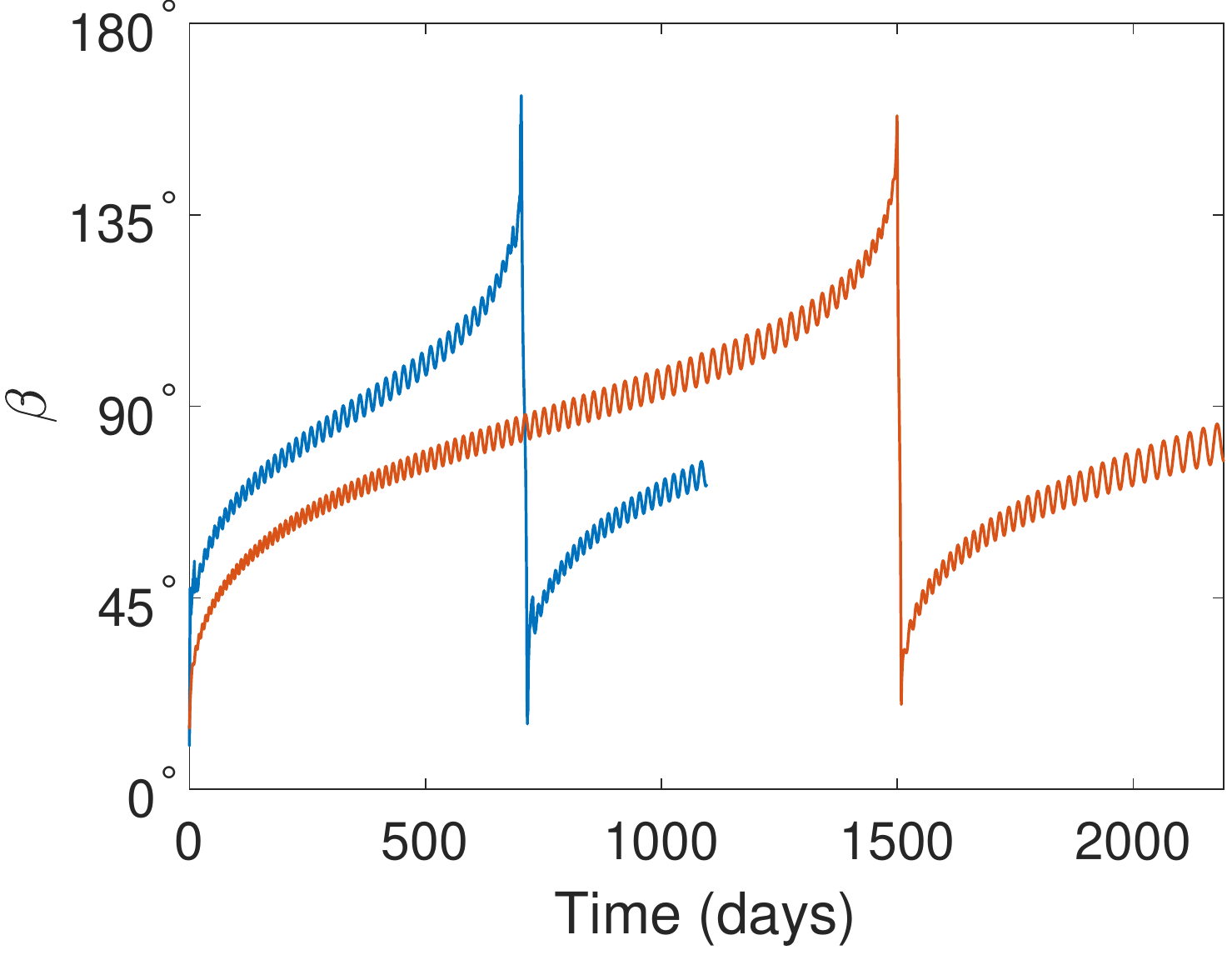}}	
	\caption{GOES 8 full and averaged dynamical evolution (initial conditions: $\overline{\alpha}=0^{\circ}$, $\overline{\beta}=15^{\circ}$, $\overline{I}\!_{d}=$ 3500 kg${\cdot}$m$^2$ SAM+, and $P_e=2\pi/\overline{\omega}_e=$ 120 min).}
\label{fig:evol_comp}
\end{figure}

We will now compare the dynamical evolution for the full and analytically averaged models by numerically integrating Eqs.~\ref{eq:alphadotavg} - \ref{eq:Iddotavg}. For both models, the same initial spin state is prescribed with $\overline{\alpha}=0^{\circ}$, $\overline{\beta}=15^{\circ}$, $\overline{I}\!_{d}=$ 3500 kg${\cdot}$m$^2$ (SAM+), and $P_e=$ 120 min.  Using MATLAB's ode113 numerical integrator with 1e-12 absolute and relative tolerances for both models, the full model was propagated for three years and the averaged model for six to show at least one tumbling cycle. The resulting evolution is provided in Figure~\ref{fig:evol_comp}. We see that the trends in the two models agree, but tumbling cycle times differ considerably with the full model progressing through the first tumbling cycle in roughly 700 days while the averaged model takes 1500 days. As the full model first passes through the 2:1 and 1:1
tumbling resonances, it is perturbed similarly to run 2 in Fig.~\ref{fig:run2_evol}. These
perturbing resonances may explain the initial jump in $\beta$ and advancement
in the tumbling cycle compared to the averaged model which
does not account for resonances. Another contributing factor to this difference is that $\overline{M_x}$ is slightly smaller for the averaged model than for the full model when $\beta<$ 90$^{\circ}$ (see Figure~\ref{fig:avg_torques}b.). This causes $\beta$ for the averaged solution to evolve more slowly, allowing $\omega_e$ (and $H$) more time to increase. In Figure~\ref{fig:evol_comp}a, the peak $\omega_e$ is 50$\%$ larger for the averaged model than the full model. The added pole "stiffness" provided by this larger spin rate further slows $\beta$ evolution for the averaged model compared to the full model. Artificially increasing the average model $\overline{M_x}$ by 20$\%$, approximately the difference between $\overline{M_x}$ for the two models, brought the averaged model's tumbling cycle time into agreement with  the full model. 

While the full and averaged models provide quantitatively different results due to our averaging assumptions (most notably the neglect of resonances and the illumination function approximation), the averaged model replicates the tumbling cycles and sun-tracking behavior of the full model. Furthermore, for the Figure~\ref{fig:evol_comp} example, the total averaged model computation time was 7 seconds, compared to 70 minutes for the full model's three year propagation. This roughly three order of magnitude decrease in computation time was consistently observed for the averaged model runs. 

\subsection{Averaged YORP-Driven Evolution}
\subsubsection{Uniform to Tumbling Transition}

\begin{figure}[H]
	\centering
	\subcaptionbox{Effective Spin Rate}{\includegraphics[width=2.8in]{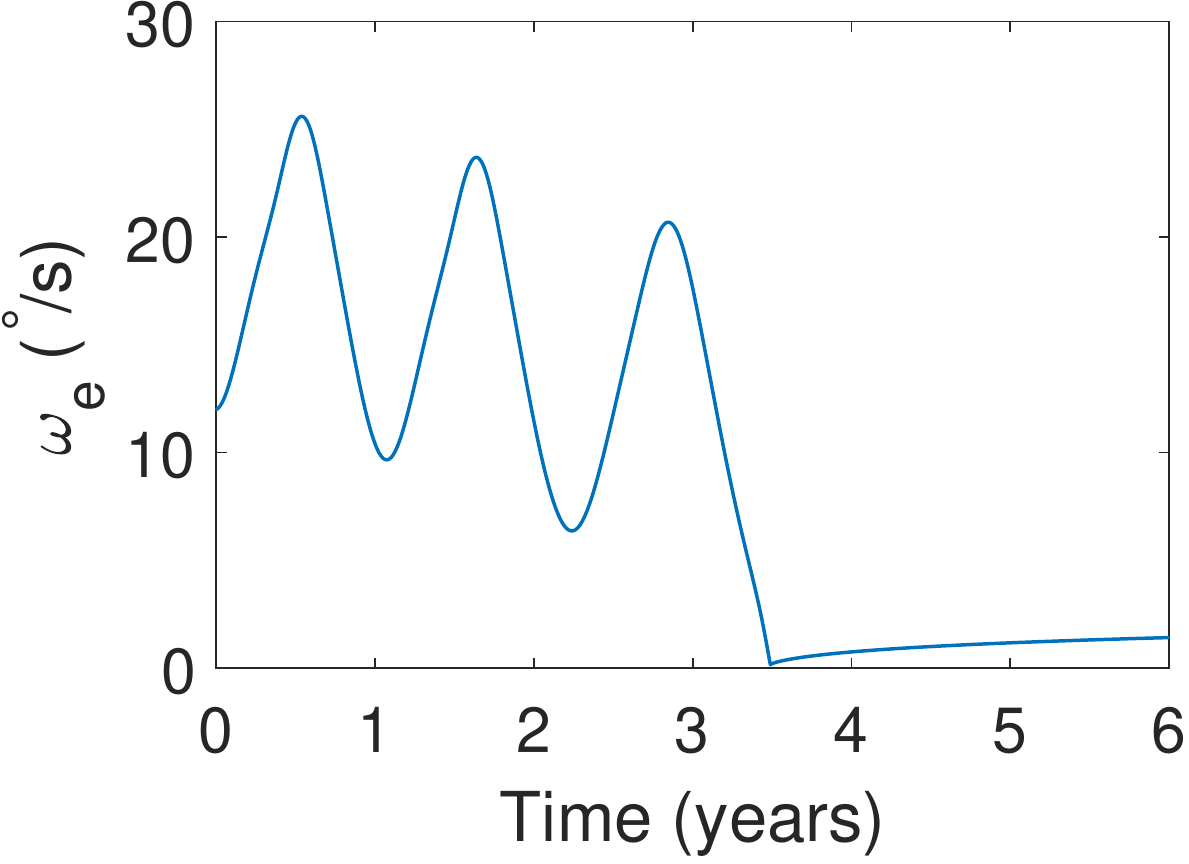}}
	\subcaptionbox{Scaled Dynamic Moment of Inertia}{\includegraphics[width=2.8in]{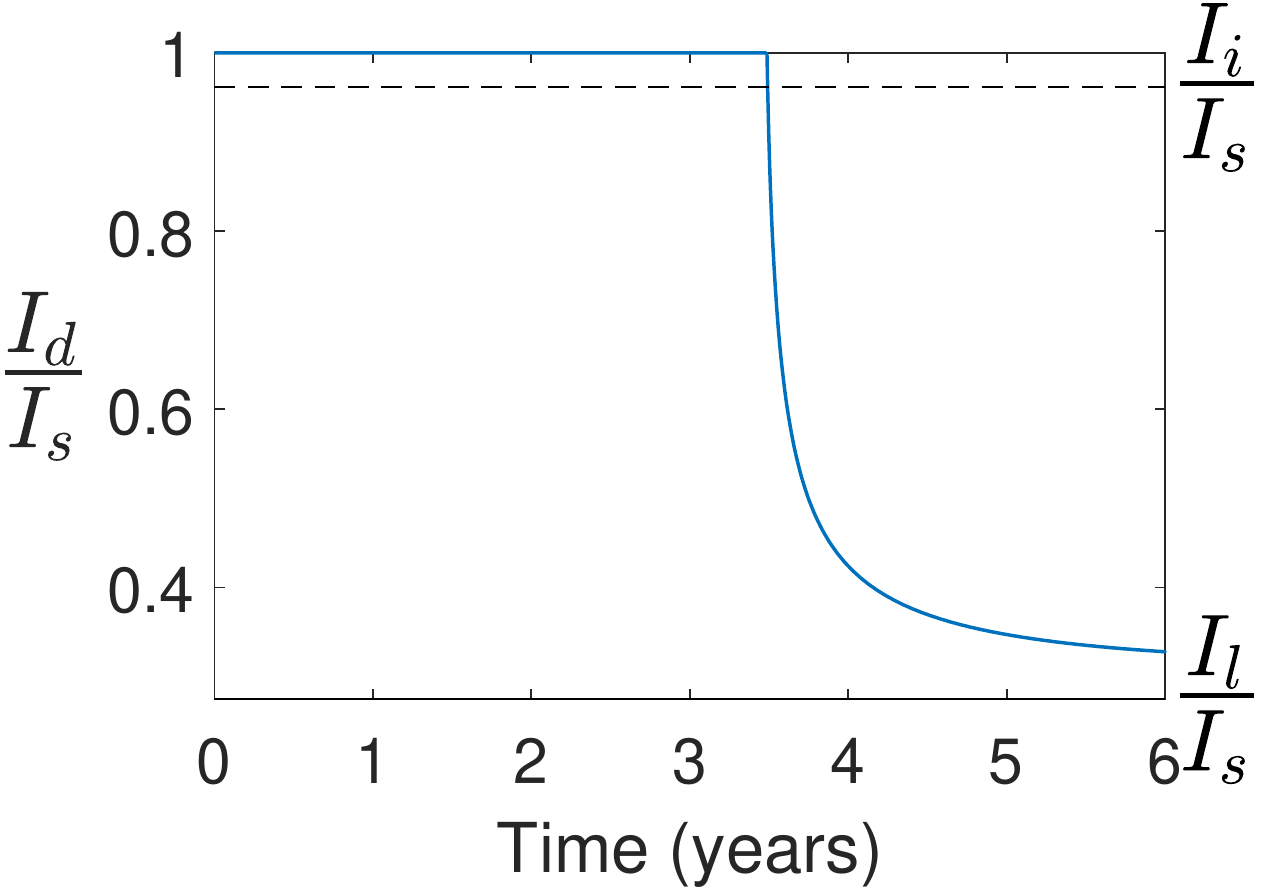}}
	\subcaptionbox{Clocking Angle}{\includegraphics[width=2.8in]{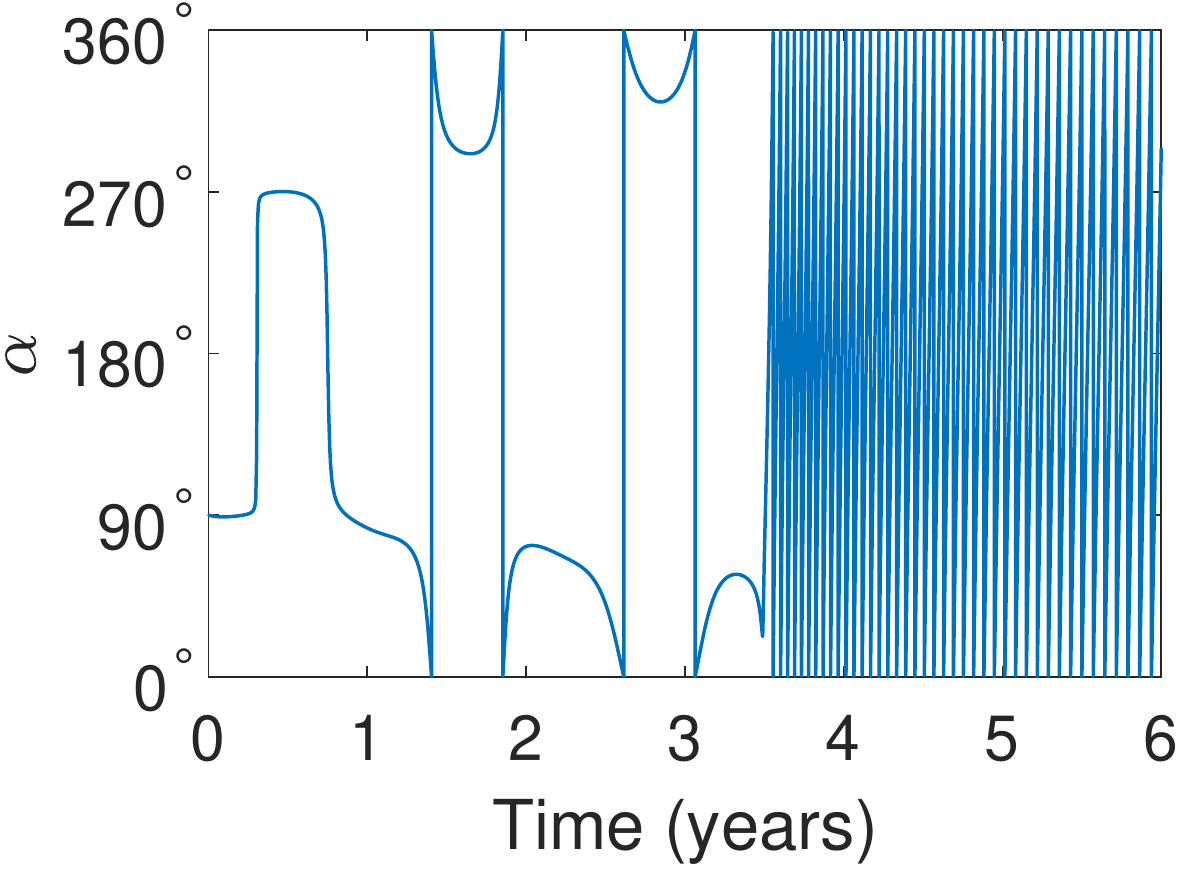}}	
	\subcaptionbox{Angle between $\bm{H}$ and $\bm{\hat{u}}$}{\includegraphics[width=2.8in]{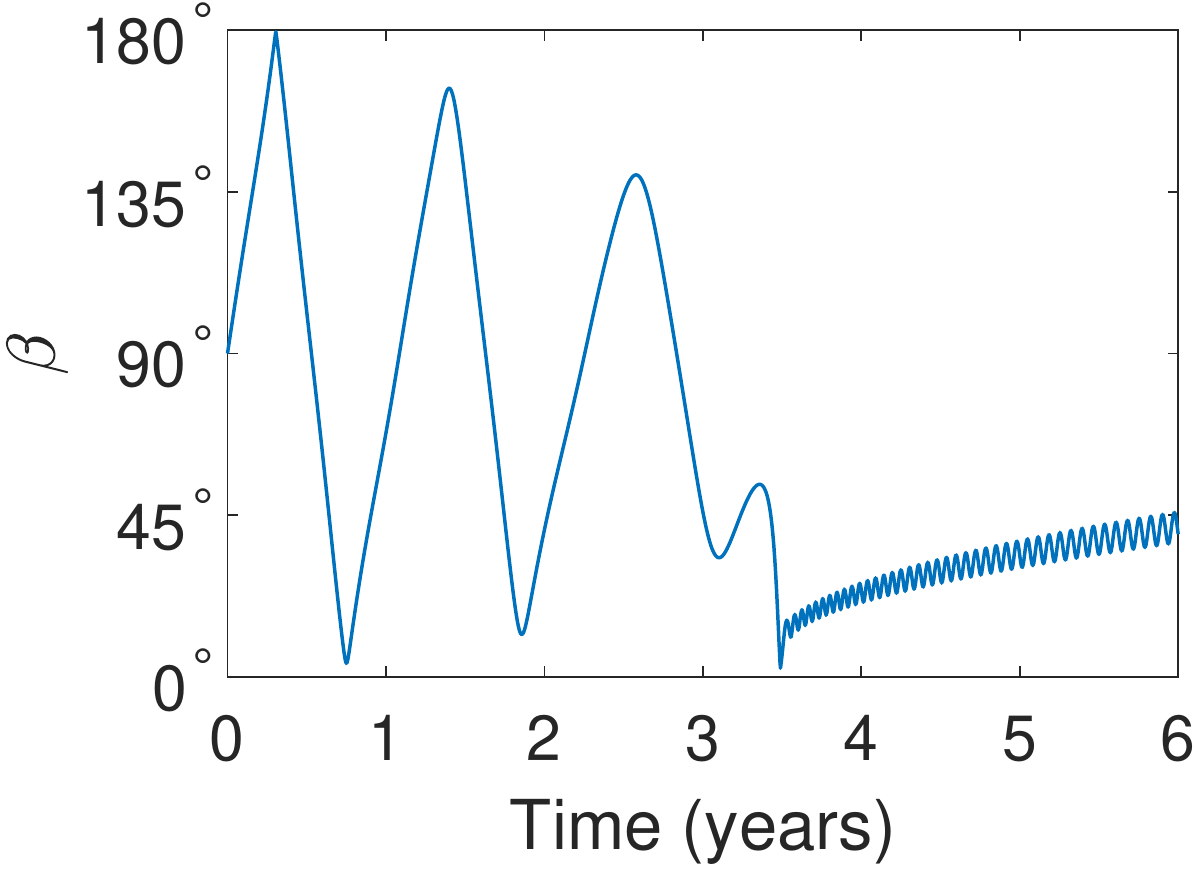}}	
	\caption{Averaged model transition from uniform rotation to tumbling for GOES 8 (initial conditions: $\overline{\alpha}=$ 90$^{\circ}$, $\overline{\beta}=$ 90$^{\circ}$, $P_e=$ 30 s, $\overline{I}_d/I_s\;{\approx}\;1-10^{-9}$).}
\label{fig:uniform_tumbling_transition}
\end{figure}

The tumbling-averaged model essentially extends the uniform spin-averaged model explored by Refs. \cite{scheeres2007,albuja2015,albuja2018,benson2020b} to general tumbling motion. Being much faster than the full dynamics model, the tumbling-averaged model readily allows for exploration of long-term uniform rotation and the transition to tumbling. Figure~\ref{fig:uniform_tumbling_transition} shows the six year evolution for GOES 8 starting in nearly uniform major axis rotation. Here we assume an initial $P_e=$ 30 s and long axis rotation angle amplitude $\psi_{\mathrm{max}}=$ 0.01$^{\circ}$. Referencing \cite{sa1991}, this $\psi_{\mathrm{max}}$ corresponds to $I_d/I_s\;{\approx}\;1-10^{-9}$. This slight negative offset from uniform rotation prevents $I_d$ from exceeding $I_s$ during propagation due to truncation error. For the first 3.5 years, the satellite remains in uniform rotation and exhibits a roughly one year periodicity in $\omega_e$. This is due to $\overline{M_z}$ and $\dot{\overline{\omega}}_e$ changing sign at $\beta=$ 90$^{\circ}$ (see Figure~\ref{fig:avg_torques}a and Figure~\ref{fig:eom_contours_17}c) as $\bm{H}$ remains nearly inertially fixed due to the fast spin rate. The same behavior can be observed with the uniform spin-averaged model results in Ref. \cite{benson2020b} (see Figure 12 in that paper). Defunct satellites including Telstar 401 and retired Glonass satellites observed by Refs. \cite{earl,rachman} exhibit similar yearly spin rate oscillations. During this initial 3.5 year period, there is also a secular decrease in $\overline{\omega}_e$. After roughly 3.5 years, the satellite reaches a maximum $P_e$ of approximately 40 min with $\overline{\beta}$ approaching 0$^{\circ}$. At this point, the satellite loses sufficient spin stability and transitions to tumbling. It then spins up about the long axis and progresses into a tumbling cycle with $\bm{H}$ precessing around the sun line.

\subsubsection{Tumbling Cycles}

\begin{figure}[H]
	\centering
\subcaptionbox{$\dot{\overline{I}}_d$}{\includegraphics[width=3in]{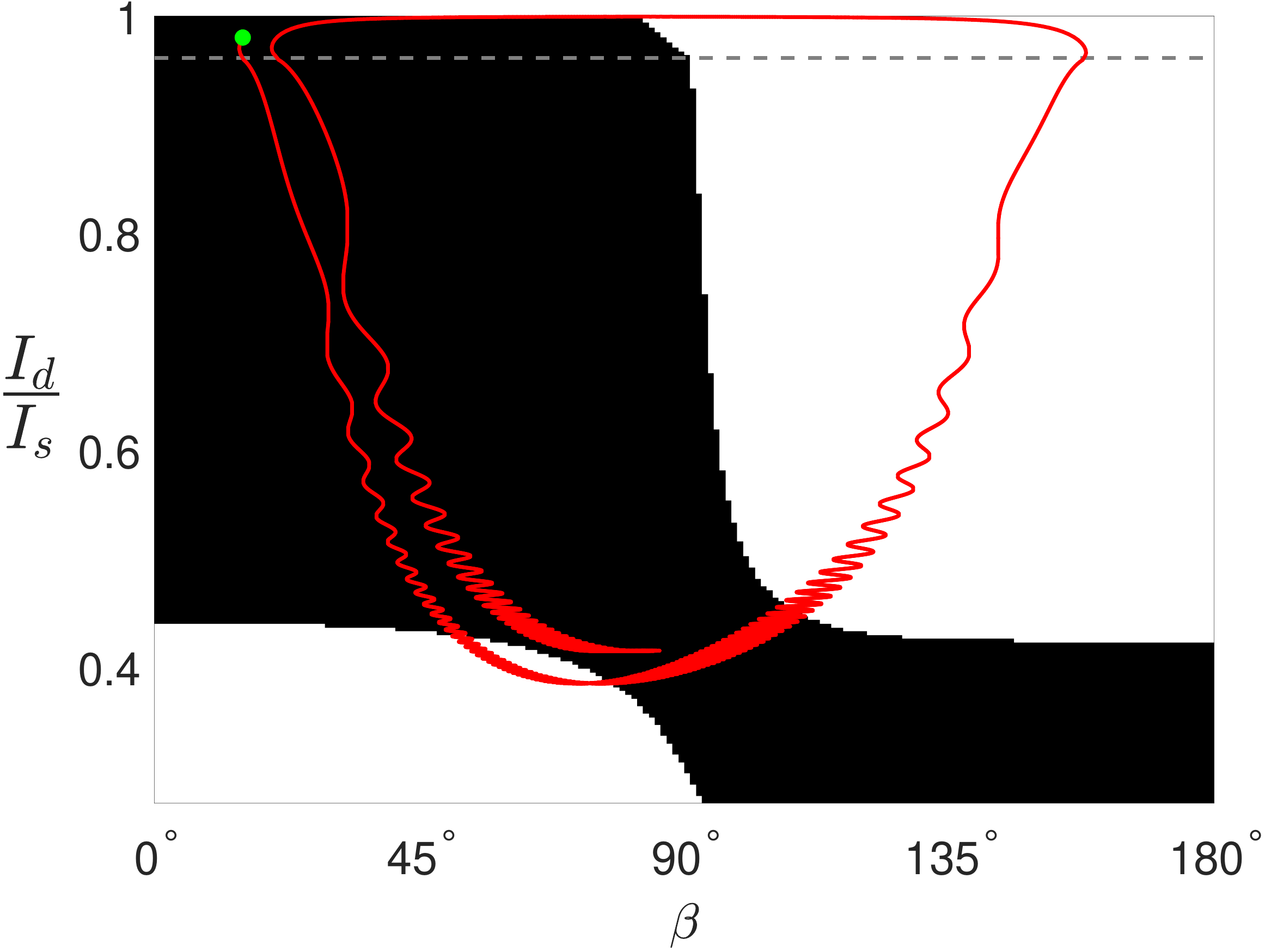}}
\subcaptionbox{$\dot{\overline{\beta}}$}{\includegraphics[width=3in]{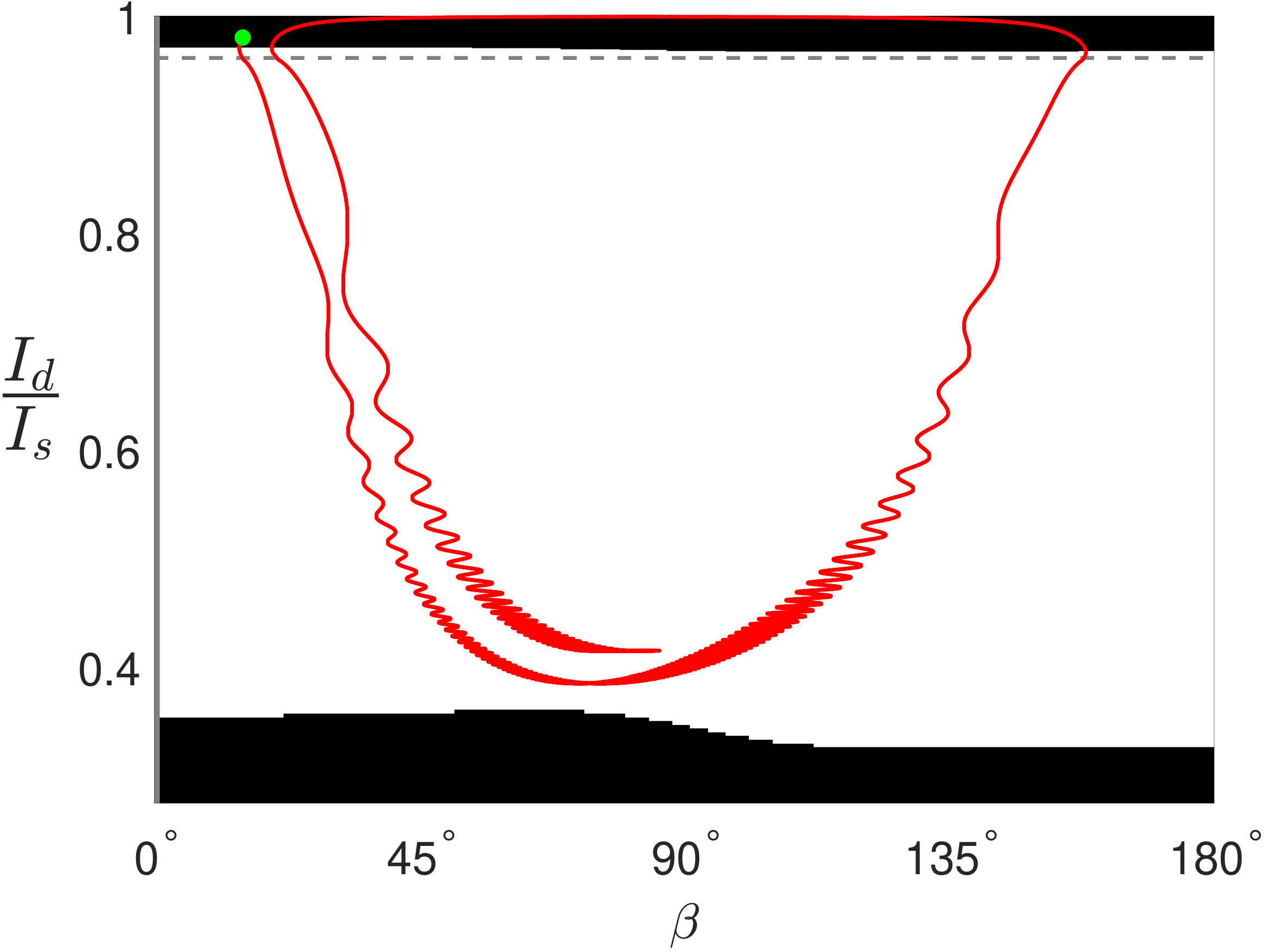}}
\subcaptionbox{$\dot{\overline{\omega}}_e$}{\includegraphics[width=3in]{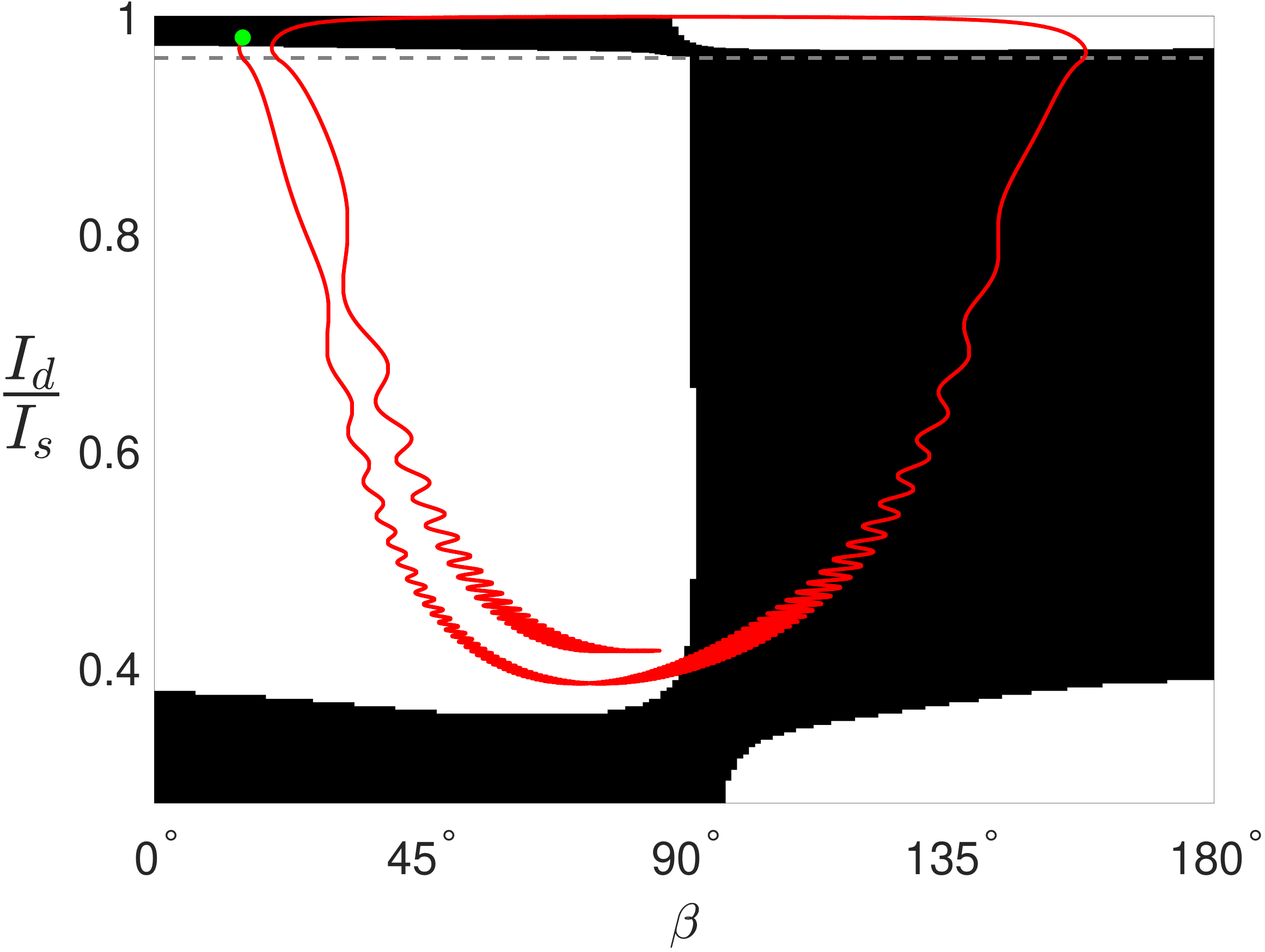}}
	\caption{Signs of averaged parameter derivatives vs. $I_d$ and $\beta$ (SAM+/LAM+) for GOES 8 with Figure~\ref{fig:evol_comp} averaged evolution overlaid in red, starting at the green dot. Black regions denotes negative values and white denotes positive values. The dashed gray line is the separatrix.}
	\label{fig:eom_contours_17}
\end{figure}

We will now leverage the averaged model to better understand the observed tumbling cycles. Figure~\ref{fig:eom_contours_17} shows the signs of $\dot{\overline{I}}_d$, $\dot{\overline{\beta}}$, and $\dot{\overline{\omega}}_e$ computed over $I_d$ and $\beta$ (the sign contours for $\dot{\overline{H}}$ are nearly identical to those for $\dot{\overline{\omega}}_e$). The black regions denote negative values and the white regions denote positive values. To simplify analysis, $\dot{\overline{\beta}}$ (Eq.~\ref{eq:betadotavg}) has been averaged over $\overline{\alpha}$ to remove dependency. This is valid because $\overline{\alpha}$ is a fast variable compared to $\overline{I}_d$, $\overline{\beta}$, and $\overline{\omega}_e$ during the tumbling cycles. The averaged model evolution from Figure~\ref{fig:evol_comp} has been overlaid on the contours in Figure~\ref{fig:eom_contours_17}. Starting at the green dot, Figure~\ref{fig:eom_contours_17}a shows that $\overline{I}_d$ will initially decrease as the satellite is pushed into more excited tumbling. As we near the separatrix (the dashed grey line), Figure~\ref{fig:eom_contours_17}b shows that $\beta$ will start increasing. At the same time, the satellite effective spin rate ($\overline{\omega}_e$) will begin increasing as well. These combined effects cause the satellite to proceed into more excited tumbling with a faster spin rate and the pole moving away from the sun. Once $\beta$ increases past 90$^{\circ}$ (i.e. pole perpendicular to the sun) the satellite begins spinning down and moving back towards uniform rotation. Upon crossing the separatrix, the signs of $\dot{\overline{\beta}}$ and $\dot{\overline{\omega}}_e$ flip. So, the satellite then spins up, entering a nearly uniform rotation phase with the pole moving back towards the sun direction. Finally, passing through $\beta=$ 90$^{\circ}$, $\dot{\overline{I}}_d$ and $\dot{\overline{\omega}}_e$ flip signs resulting in spin down and progression back towards tumbling. At this point, the next tumbling cycle can begin. From Eqs.~\ref{eq:alphadotavg}, \ref{eq:betadotavg}, and \ref{eq:Iddotavg}, we note that the tumbling cycle duration will be driven directly by $\overline{H}$. The larger the initial $\overline{H}$, the slower the satellite will progress through the tumbling cycle.  For GOES 8, any escape to long-term uniform rotation from these tumbling cycles will likely occur in the upper right (after passing upward across the separatrix). To escape, the satellite must spin up sufficiently before $\beta$ decreases below 90$^{\circ}$. Alternatively, capture into these tumbling cycles from uniform rotation ($I_d=I_s$) requires $\beta<$ 90$^{\circ}$ so that $\dot{\overline{I}}_d$ and $\dot{\overline{\omega}}_e$ are negative. If the spin rate is small enough, $\bm{H}$ will be pulled towards the sun line and the satellite will spin down and transition into a tumbling cycle.  

\subsubsection{Sun-Tracking Behavior}

We will now discuss the sun-tracking precession behavior observed during tumbling cycles. The foundation of the following analysis is that $\bm{M}$ is nearly aligned with $\bm{\hat{Z}}\times\bm{\hat{H}}$ for the majority of the $I_d$ - $\beta$ phase space. To show this, we first calculate the component of $\bm{M}$ along $\bm{\hat{B}}=\bm{\hat{Z}}\times\bm{\hat{H}}/|\bm{\hat{Z}}\times\bm{\hat{H}}|$,

\begin{equation}
\bm{\hat{B}}\cdot\bm{M}=M_y
\label{eq:Bhat}
\end{equation}
and the angle between $\bm{M}$ and $\bm{\hat{B}}$ is then given by, 
\begin{equation}
\cos\theta_{BM}=\bm{\hat{B}}\cdot\bm{\hat{M}}=\frac{M_y}{\sqrt{M^2_x+M^2_y+M^2_z}}
\label{eq:thetazxh}
\end{equation}

\begin{figure}[H]
	\centering
	\subcaptionbox{$\theta_{BM}$}{\includegraphics[width=3.2in]{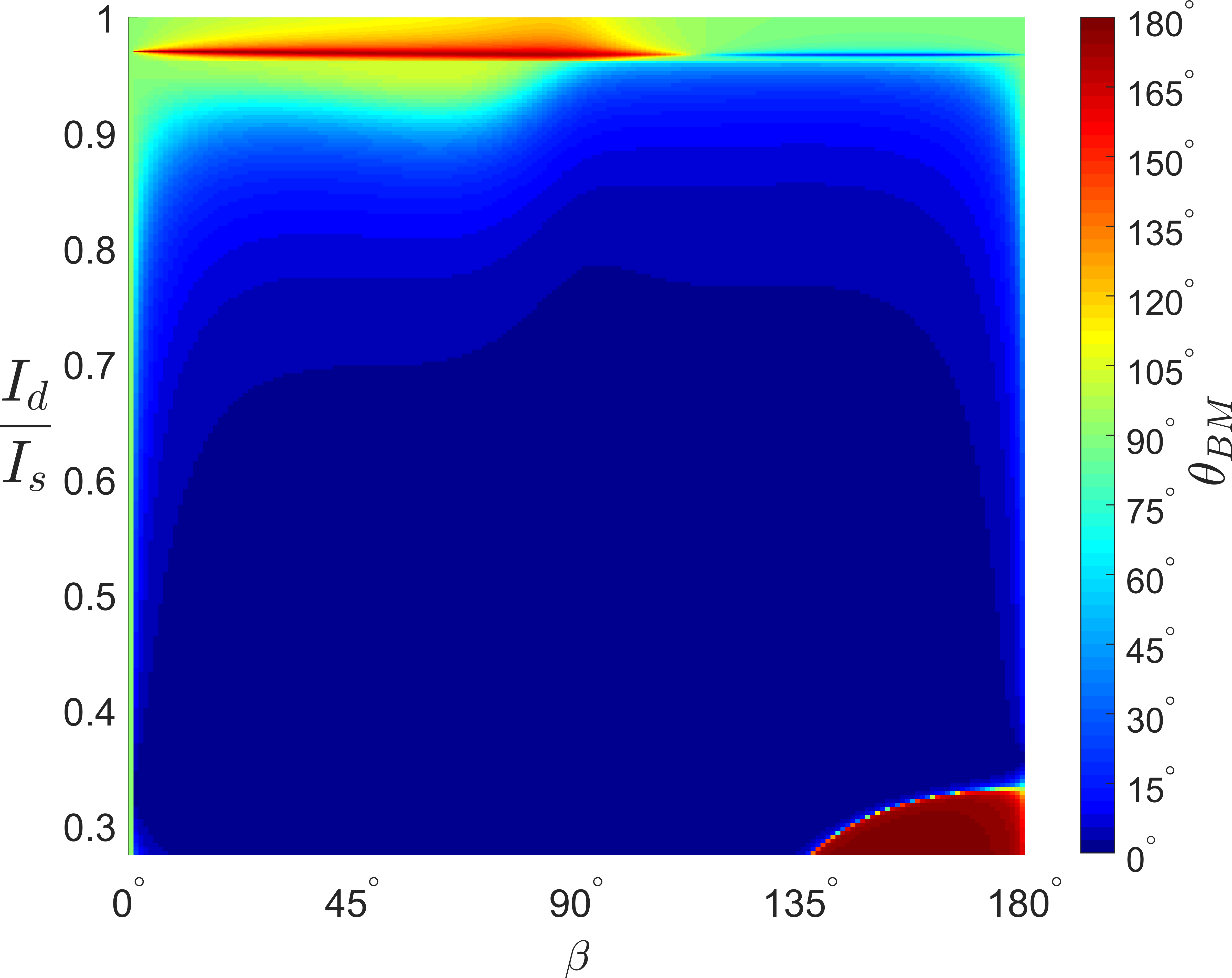}}
	\subcaptionbox{$\overline{M_y}$ and $M\sin{\beta}$ Approximation}{\includegraphics[width=3.2in]{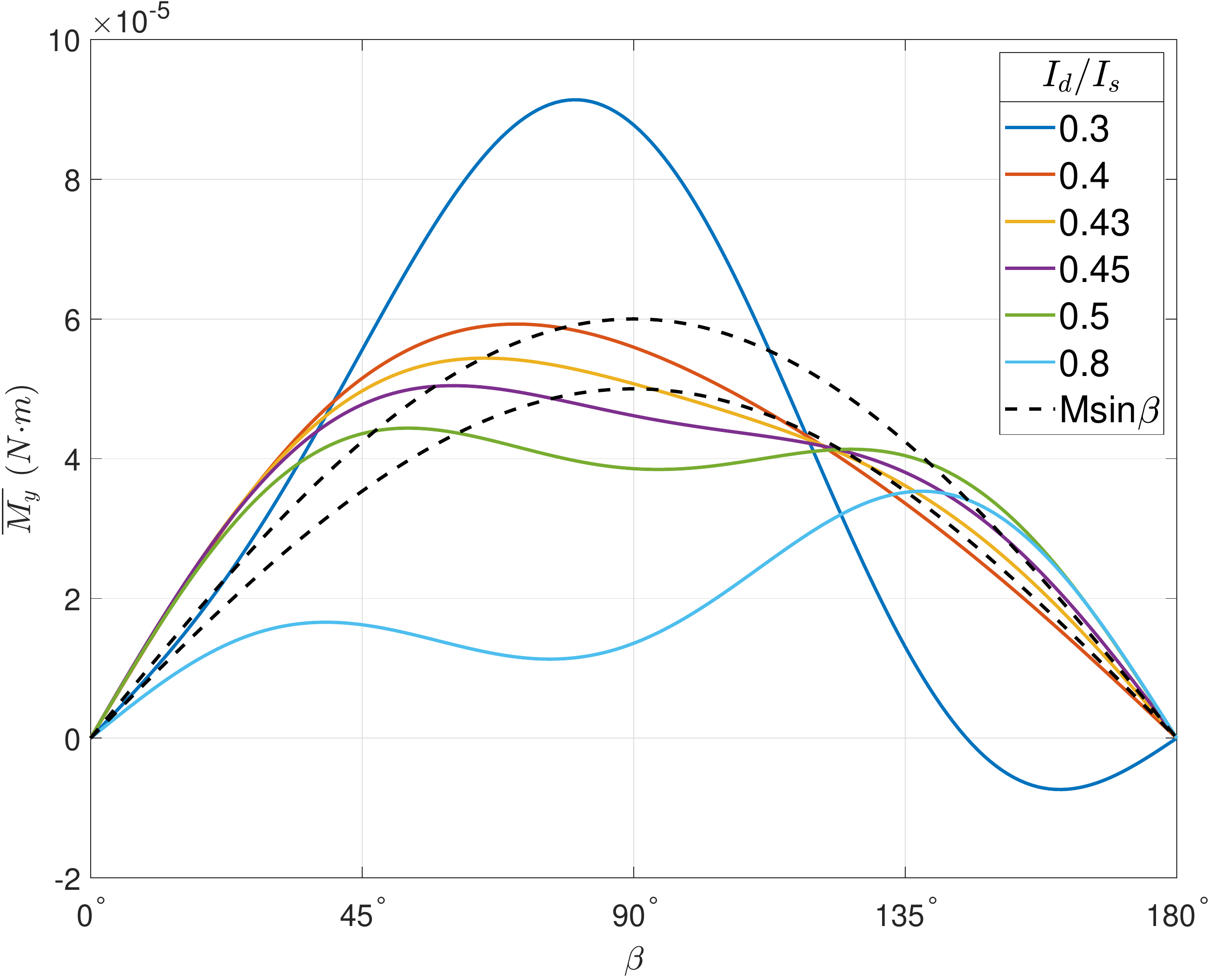}}
	\caption{Structure of $\overline{M_y}$ for GOES 8 (SAM+/LAM+).}
	\label{fig:zhatxhhat}
\end{figure}

Plotting Eq.~\ref{eq:thetazxh} over $I_d$ and $\beta$, the resulting values are provided in Figure~\ref{fig:zhatxhhat}a for GOES 8. From the small $\theta_{BM}$, we see that $\bm{M}$ is closely aligned with $\bm{\hat{B}}$ for most LAM $I_d$, $\beta$ values and therefore nearly perpendicular to both $\bm{\hat{Z}}$ and $\bm{\hat{H}}$. This makes sense given the large relative magnitude of $M_y$ to $M_x$ and $M_z$ in Figures~\ref{fig:avg_torques}b,d. Calculating $\overline{M_y}$ for a number of LAM $I_d$ values for GOES 8, Figure~\ref{fig:zhatxhhat}b shows that $\overline{M_y}\;{\approx}\;M\sin{\beta}$ for $I_d/I_s$ values near 0.4 - 0.5 (where $M$ is the arbitrary torque amplitude). From Figure~\ref{fig:evol_comp}b, we see that the satellite spends most of the tumbling cycle near $I_d/I_s=$ 0.45, where this $M\sin{\beta}$ approximation agrees best.

Given this near orthogonality, we can develop an approximate system to better understand the sun-tracking precession. Approximating the torque as $\bm{M}=M_y\bm{\hat{B}}$, we can calculate $\frac{{^\mathcal{O}}d}{dt}(\bm{H})$ using the transport theorem,
\begin{equation}
\frac{^\mathcal{O}d}{dt}(\bm{H})=M_y\bm{\hat{B}}-\boldsymbol{\omega}_{\mathcal{O}/\mathcal{N}}\times\bm{H}
\label{eq:Hvecdotorb}
\end{equation}

Then assuming $M_y=M\sin{\beta}$ and noting that $\sin{\beta}=|\bm{\hat{Z}}\times\bm{\hat{H}}|$, we can simplify Eq.~\ref{eq:Hvecdotorb} to find,
\begin{equation}
\frac{^\mathcal{O}d}{dt}(\bm{H})=\Bigg(\frac{M}{H}\bm{\hat{Z}}-n\bm{\hat{X}}\Bigg)\times\bm{H}
\label{eq:Hvecdotorb2}
\end{equation}

\begin{figure}[H]
	\centering
	\includegraphics[width=3.25in]{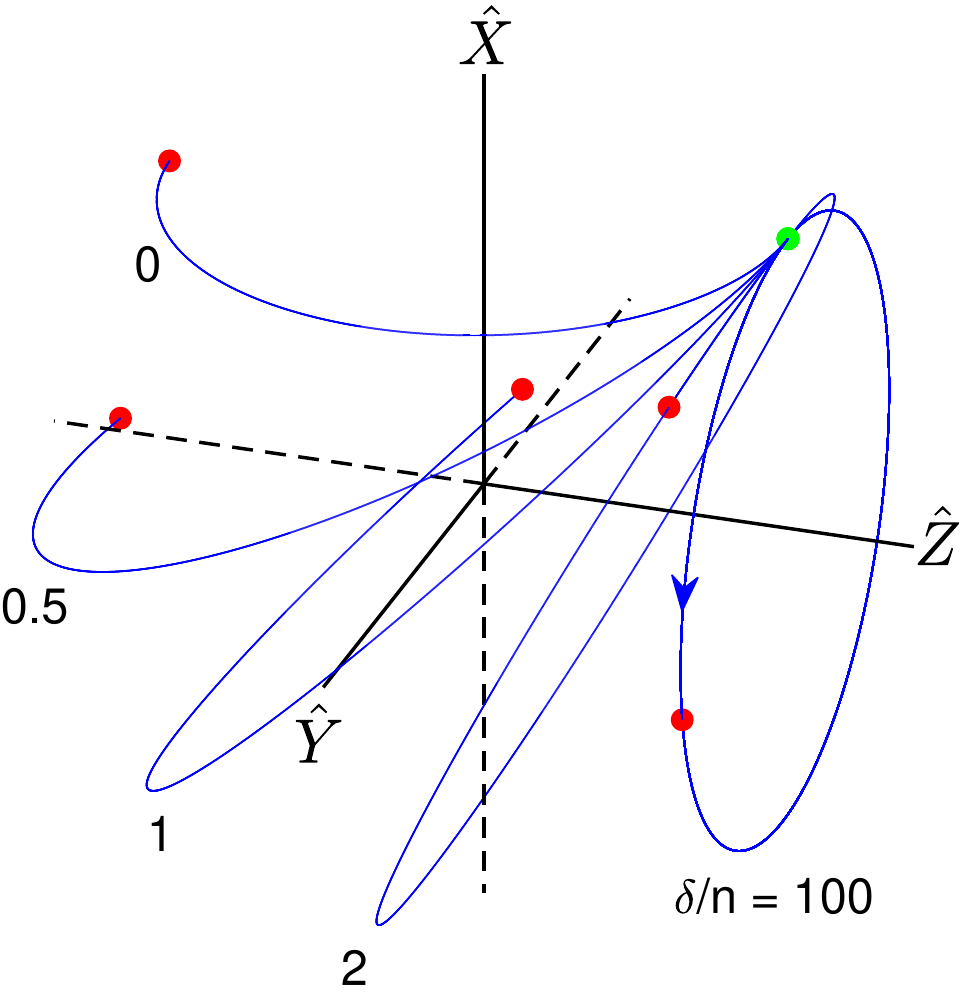}
	\caption{$\bm{\hat{H}}(t)$ from Eq.~\ref{eq:H(t)} over 180 days with varying $\delta/n$ (0, 0.5, 1, 2, 100). The green dot denotes the initial state ($\alpha$= 0$^{\circ}$, $\beta=$ 45$^{\circ}$) and the red dots denote the final states for each $\delta/n$.}
	\label{fig:H(t)}
\end{figure}

Since we assume $\bm{M}\cdot\bm{H}=0$, $H$ is constant. Therefore, Eq.~\ref{eq:Hvecdotorb2} is a linear system with constant coefficients. Solving the initial value problem with $\bm{H}(t=0)=H[\sin{\beta_o},0,\cos{\beta_o}]^T$, we find,
\begin{singlespace}
\begin{equation}
\bm{H}(t)=\frac{H}{\omega^2}\begin{bmatrix}
{\delta}(n\cos{\beta_o}+\delta\sin{\beta_o})\cos{\omega}t-n(\delta\cos{\beta_o}-n\sin{\beta_o}) \\
{\omega}(n\cos{\beta_o}+\delta\sin{\beta_o})\sin{\omega}t\\
n(n\cos{\beta_o}+\delta\sin{\beta_o})\cos{\omega}t+{\delta}(\delta\cos{\beta_o}-n\sin{\beta_o}) \\
\end{bmatrix}
\label{eq:H(t)}
\end{equation}
\end{singlespace}
\noindent where $\delta=M/H$ and $\omega = \sqrt{\delta^2+n^2}$. Note that $\bm{H}(t)$ is periodic with period $2\pi/\omega$. Taking the time derivative of Eq.~\ref{eq:H(t)}, we find,
\begin{singlespace}
\begin{equation}
\frac{^\mathcal{O}d}{dt}(\bm{H})=H(n\cos{\beta_o}+\delta\sin{\beta_o})\begin{bmatrix}
-\frac{\delta}{\omega}\sin{\omega}t \\
\cos{\omega}t\\
-\frac{n}{\omega}\sin{\omega}t\\
\end{bmatrix}
\label{eq:Hdot(t)}
\end{equation}
\end{singlespace}
For $\delta>>n$, $\omega\approx\delta$, so $\dot{H}_Z$ is relatively small and evolution occurs mostly parallel to the the $\bm{\hat{X}}$ - $\bm{\hat{Y}}$ plane (i.e. sun-tracking precession). Here, precession occurs much faster than the mean motion $n$ because $\omega>>n$. As $\delta/n$ decreases, the precession rate slows and motion transitions more towards the $\bm{\hat{Y}}$ - $\bm{\hat{Z}}$ plane. As $\delta/n\rightarrow0$, $\dot{H}_X\rightarrow0$ and motion becomes confined parallel to the $\bm{\hat{Y}}$ - $\bm{\hat{Z}}$ plane with $\omega{\rightarrow}n$. Here, the torque is not sufficient to turn $\bm{H}$ which remains inertially fixed. Figure~\ref{fig:H(t)} illustrates this transition from sun-tracking precession to inertially fixed $\bm{H}$ for a number of $\delta/n$ values. Proceeding clockwise from lower right to upper left, $\delta/n$ decreases and circulation gradually transitions from $\bm{\hat{Z}}$ to $\bm{\hat{X}}$.

\subsubsection{Influence of End of Life Configurations}

\begin{figure}[H]
	\centering
\subcaptionbox{$\overline{M_x}$}{\includegraphics[width=3in]{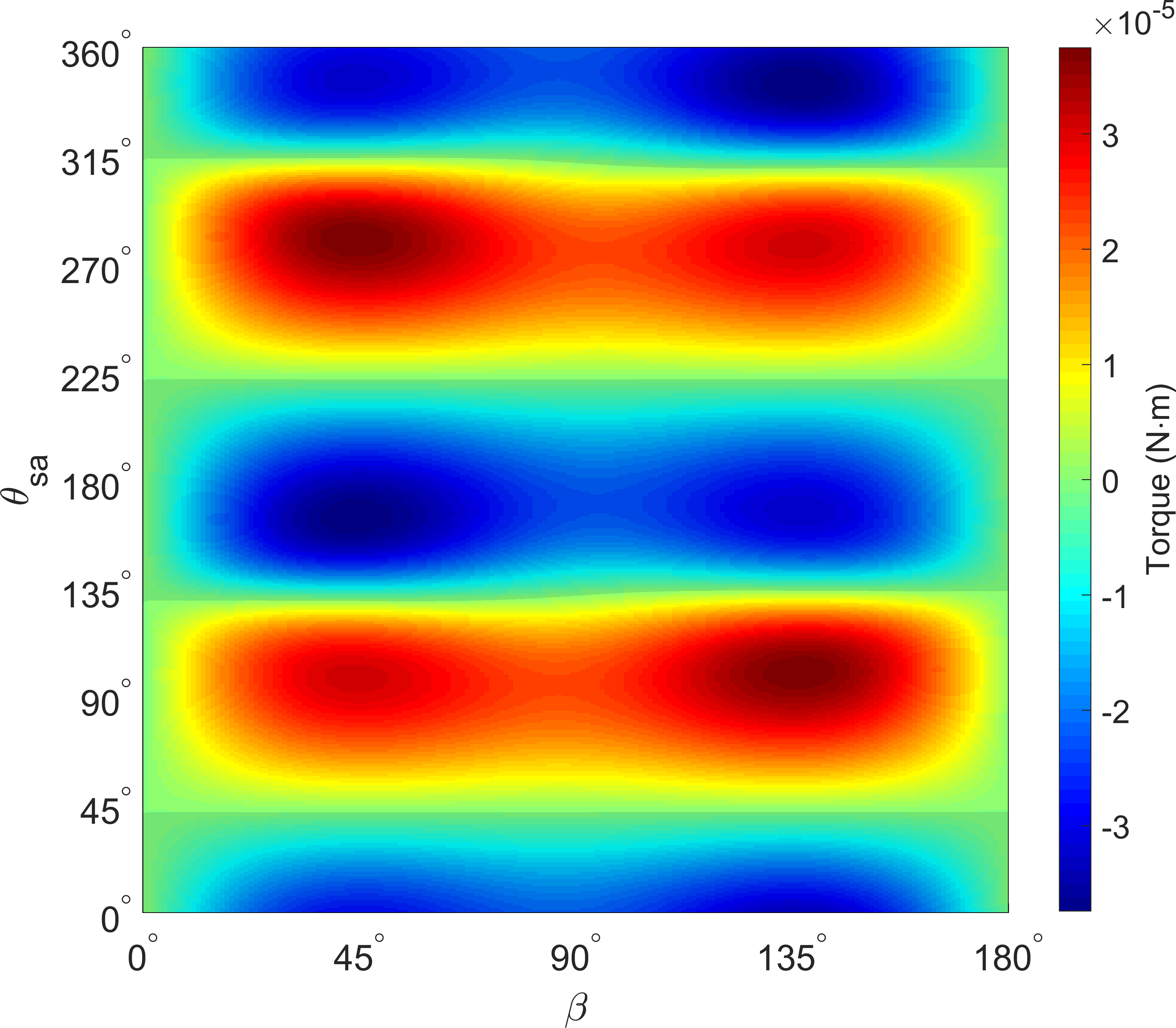}}
\subcaptionbox{$\overline{M_y}$}{\includegraphics[width=3in]{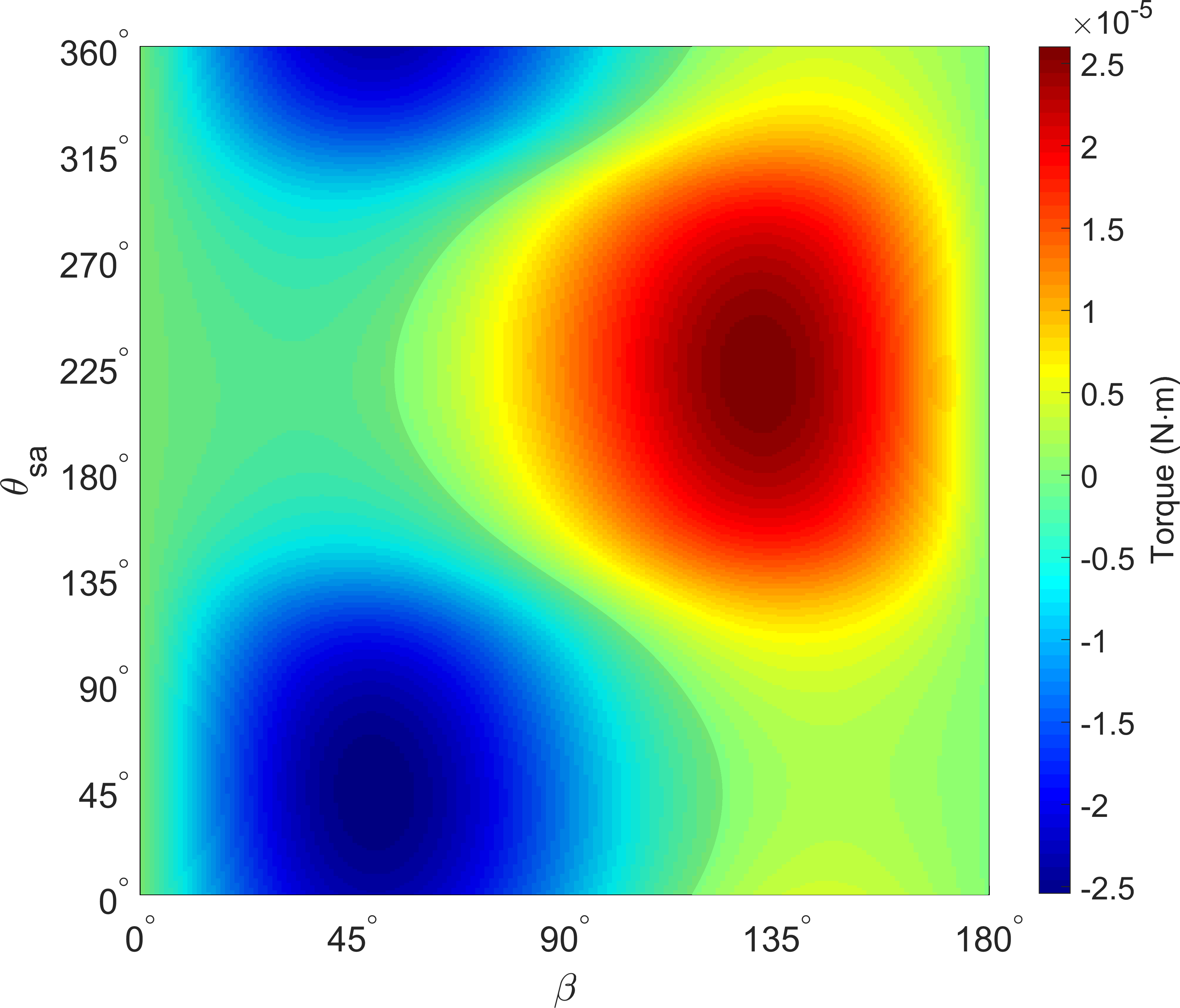}}
\subcaptionbox{$\overline{M_z}$}{\includegraphics[width=3in]{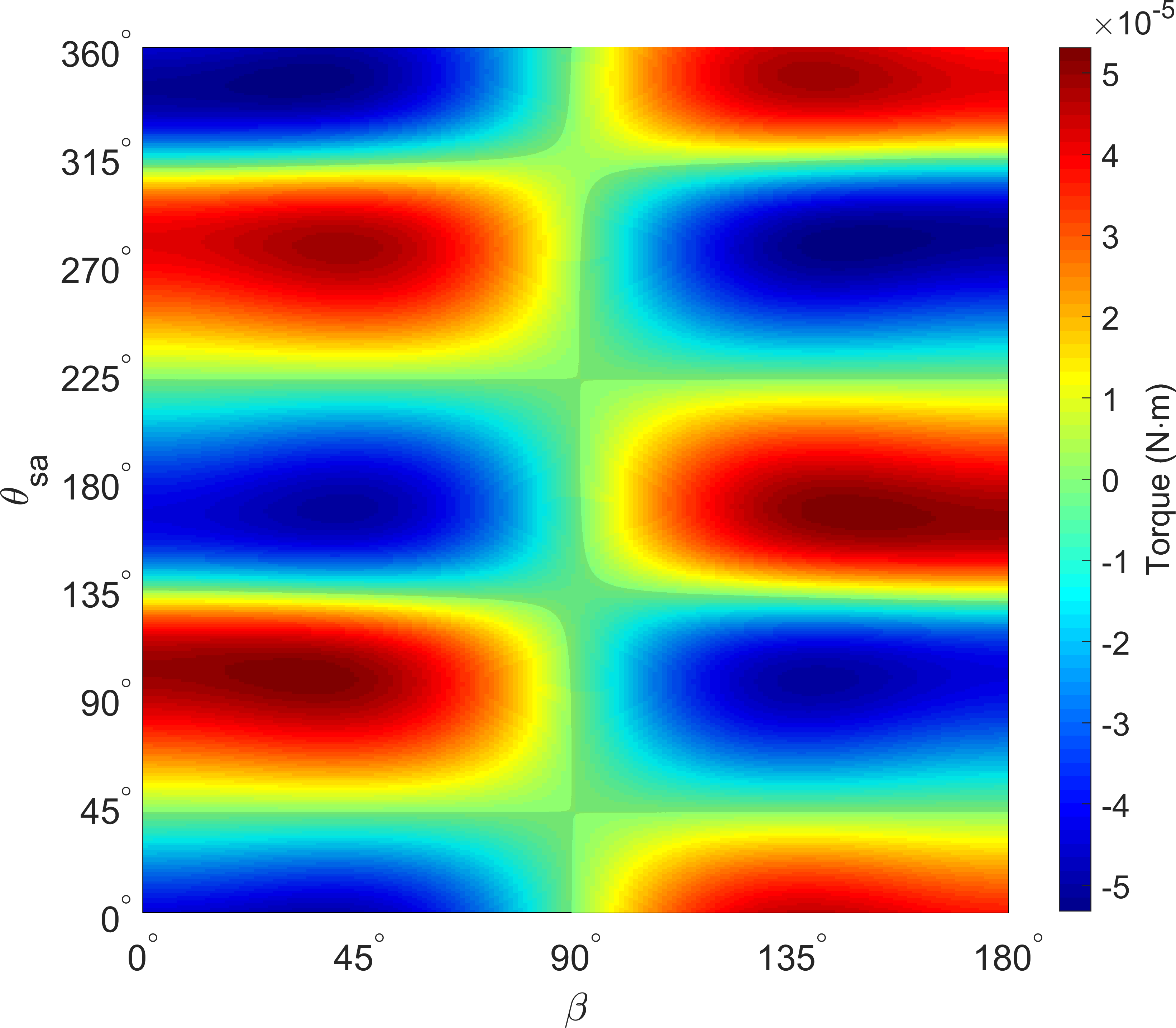}}
\subcaptionbox{$\dot{\overline{I}}_d$}{\includegraphics[width=3in]{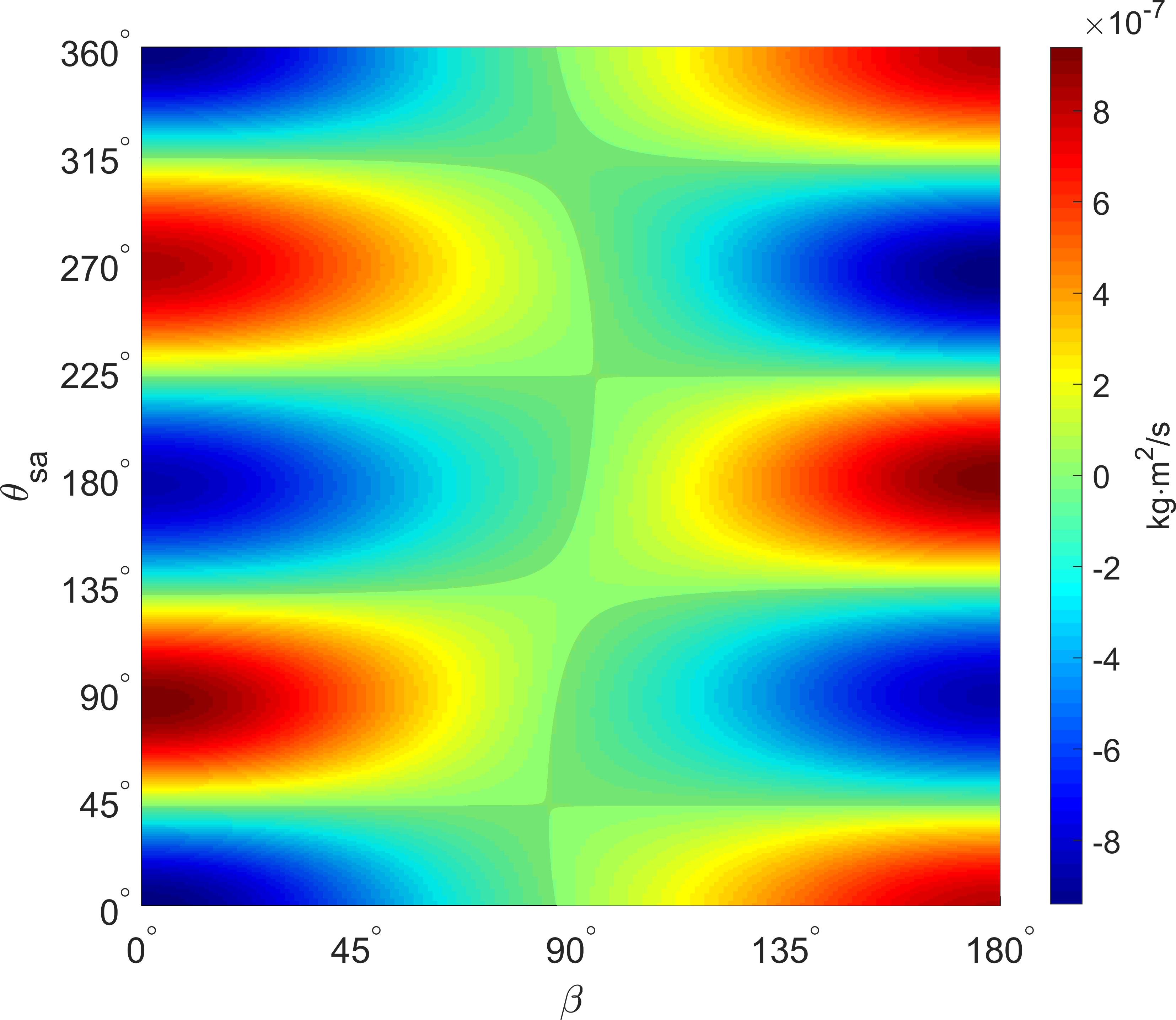}}
	\caption{GOES 8 Averaged Terms vs. $\beta$ and Solar Array Angle $\theta_{sa}$ (SAM+ $I_d=$ 3500 kg${\cdot}$m$^2$)}
	\label{fig:torque_beta_thetasp_sam}
\end{figure}

It is important to note that the counter-clockwise ($I_d$, $\beta$) motion in Figure~\ref{fig:eom_contours_17} is just one of the possible evolutionary scenarios. In Benson et al. \cite{benson2020b}, we found that long-term uniform GOES evolution strongly depends on the end of life solar array angle $\theta_{sa}$ (see Figures 8-12 in that paper and the associated discussion). Computing $\overline{M_x}$, $\overline{M_y}$, $\overline{M_z}$, and $\dot{\overline{I}}_d$  over all possible end of life GOES 8 solar array angles with the averaged model, we find the following contours in Figure~\ref{fig:torque_beta_thetasp_sam}. For $\dot{\overline{I}}_d$, $\omega_e=$ 2$\pi$ rad/s was again assumed. Sweeping over $\theta_{sa}$, the averaged components change significantly in sign and magnitude, indicating that $\theta_{sa}$ greatly affects general long-term satellite evolution. The results in Figure~\ref{fig:torque_beta_thetasp_sam} are analogous to the uniform spin-averaged coefficients in Ref. \cite{benson2020b}. The most easily comparable are $\overline{M_z}$ and $\mathcal{C}_{0,z}$ which share very similar structure (see Figures 8 and 11 in that paper). In addition, for $\theta_{sa}$ near odd multiples of 42$^{\circ}$, we find that $\overline{M_x}$, $\overline{M_z}$, and $\dot{\overline{I}}_d$ are approximately zero. These critical $\theta_{sa}$ values also hold for the uniform spin-averaged results in Ref. \cite{benson2020b}. Obviously, these negligible torque configurations are specific to GOES 8's geometry and mass distribution. For other satellites, the averaged framework will allow for fast and efficient studies of the parameter space to identify any similar configurations. These GOES findings illustrate the potential to reduce long-term spin state variation by properly setting end of life configurations. 

\begin{figure}[H]
	\centering
\subcaptionbox{$\dot{\overline{I}}_d$}{\includegraphics[width=3in]{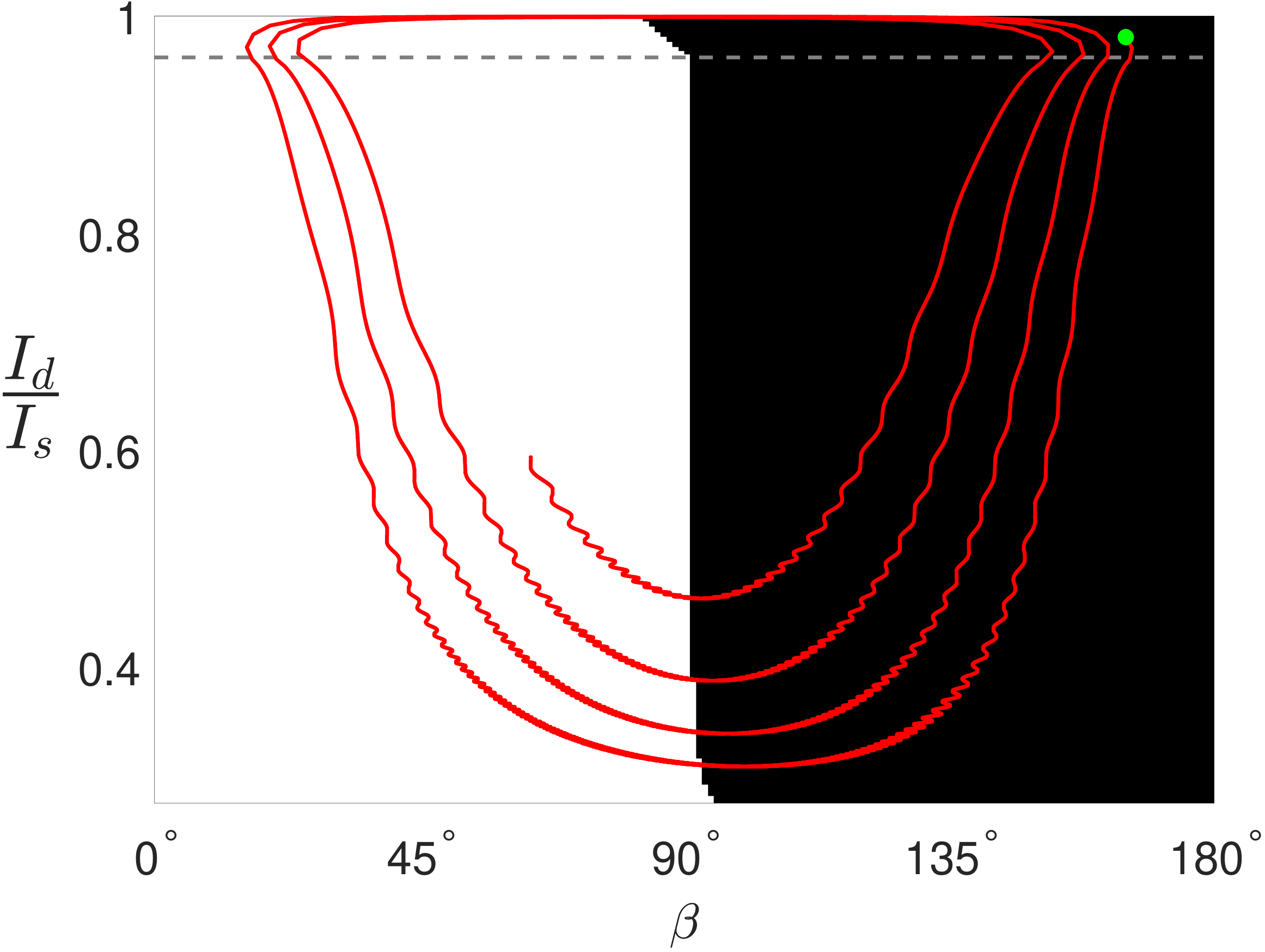}}
\subcaptionbox{$\dot{\overline{\beta}}$}{\includegraphics[width=3in]{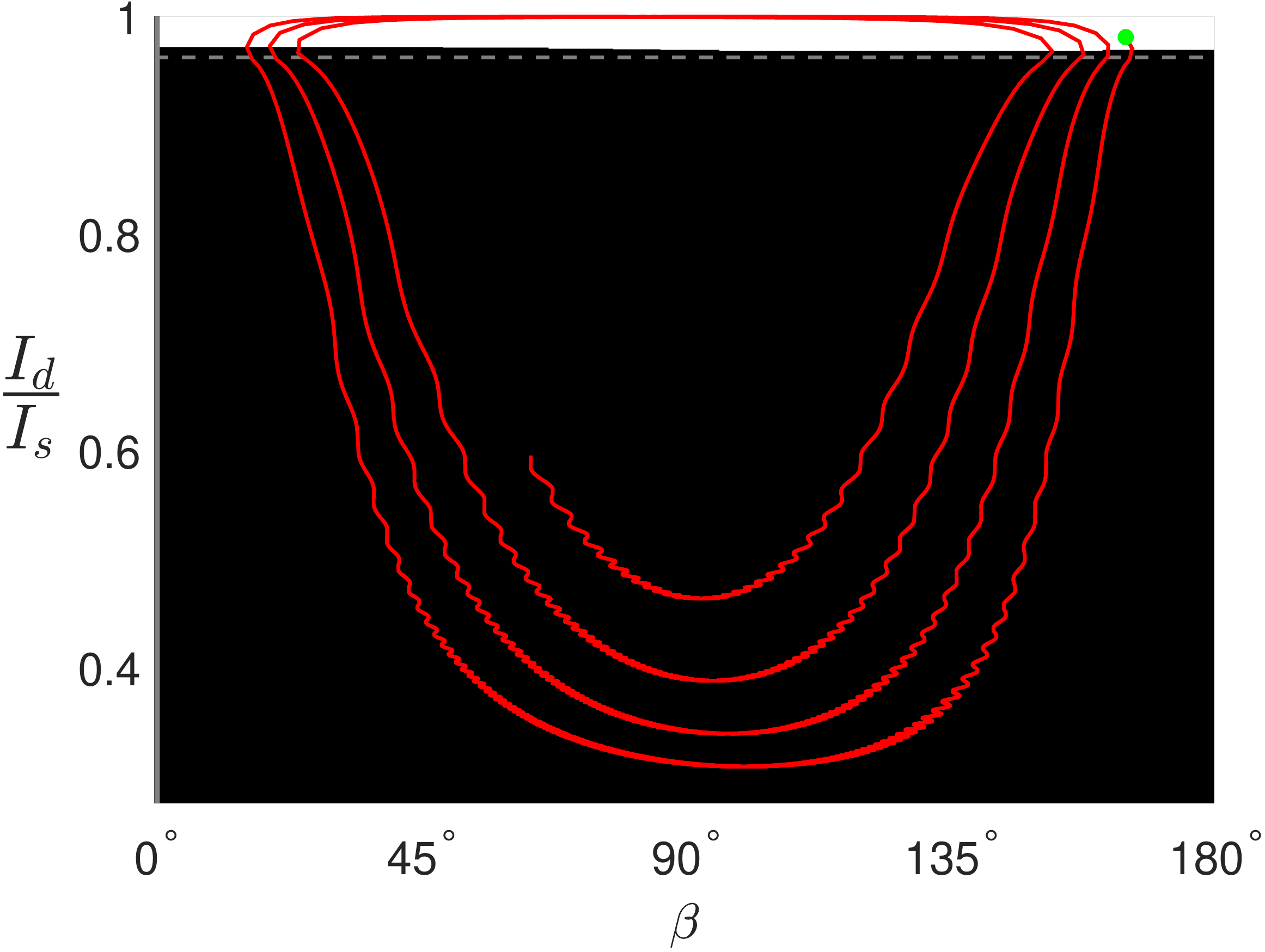}}
\subcaptionbox{$\dot{\overline{\omega}}_e$}{\includegraphics[width=3in]{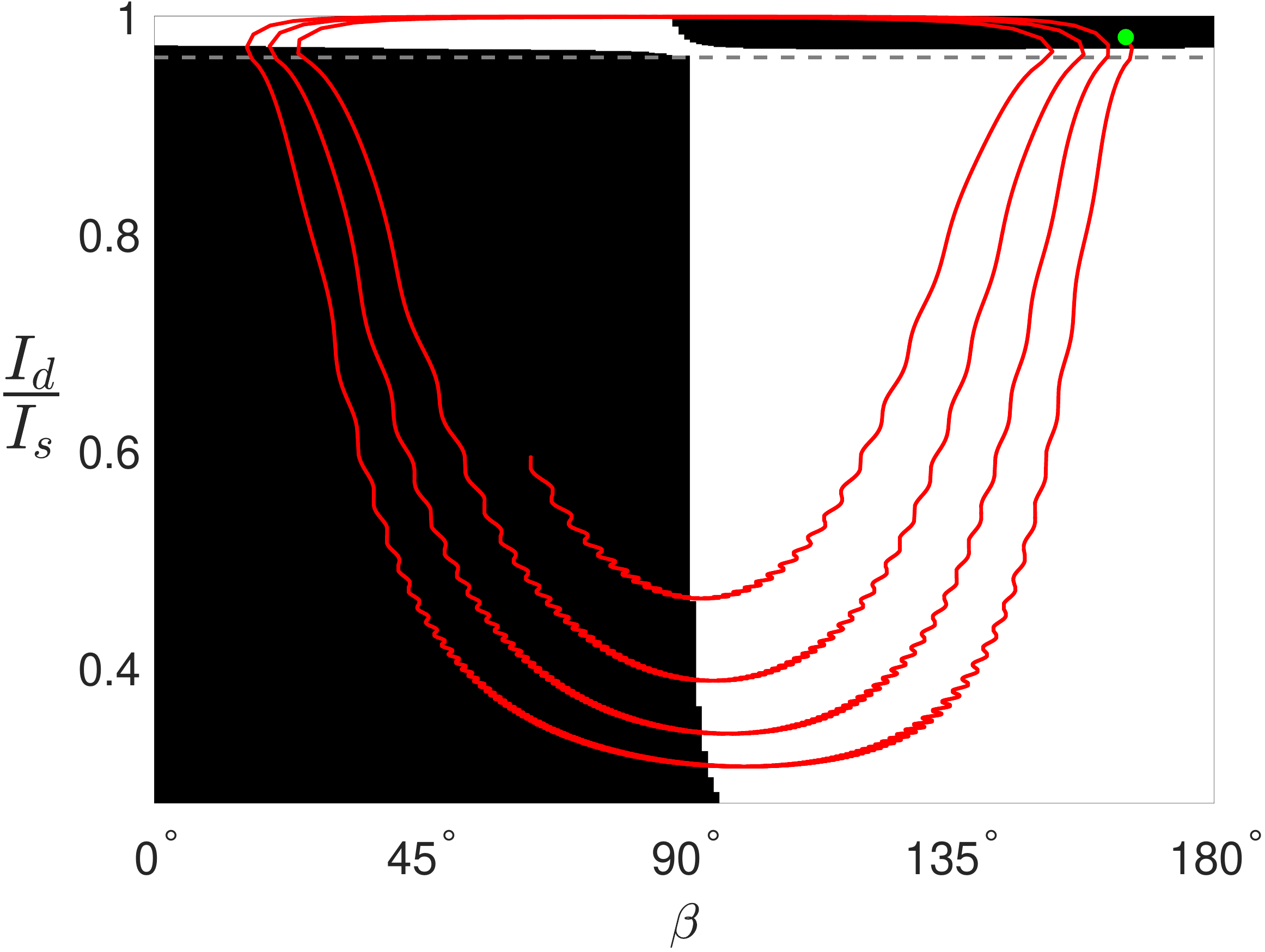}}
	\caption{Same as Figure~\ref{fig:eom_contours_17} except with $\theta_{sa}=$ 70$^{\circ}$ and corresponding averaged evolution.}
	\label{fig:eom_contours_70}
\end{figure}

We will now briefly consider the long-term evolution for GOES 8 with a different solar array angle. Changing GOES 8's  $\theta_{sa}$ from 17$^{\circ}$ to 70$^{\circ}$ yields the contours in Figure~\ref{fig:eom_contours_70}. Here, the signs of $\dot{\overline{I}}_d$ and $\dot{\overline{\omega}}_e$ are essentially mirrored about $\beta=$ 90$^{\circ}$ as compared to Figure~\ref{fig:eom_contours_17}. For $\dot{\overline{\beta}}$, the sign is mirrored about the separatrix. Complementing the contours is the six year averaged evolution given by the following initial conditions: $\overline{\alpha}=$ 0$^{\circ}$, $\overline{\beta}=$ 165$^{\circ}$, $\overline{I}\!_{d}=$ 3500 kg${\cdot}$m$^2$ (SAM+), and $P_e=$ 240 min. The satellite goes through several tumbling cycles as in Figure~\ref{fig:eom_contours_17} except that ($I_d$, $\beta$) evolution instead proceeds clockwise with $\beta$ now decreasing over the course of each tumbling cycle. 

\section{Discussion}

Comparing the full and averaged dynamical models in Section IV, we found that the averaged model captures the tumbling cycles and sun-tracking behavior of the full model. Nevertheless, there were quantitative differences between the two models due to our averaging assumptions. Most notable are the neglect of resonances and the second order Fourier series illumination function approximation. Higher order Fourier series approximations of $\mathrm{max}(0,\bm{\hat{u}}\cdot{\bm{\hat{n}}})$ would yield better agreement with the full model at the expense of increased average model complexity. Another shortfall of the current semi-analytical averaged model is that $\dot{\alpha}$ is singular for $\beta=$ 0$^{\circ}$ and 180$^{\circ}$. Again this could be remedied by replacing $\alpha$ and $\beta$ with an alternate coordinate set when very close to these singularities (e.g. $v=\sin{\alpha}\sin{\beta}$ and $w=\cos{\alpha}\sin{\beta}$ which has a $\beta$ ambiguity). In practice though, these singularities were never encountered during averaged model propagation, so this approach was not implemented in our model. Finally, while this paper only considered solar torques, the averaged model could be readily expanded to include energy dissipation as well as averaged gravity gradient and magnetic torques. 

Given the transition from uniform rotation to non-principal axis tumbling observed for the GOES model, it is possible that other satellites could undergo similar transitions. There is clear indication that defunct satellites are exhibiting large amplitude, secular period variations consistent with Figure~\ref{fig:uniform_tumbling_transition} \citep{earl,rachman}. From the active debris removal (ADR)/servicing perspective, this implies that a satellite may not remain in uniform rotation indefinitely. In general, a uniform to tumbling transition would require a secular decrease in uniform spin rate with $\dot{\overline{I_d}}<0$. Furthermore, the results in Figure~\ref{fig:uniform_tumbling_transition} demonstrate that the transition to tumbling could occur quickly, in a couple of weeks or less. From Figures~\ref{fig:eom_contours_17} and \ref{fig:eom_contours_70}, it seems possible for a satellite to escape these tumbling cycles and enter fast uniform rotation, a process that could occur as rapidly. As a result, target satellite spin state monitoring and prediction will be crucial for ADR and servicing. The possible existence of tumbling cycles would have additional implications for ADR and servicing missions. Slow, uniform rotation would be ideal for rendezvous, capture, and de-spin procedures. Even for proposed "touchless" electromagnetic detumbling approaches \cite{gomez2015}, leveraging YORP to partially de-spin a target satellite would reduce the time, energy, and risk required by the ADR/servicing spacecraft. So predicting future windows of slow, uniform rotation between the tumbling phases would be valuable. The above analysis shows that the primary driver of sun-tracking for GOES is the near orthogonality of the solar torque and the sun line. It would be valuable to determine how often this orthogonality holds for different satellites and rocket bodies. In terms of satellite size $r$, $I_d\;{\propto}\;r^5$ and the solar torque $\bm{M}\;{\propto}\;r^3$. So Eqs.~\ref{eq:alphadotavg}, \ref{eq:betadotavg}, \ref{eq:Iddotavg} ($\overline{I}_d$ normalized), and \ref{eq:wedotavg} are proportional to $1/r^2$. In other words, reducing satellite size by a factor of ten (maintaining  density and optical properties), will cause it to evolve 100 times faster. Similarly, $\delta/n\;{\propto}\;1/r^2$, so sun-tracking precession is equally more effective for smaller satellites.

The importance of solar array angle on long-term GOES evolution demonstrates the potential for dictating the post-disposal spin state evolution of defunct satellites by carefully setting their end of life configurations. For example, configurations that minimize $|\dot{\overline{\beta}}|$ could be used to shut off or greatly slow the observed tumbling cycles, facilitating debris removal and servicing missions. Also, minimizing $|\overline{M_z}|$ would reduce spin rates and their variation amplitude, making satellites easier to capture and reducing potential for material shedding.

Here, it is also worthwhile to briefly discuss the implications of our findings for natural small bodies such as asteroids. In YORP simulations, Vokrouhlicky et al. found that small asteroids can exhibit transitions from uniform rotation to tumbling and subsequent tumbling spin up \cite{vok2007}.  Given these similarities, it is possible that the tumbling cycles, angular momentum sun-tracking, and tumbling resonances observed here for artificial satellites hold for some asteroids as well. Since solar radiation pressure goes as $1/a^2$, where $a$ is the heliocentric semi-major axis, Eqs.~\ref{eq:alphadotavg} - \ref{eq:wedotavg} will go as the same. Furthermore, the mean motion $n$ goes as $1/\sqrt{a^3}$, so $\delta/n\;{\propto}\;1/\sqrt{a}$. This implies that uniform to tumbling transitions, tumbling cycles, and sun-tracking precession would be more likely for smaller asteroids in the inner solar system (all else equal). Again, dedicated study of these natural small bodies is needed to determine whether the tumbling-averaged torques provide the necessary structure for this behavior (e.g. near orthogonality of $\bm{\hat{B}}$ and $\bm{M}$).

\section{Conclusions}

This paper illustrates the complex, yet structured Yarkovsky-O'Keefe-Radzievskii-Paddack (YORP) effect behavior for defunct satellites including transitions from uniform rotation to non-principal axis tumbling, angular momentum sun-tracking, tumbling cycles, and resonant YORP tumbling states. To help understand these behaviors, we developed a semi-analytical tumbling-averaged YORP model. This model captures the uniform to tumbling transition, sun-tracking, and tumbling cycle behavior observed with the full dynamics while being roughly three orders of magnitude faster to propagate. Furthermore, the averaged model uncovers the mechanics behind the observed tumbling transition, sun-tracking, and tumbling cycles. Overall, the greater computational efficiency and reduced state space of the averaged framework allows us to more easily classify and analyze the general YORP evolution of different satellites and rocket bodies with various end of life configurations.

\section*{Appendix A: Torque Free Solutions}

Here we summarize the analytical solutions for torque-free rotation. We assume the long axis convention where the $\bm{\hat{b}}_1$, $\bm{\hat{b}}_2$ and $\bm{\hat{b}}_3$ body axes are aligned with the intermediate ($I_i$), maximum ($I_s$), and minimum ($I_l$) principal moments of inertia respectively. 3-1-3 ($\phi$-$\theta$-$\psi$) Euler angles are used to rotate between the $\mathcal{H}$ and $\mathcal{B}$ frames. This is the same convention used in Refs. \cite{sa1991,bensonda14}. 

Equating $\bm{H}$ in the $\mathcal{H}$ and $\mathcal{B}$ frames with Eq.~\ref{eq:BH}, we find,
\begin{equation}
a_{z1}=\sin{\theta}\sin{\psi}=\frac{I_i\omega_1}{I_d\omega_e}\;\;\;\;\;\;\;a_{z2}=\sin{\theta}\cos{\psi}=\frac{I_s\omega_2}{I_d\omega_e}\;\;\;\;\;\;a_{z3}=\cos{\theta}=\frac{I_l\omega_3}{I_d\omega_e}
\label{eq:sthetaspsi}
\end{equation}

The angles $\theta$ and $\psi$ can be unambiguously calculated using Eq.~\ref{eq:sthetaspsi} with Eq.~\ref{eq:w_lam} for LAMs or Eq.~\ref{eq:w_sam} for SAMs. The equations for $\phi$ are much more complicated and are provided below. 

\subsection*{Long Axis Modes}

For long axis modes (LAMs), the body frame angular velocity $\boldsymbol{\omega}=[\omega_1,\omega_2,\omega_3]^T$ is given by, 

\begin{equation}
\omega_1=\pm\omega_e\sqrt{\frac{I_d(I_d-I_l)}{I_i(I_i-I_l)}}\sn{\tau}\;\;\;\;\;\;\omega_2=\omega_e\sqrt{\frac{I_d(I_d-I_l)}{I_s(I_s-I_l)}}\cn{\tau}\;\;\;\;\;\;\omega_3=\pm\omega_e\sqrt{\frac{I_d(I_s-I_d)}{I_l(I_s-I_l)}}\dn{\tau}
\label{eq:w_lam}
\end{equation}

where $\sn\tau$, $\cn\tau$, and $\dn\tau$ are Jacobi elliptic functions \citep{sa1991,numericalrecipes,friedman}. The $\pm$ distinguishes between the two possible LAM regions: $+$ for ${\omega}_3>0$ (LAM+) and $-$ for $\omega_3<0$ (LAM-). For LAMs, $\tau$ is given by, 
\begin{equation}
\label{eq:tauLAM}
\tau = \tau_o + \omega_e\sqrt{\frac{I_d(I_i-I_l)(I_s-I_d)}{I_lI_iI_s}}(t-t_o)
\end{equation}
where $t$ is the time and $t_o$, $\tau_o$ are the initial values. The period of $\sn\tau$ and $\cn\tau$ is $4K(k)$ while $\dn\tau$ is periodic on $2K(k)$ where $K(k)$ is the complete elliptic integral of the first kind \citep{landau,numericalrecipes},
\begin{equation}
\label{K}
K(k)={\int_0^{\pi/2}}\frac{du}{\sqrt{1-k^{2}\sin^{2}\!u}}
\end{equation}
and $k$ is the modulus. The parameter $n$ features in the torque-free solutions for $\phi$ and $P_{\bar{\phi}}$. For LAMs, $k$ and $n$ are given by, 
\begin{equation}
\label{eq:knLAM}
k^2=\frac{(I_s-I_i)(I_d-I_l)}{(I_i-I_l)(I_s-I_d)}\;\;\;\;\;\;n=\frac{I_l}{I_s}\frac{(I_s-I_i)}{(I_i-I_l)}
\end{equation}

For LAMs, the Euler angle $\phi$ is given by, 
\begin{equation}
\label{phiLAMPibar}
\phi = \phi_o + \frac{H}{I_l}(t-t_o)-(I_s-I_l)\sqrt{\frac{I_iI_d}{I_lI_s(I_i-I_l)(I_s-I_d)}}\Big[\bar{\Pi}(\tau,n)-\bar{\Pi}(\tau_o,n)\Big]
\end{equation}
where $\bar{\Pi}(\tau,n)$ is the modified incomplete elliptic integral of the third kind. Most routines for calculating the incomplete elliptic integral of the third kind $\Pi(\tau,n)$ (e.g. Ref. \cite{numericalrecipes}) only accept $0\leq\tau\leq{K(k)}$ even though $\tau$ increases unbounded with $t$. To calculate $\bar{\Pi}(\tau,n)$ correctly, we use the following algorithm \citep{bensonda14}. Dropping the implied dependence of $k$ on $K$ for brevity,

\begin{enumerate}
\item If $\tau$ has most recently passed through an even multiple of $K$, i.e. if $\mathrm{mod}(m,2)=0$,
\begin{equation}
\label{Pibareven}
\bar{\Pi}(\tau,n) = m\Pi(K,n)+\Pi(\tau-mK,n) 
\end{equation}
\item Instead, if $\tau$ has most recently passed through an odd multiple of $K$, i.e. if $\mathrm{mod}(m,2)=1$,
\begin{equation}
\label{Pibarodd}
\bar{\Pi}(\tau,n) = (m+1)\Pi(K,n)-\Pi\Big((m+1)K-\tau,n\Big) 
\end{equation} 
\end{enumerate}
Here, the integer multiple $m=\mathrm{int}(\tau/K)$ and $\mathrm{mod}$ is the remainder after division modulo operator. 

For LAMs, the average period of $\phi$ ($P_{\bar{\phi}}$) and the constant period of $\psi$ ($P_\psi$) are given by,

\begin{equation}
\label{PphiLAM}
P_{\bar{\phi}}=\frac{2\pi}{\omega_e}\frac{I_l}{I_d}\Bigg[1-\frac{(I_s-I_l)}{I_s}\frac{\varPi(K,n)}{K}\Bigg]^{-1}
\end{equation}
\begin{equation}
\label{Ppsi_lam}
P_{\psi}=\frac{4}{\omega_e}\sqrt{\frac{I_{l}I_{i}I_{s}}{I_{d}(I_i-I_l)(I_s-I_d)}}K
\end{equation}

\subsection*{Short Axis Modes}

For short axis modes (SAMs), the body frame angular velocity $\boldsymbol{\omega}=[\omega_1,\omega_2,\omega_3]^T$ is given by, 

\begin{equation}
\label{eq:w_sam}
\omega_1 = \omega_e\sqrt{\frac{I_d(I_s-I_d)}{I_i(I_s-I_i)}}\sn{\tau}\;\;\;\;\;\;\omega_2 = \pm\omega_e\sqrt{\frac{I_d(I_d-I_l)}{I_s(I_s-I_l)}}\dn{\tau}\;\;\;\;\;\;\omega_3 = \pm\omega_e\sqrt{\frac{I_d(I_s-I_d)}{I_l(I_s-I_l)}}\cn{\tau}
\end{equation}

Again $+$ holds for ${\omega}_2>0$ and $-$ holds for $\omega_2<0$ (SAM$+$ and SAM$-$). For SAMs, $\tau$, $k$, and $n$ are,
\begin{equation}
\label{eq:tauSAM}
\tau = \tau_o + \omega_e\sqrt{\frac{I_d(I_s-I_i)(I_d-I_l)}{I_lI_iI_s}}(t-t_o)
\end{equation}

\begin{equation}
\label{eq:knSAM}
k^2 = \frac{(I_i-I_l)(I_s-I_d)}{(I_s-I_i)(I_d-I_l)}\;\;\;\;\;\;n=\frac{I_l}{I_s}\frac{(I_s-I_d)}{(I_d-I_l)}
\end{equation}

For SAMs, $\phi$ is instead given by,
\begin{equation}
\label{phiSAMPibar}
\phi = \phi_o + \frac{H}{I_l}(t-t_o)-(I_s-I_l)\sqrt{\frac{I_iI_d}{I_lI_s(I_s-I_i)(I_d-I_l)}}\Big[\bar{\Pi}(\tau,n)-\bar{\Pi}(\tau_o,n)\Big]
\end{equation}

For SAMs, $P_{\bar{\phi}}$ is also given by Eq.~\ref{PphiLAM} with $n$ from Eq.~\ref{eq:knSAM}. Finally, $P_\psi$ for SAMs is given by, 

\begin{equation}
\label{Ppsi_s}
P_{\psi}=\frac{4}{\omega_e}\sqrt{\frac{I_{l}I_{i}I_{s}}{I_{d}(I_s-I_i)(I_d-I_l)}}K
\end{equation}

\section*{Appendix B: Averaged Quantities}

From Ref. \cite{friedman}, we can obtain the following elliptic function averages,
\begin{equation}
\frac{1}{4K}\int_0^{4K}\sn\tau{d}\tau = 0
\label{eq:az1avgLAM}
\end{equation}
\begin{equation}
\frac{1}{4K}\int_0^{4K}\cn\tau{d}\tau=0
\label{eq:az2avgLAM}
\end{equation}
\begin{equation}
\frac{1}{4K}\int_0^{4K}\dn\tau{d}\tau = \frac{\pi}{2K}
\label{eq:az3avgLAM}
\end{equation}
\begin{equation}
\frac{1}{4K}\int_0^{4K}\sn^2\!\tau{d}\tau = \frac{K - E}{k^2K}
\label{eq:az12avgLAM}
\end{equation}
\begin{equation}
\frac{1}{4K}\int_0^{4K}\cn^2\!\tau{d}\tau = \frac{E - k'^2K}{k^2K}
\label{eq:az22avgLAM}
\end{equation}
\begin{equation}
\frac{1}{4K}\int_0^{4K}\dn^2\!\tau{d}\tau = \frac{E}{K}
\label{eq:az32avgLAM}
\end{equation}
\begin{equation}
\frac{1}{4K}\int_0^{4K}\sn^2\!\tau\dn\tau{d}\tau = \frac{\pi}{4K}
\label{eq:az12az3avgLAM}
\end{equation}
\begin{equation}
\frac{1}{4K}\int_0^{4K}\cn^2\!\tau\dn\tau{d}\tau = \frac{\pi}{4K}
\label{eq:az22az3avgLAM}
\end{equation}
\begin{equation}
\frac{1}{4K}\int_0^{4K}\dn^3\!\tau{d}\tau = \frac{(k'^2+1)\pi}{4K}
\label{eq:az33avgLAM}
\end{equation}
\begin{equation}
\frac{1}{4K}\int_0^{4K}\sn^4\!\tau{d}\tau = \frac{(k^2+2)K-2(k^2+1)E}{3k^4K}
\label{eq:az14avgLAM}
\end{equation}
\begin{equation}
\frac{1}{4K}\int_0^{4K}\cn^4\!\tau{d}\tau = \frac{(4k^2-2)E-k'^2(3k^2-2)K}{3k^4K}
\label{eq:az24avgLAM}
\end{equation}
\begin{equation}
\frac{1}{4K}\int_0^{4K}\dn^4\!\tau{d}\tau = \frac{2(k'^2+1)E-k'^2K}{3K}
\end{equation}
\begin{equation}
\frac{1}{4K}\int_0^{4K}\sn^2\!\tau\cn^2\!\tau{d}\tau = \frac{(1+k'^2)E-2k'^2K}{3k^4K}
\label{eq:az12az22avgavgLAM}
\end{equation}
\begin{equation}
\frac{1}{4K}\int_0^{4K}\sn^2\!\tau\dn^2\!\tau{d}\tau = \frac{(2k^2-1)E+k'^2K}{3k^2K}
\label{eq:az12az32avgavgLAM}
\end{equation}
\begin{equation}
\frac{1}{4K}\int_0^{4K}\cn^2\!\tau\dn^2\!\tau{d}\tau = \frac{(1+k^2)E-k'^2K}{3k^2K}
\label{eq:az22az32avgavgLAM}
\end{equation}
\normalsize
where $E$ is the complete elliptic integral of the second kind \citep{numericalrecipes} and $k'^2=1-k^2$. 

\subsection*{Long Axis Modes}

After averaging over $\phi$, we follow Ref. \cite{cicalo} and write all averaged expressions in terms of $\overline{a_{z1}}$, $\overline{a_{z2}}$, and $\overline{a_{z3}}$ because they are independent of $\phi$. Following the notation of Eq.~\ref{eq:psiavgK}, for LAMs we have the following expressions with ($1$, $2$, $3$) subscripts denoting the $\mathcal{B}$ frame vector components and using $f$ as a placeholder for $d$ and $r$. 
\small
\begin{equation}
\overline{f_z} = f_3\overline{a_{z3}}
\label{eq:l_fz}
\end{equation}
\begin{equation}
\overline{f_xn_x} = \frac{1}{2}\Big(f_3n_3 - f_1n_1\Big)\overline{a^2_{z1}} + \frac{1}{2}\Big(f_3n_3 - f_2n_2\Big)\overline{a^2_{z2}} + \frac{1}{2}\Big(f_1n_1+f_2n_2\Big)
\label{eq:l_fxnx}
\end{equation}
\begin{equation}
\overline{f_yn_x} = \frac{1}{2}\Big(f_2n_1 - f_1n_2\Big)\overline{a_{z3}}
\label{eq:l_fynx}
\end{equation}
\begin{equation}
\overline{f_zn_z} = f_1n_1\overline{a^2_{z1}} + f_2n_2\overline{a^2_{z2}} + f_3n_3\overline{a^2_{z3}}
\label{eq:l_fznz}
\end{equation}
\begin{equation}
\overline{f_xn_xn_z} = \frac{1}{2}\Big(f_1n_1n_3 + f_2n_2n_3\Big)\overline{a_{z3}} - \frac{1}{2}\Big(f_3n^2_1 + 2f_1n_1n_3 - f_3n^2_3\Big)\overline{a^2_{z1}a_{z3}} - \frac{1}{2}\Big(f_3n^2_2 + 2f_2n_2n_3 - f_3n^2_3\Big)\overline{a^2_{z2}a_{z3}}
\label{eq:l_fxnxnz}
\end{equation}
\begin{equation}
\overline{f_yn_xn_z} = \frac{1}{2}\Big(f_3n_1n_2 - f_2n_1n_3\Big)\overline{a^2_{z1}} + \frac{1}{2}\Big(f_1n_2n_3 - f_3n_1n_2\Big)\overline{a^2_{z2}} + \frac{1}{2}\Big(f_2n_1n_3 - f_1n_2n_3\Big)\overline{a^2_{z3}}
\label{eq:l_fynxnz}
\end{equation}
\begin{equation}
\overline{f_zn^2_x} = \frac{1}{2}\Big(f_3n^2_1 + f_3n^2_2\Big)\overline{a_{z3}} - \frac{1}{2}\Big(f_3n^2_1 + 2f_1n_1n_3 - f_3n^2_3\Big)\overline{a^2_{z1}a_{z3}} - \frac{1}{2}\Big(f_3n^2_2 + 2f_2n_2n_3 - f_3n^2_3\Big)\overline{a^2_{z2}a_{z3}}
\label{eq:l_fznx2}
\end{equation}
\begin{equation}
\overline{f_zn^2_z} = \Big(f_3n^2_1 + 2f_1n_1n_3\Big)\overline{a^2_{z1}a_{z3}} + \Big(f_3n^2_2 + 2f_2n_2n_3\Big)\overline{a^2_{z2}a_{z3}} + f_3n^2_3\overline{a^3_{z3}}
\label{eq:l_fznz2}
\end{equation}
\begin{equation}
\begin{split}
\overline{f_xn^3_x} = & +\frac{3}{8}\Big(3f_3n^2_1n_3 -2f_1n^3_1 - 3f_2n^2_1n_2 - f_1n_1n^2_2 + 3f_1n_1n^2_3 +f_3n^2_2n_3 + f_2n_2n^2_3\Big)\overline{a^2_{z1}} \\ & 
+ \frac{3}{8}\Big(f_2n^2_1n_2 + f_3n^2_1n_3 - f_1n_1n^2_2 + f_1n_1n^2_3 - 2f_2n^3_2 + 3f_3n^2_2n_3 + 3f_2n_2n^2_3\Big)\overline{a^2_{z2}} \\ &
+ \frac{3}{8}\Big(f_1n^3_1 - 3f_3n^2_1n_3 - 3f_1n_1n^2_3 + f_3n^3_3\Big)\overline{a^4_{z1}}
+ \frac{3}{8}\Big(f_2n^3_2- 3f_3n^2_2n_3 - 3f_2n_2n^2_3 + f_3n^3_3\Big)\overline{a^4_{z2}} \\ &
+ \frac{3}{8}\Big(3f_2n^2_1n_2 - 3f_3n^2_1n_3 + 3f_1n_1n^2_2 - 3f_1n_1n^2_3 - 3f_3n^2_2n_3 - 3f_2n_2n^2_3 + 2f_3n^3_3\Big)\overline{a^2_{z1}a^2_{z2}} \\ & 
+ \frac{3}{8}\Big(f_1n^3_1 + f_2n^2_1n_2 + f_1n_1n^2_2 + f_2n^3_2\Big)
\end{split}
\label{eq:l_fxnx3}
\end{equation}
\begin{equation}
\begin{split}
\overline{f_xn_xn^2_z} = & + \frac{1}{2}\Big(f_1n^3_1 - 2f_3n^2_1n_3 + f_2n_2n^2_1 - 4f_1n_1n^2_3 + f_3n^3_3 - f_2n_2n^2_3\Big)\overline{a^2_{z1}} \\ & 
+ \frac{1}{2}\Big(f_2n^3_2 - 2f_3n^2_2n_3 + f_1n_1n^2_2 - 4f_2n_2n^2_3 + f_3n^3_3 - f_1n_1n^2_3\Big)\overline{a^2_{z2}} \\ & 
+ \frac{1}{2}\Big(-f_1n^3_1 + 3f_3n^2_1n_3 + 3f_1n_1n^2_3 - f_3n^3_3\Big)\overline{a^4_{z1}} \\ &
+ \frac{3}{2}\Big(f_2n^2_1n_2 + f_3n^2_1n_3 - f_1n_1n^2_2 + f_1n_1n^2_3 +f_3n^2_2n_3 + f_2n_2n^2_3 - \frac{2}{3}f_3n^3_3\Big)\overline{a^2_{z1}a^2_{z2}} \\ & 
+ \frac{1}{2}\Big(-f_2n^3_2 + 3f_3n^2_2n_3 + 3f_2n_2n^2_3 - f_3n^3_3\Big)\overline{a^4_{z2}}+ \frac{1}{2}\Big(f_1n_1n^2_3 + f_2n_2n^2_3\Big)
\end{split}
\label{eq:l_fxnxnz2}
\end{equation}
\begin{equation}
\begin{split}
\overline{f_yn^3_x} = & - \frac{3}{8}\Big((f_2n^3_1 - f_1n_2n^2_1 - 3f_2n_1n^2_3 + 2f_3n_2n_1n_3 + f_1n_2n^2_3\Big)\overline{a^2_{z1}a_{z3}} \\ & 
+ \frac{3}{8}\Big(f_1n^3_2 - f_2n_1n^2_2 - 3f_1n_2n^2_3 + 2f_3n_1n_2n_3 + f_2n_1n^2_3\Big)\overline{a^2_{z2}a_{z3}} \\ & 
- \frac{3}{8}\Big((- f_2n^3_1 + f_1n^2_1n_2 - f_2n_1n^2_2 + f_1n^3_2)\overline{a_{z3}}
\end{split}
\label{eq:l_fynx3}
\end{equation}
\begin{equation}
\begin{split}
\overline{f_yn_xn^2_z} = & + \frac{1}{2}\Big(f_2n^3_1 - f_1n_2n^2_1 - 3f_2n_1n^2_3 + 2f_3n_2n_1n_3 + f_1n_2n^2_3\Big)\overline{a^2_{z1}a_{z3}} \\ &
 - \frac{1}{2}\Big(f_1n^3_2 - f_2n_1n^2_2 - 3f_1n_2n^2_3 + 2f_3n_1n_2n_3 + f_2n_1n^2_3\Big)\overline{a^2_{z2}a_{z3}}\\ & 
 - \frac{1}{2}\Big(f_1n_2 - f_2n_1)n^2_3\overline{a_{z3}}
 \end{split}
 \label{eq:l_fynxnz2}
\end{equation}
\begin{equation}
\begin{split}
\overline{f_zn^2_xn_z} = & + \frac{1}{2}\Big(f_1n^3_1 - 4f_3n^2_1n_3 + f_1n_1n^2_2 - 2f_1n_1n^2_3 - f_3n^2_2n_3 + f_3n^3_3\Big)\overline{a^2_{z1}}\\ &
+ \frac{1}{2}\Big(f_2n_1^2n_2 - f_3n_1^2n_3 + f_2n_2^3 - 4f_3n_2^2n_3 - 2f_2n_2n_3^2 + f_3n_3^3\Big)\overline{a_{z2}^2} 
+ \frac{1}{2}\Big(f_3n_3n_1^2+f_3n_3n_2^2\Big) \\ &
+ \frac{1}{2}\Big(- f_1n^3_1 + 3f_3n^2_1n_3 + 3f_1n_1n^2_3 - f_3n^3_3\Big)\overline{a^4_{z1}}
+ \frac{1}{2}\Big(-f_2n_2^3 + 3f_3n_2^2n_3 + 3f_2n_2n_3^2 - f_3n_3^3\Big)\overline{a_{z2}^4} \\ & 
+ \frac{3}{2}\Big(-f_2n^2_1n_2 + f_3n^2_1n_3 - f_1n_1n^2_2 + f_1n_1n^2_3 + f_3n^2_2n_3 + f_2n_2n^2_3 - \frac{2}{3}f_3n^3_3\Big)\overline{a^2_{z1}a^2_{z2}}
\end{split}
\label{eq:l_fznx2nz}
\end{equation}
\begin{equation}
\begin{split}
\overline{f_zn^3_z} = & + \Big(3f_3n_1^2n_3 + 3f_1n_1n_3^2 - 2f_3n_3^3\Big)\overline{a_{z1}^2}
+ \Big(3f_3n_2^2n_3 + 3f_2n_2n_3^2 - 2f_3n_3^3\Big)\overline{a_{z2}^2} \\ &  
+ \Big(f_1n_1^3 - 3f_3n_1^2n_3 - 3f_1n_1n_3^2 + f_3n_3^3\Big)\overline{a_{z1}^4}
+ \Big(f_2n_2^3 - 3f_3n_2^2n_3 - 3f_2n_2n_3^2 + f_3n_3^3\Big)\overline{a_{z2}^4} \\ &
+ 3\Big(f_2n_1^2n_2 - f_3n_1^2n_3 + f_1n_1n_2^2 - f_1n_1n_3^2 - f_3n_2^2n_3 - f_2n_2n_3^2 + \frac{2}{3}f_3n_3^3\Big)\overline{a_{z1}^2a_{z2}^2} 
+ f_3n_3^3
\end{split}
\label{eq:l_fznz3}
\end{equation}
\begin{equation}
\begin{split}
\overline{ga_{z1}\delta_1}= & +\frac{2}{3\pi}\Big(6n_1n_3r_2u_x^2u_z - 4n_1n_3r_2u_z^3\Big)\overline{a_{z1}^4} 
+ \frac{1}{4}\Big(2n_1r_2u_z^2 - n_1r_2u_x^2\Big)\overline{a_{z1}^2a_{z3}} \\ &
+ \frac{2}{3\pi}\Big(6n_1n_2r_3u_x^2u_z - 4n_1n_3r_2u_z^3 - 4n_1n_2r_3u_z^3 + 6n_1n_3r_2u_x^2u_z\Big)\overline{a_{z1}^2a_{z2}^2} \\ &
+ \frac{4}{3\pi}\Big(2n_1n_3r_2u_z^3 - n_1n_2r_3u_x^2u_z - 2n_1n_3r_2u_x^2u_z\Big)\overline{a_{z1}^2}
\end{split}
\end{equation}
\begin{equation}
\begin{split}
\overline{ga_{z2}\delta_2}= & +\frac{2}{3\pi}\Big(4n_1n_2r_3u_z^3 + 4n_2n_3r_1u_z^3 - 6n_1n_2r_3u_x^2u_z - 6n_2n_3r_1u_x^2u_z\Big)\overline{a_{z1}^2a_{z2}^2} \\ &
+ \frac{2}{3\pi}\Big(4n_2n_3r_1u_z^3 - 6n_2n_3r_1u_x^2u_z\Big)\overline{a_{z2}^4}
+ \frac{1}{4}\Big(n_2r_1u_x^2 - 2n_2r_1u_z^2\Big)\overline{a_{z2}^2a_{z3}} \\ &
+ \frac{4}{3\pi}\Big(n_1n_2r_3u_x^2u_z - 2n_2n_3r_1u_z^3 + 2n_2n_3r_1u_x^2u_z\Big)\overline{a_{z2}^2}
\end{split}
\end{equation}
\begin{equation}
\begin{split}
\overline{ga_{z3}\delta_3}= & +\frac{2}{3\pi}\Big(6n_1n_3r_2u_x^2u_z - 4n_1n_3r_2u_z^3\Big)\overline{a_{z1}^2a_{z3}^2}
+ \frac{1}{4}\Big(n_1r_2u_x^2 - 2n_1r_2u_z^2\Big)\overline{a_{z1}^2a_{z3}} \\ &
+ \frac{2}{3\pi}\Big(4n_2n_3r_1u_z^3 - 6n_2n_3r_1u_x^2u_z\Big)\overline{a_{z2}^2a_{z3}^2}
- \frac{1}{4}\Big(n_2r_1u_x^2 - 2n_2r_1u_z^2\Big)\overline{a_{z2}^2a_{z3}} \\ &
 + \frac{4}{3\pi}\Big(n_2n_3r_1u_x^2u_z - n_1n_3r_2u_x^2u_z\Big)\overline{a_{z3}^2}
- \frac{1}{4\pi}\Big(n_1r_2u_x^2 - n_2r_1u_x^2\Big)\overline{a_{z3}}
\end{split}
\end{equation}

\subsection*{Short Axis Modes}

\normalsize
The following averaged expressions hold for SAMs,
\small
\begin{equation}
\overline{f_z} = f_2\overline{a_{z2}}
\label{eq:s_fz}
\end{equation}
\begin{equation}
\overline{f_xn_x} = \frac{1}{2}\Big(f_3n_3 - f_1n_1\Big)\overline{a^2_{z1}} + \frac{1}{2}\Big(f_3n_3 - f_2n_2\Big)\overline{a^2_{z2}} + \frac{1}{2}\Big(f_1n_1+f_2n_2\Big)
\label{eq:s_fxnx}
\end{equation}
\begin{equation}
\overline{f_yn_x} = \frac{1}{2}(f_1n_3 - f_3n_1)\overline{a_{z2}}
\label{eq:s_fynx}
\end{equation}
\begin{equation}
\overline{f_zn_z} = f_1n_1\overline{a^2_{z1}} + f_2n_2\overline{a^2_{z2}} + f_3n_3\overline{a^2_{z3}}
\label{eq:s_fznz}
\end{equation}
\begin{equation}
\begin{split}
\overline{f_xn_xn_z} = & + \frac{1}{2}\Big(f_2n_3^2-f_2n_1^2 - 2f_1n_2n_1  + 2f_3n_2n_3\Big)\overline{a_{z1}^2a_{z2}} \\&
+ \frac{1}{2}\Big(f_2n_3^2 - f_2n_2^2 + 2f_3n_2n_3 \Big)\overline{a_{z2}^3}
+ \frac{1}{2}\Big(f_2n_2^2 - f_3n_2n_3 + f_1n_1n_2 - f_2n_3^2\Big)\overline{a_{z2}}
\end{split}
\label{eq:s_fxnxnz}
\end{equation}
\begin{equation}
\begin{split}
\overline{f_yn_xn_z} = & + \frac{1}{2}\Big(f_1n_2n_3 - 2f_2n_1n_3 + f_3n_1n_2\Big)\overline{a_{z1}^2} + \frac{1}{2}\Big(2f_1n_2n_3 - f_2n_1n_3 - f_3n_1n_2\Big)\overline{a_{z2}^2} \\ & + \frac{1}{2}\Big(f_2n_1n_3-f_1n_2n_3\Big)
\end{split}
\label{eq:s_fynxnz}
\end{equation}
\begin{equation}
\begin{split}
\overline{f_zn^2_x} = & + \frac{1}{2}\Big(f_2n_3^2-f_2n_1^2 - 2f_1n_2n_1 + 2f_3n_2n_3\Big)\overline{a_{z1}^2a_{z2}}
+ \frac{1}{2}\Big(f_2n_3^2 - f_2n_2^2 + 2f_3n_2n_3 \Big)\overline{a_{z2}^3} \\ & 
+ \frac{1}{2}\Big(f_2n_1^2 + f_2n_2^2 - 2f_3n_3n_2\Big)\overline{a_{z2}} 
\end{split}
\label{eq:s_fznx2}
\end{equation}
\begin{equation}
\overline{f_zn^2_z} = \Big(f_2n_1^2 + 2f_1n_2n_1 - f_2n_3^2 - 2f_3n_2n_3\Big)\overline{a_{z1}^2a_{z2}} + \Big(f_2n_2^2 - 2f_3n_2n_3 - f_2n_3^2\Big)\overline{a_{z2}^3} + \Big(f_2n_3^2 + 2f_3n_2n_3\Big)\overline{a_{z2}}
\label{eq:s_fznz2}
\end{equation}
\begin{equation}
\begin{split}
\overline{f_xn^3_x} = & + \frac{1}{8}(3f_1n_1^3 - 9f_3n_1^2n_3 - 9f_1n_1n_3^2 + 3f_3n_3^3\Big)\overline{a_{z1}^4}
+ \frac{3}{8}(3f_2n_2^3 - 3f_3n_2^2n_3 - 3f_2n_2n_3^2 + f_3n_3^3\Big)\overline{a_{z2}^4} \\ & 
+ \frac{9}{8}(9f_2n_1^2n_2 - f_3n_1^2n_3 + f_1n_1n_2^2 - 9f_1n_1n_3^2 - 9f_3n_2^2n_3 - 9f_2n_2n_3^2 + \frac{2}{3}f_3n_3^3\Big)\overline{a_{z1}^2a_{z2}^2} \\ & 
+ \frac{3}{8}(- 2f_1n_1^3 - f_2n_1^2n_2 + 3f_3n_1^2n_3 - f_1n_1n_2^2 + 3f_1n_1n_3^2 + f_3n_2^2n_3 + f_2n_2n_3^2\Big)\overline{a_{z1}^2} \\ & 
+ \frac{3}{8}(- f_2n_1^2n_2 + f_3n_1^2n_3 - f_1n_1n_2^2 + f_1n_1n_3^2 - 2f_2n_2^3 + 3f_3n_2^2n_3 + 3f_2n_2n_3^2\Big)\overline{a_{z2}^2} \\ & 
+ \frac{3}{8}(3f_1n_1^3 + 3f_2n_1^2n_2 + 3f_1n_1n_2^2 + 3f_2n_2^3\Big)
\end{split}
\label{eq:s_fxnx3}
\end{equation}
\begin{equation}
\begin{split}
\overline{f_xn_xn^2_z} = & + \frac{1}{2}\Big(-f_1n_1^3 + 3f_3n_1^2n_3 + 3f_1n_1n_3^2 - f_3n_3^3\Big)\overline{a_{z1}^4}
+ \frac{1}{2}\Big(- f_2n_2^3 + 3f_3n_2^2n_3 + 3f_2n_2n_3^2 - f_3n_3^3\Big)\overline{a_{z2}^4} \\ & 
+ \frac{3}{2}\Big(- f_2n_1^2n_2 + f_3n_1^2n_3 - f_1n_1n_2^2 + f_1n_1n_3^2 + f_3n_2^2n_3 + f_2n_2n_3^2 -\frac{2}{3} f_3n_3^3\Big)\overline{a_{z1}^2a_{z2}^2} \\ & 
+ \frac{1}{2}\Big(f_1n_1^3 - 2f_3n_1^2n_3 + f_2n_2n_1^2 - 4f_1n_1n_3^2 + f_3n_3^3 - f_2n_2n_3^2\Big)\overline{a_{z1}^2} \\ & 
+ \frac{1}{2}\Big(f_2n_2^3 - 2f_3n_2^2n_3 + f_1n_1n_2^2 - 4f_2n_2n_3^2 + f_3n_3^3 - f_1n_1n_3^2\Big)\overline{a_{z2}^2} 
+ \frac{1}{2}\Big(f_1n_1n_3^2+f_2n_2n_3^2\Big)
\end{split}
\label{eq:s_fxnxnz2}
\end{equation}
\begin{equation}
\begin{split}
\overline{f_yn^3_x} = & + \frac{3}{8}\Big(f_3n_1^3 - f_1n_1^2n_3 - 2f_3n_1n_2^2 + 4f_2n_1n_2n_3 - f_3n_1n_3^2 - 2f_1n_2^2n_3 + f_1n_3^3\Big)\overline{a_{z1}^2a_{z2}} \\ & 
+ \frac{3}{8}\Big(- 3f_1n_2^2n_3 + f_3n_1n_2^2 + 2f_2n_1n_2n_3 + f_1n_3^3 - f_3n_1n_3^2\Big)\overline{a_{z2}^3} \\ & 
+ \frac{3}{8}\Big(- f_3n_1^3 + f_1n_3n_1^2 - f_3n_1n_2^2 - 2f_2n_3n_1n_2 + 3f_1n_3n_2^2\Big)\overline{a_{z2}}
\end{split}
\label{eq:s_fynx3}
\end{equation}
\begin{equation}
\begin{split}
\overline{f_yn_xn^2_z} = & + \frac{1}{2}\Big(- f_3n_1^3 + f_1n_1^2n_3 + 2f_3n_1n_2^2 - 4f_2n_1n_2n_3 + f_3n_1n_3^2 + 2f_1n_2^2n_3 - f_1n_3^3\Big)\overline{a_{z1}^2a_{z2}} \\ &
+ \frac{1}{2}\Big(3f_1n_2^2n_3 - f_3n_1n_2^2 - 2f_2n_1n_2n_3 - f_1n_3^3 + f_3n_1n_3^2\Big)\overline{a_{z2}^3} \\ & 
+ \frac{1}{2}\Big(- 2f_1n_2^2n_3 + 2f_2n_1n_2n_3 + f_1n_3^3 - f_3n_1n_3^2\Big)\overline{a_{z2}}
 \end{split}
 \label{eq:s_fynxnz2}
\end{equation}
\begin{equation}
\begin{split}
\overline{f_zn^2_xn_z} = & + \frac{1}{2}\Big(- f_1n_1^3 + 3f_3n_1^2n_3 + 3f_1n_1n_3^2 - f_3n_3^3\Big)\overline{a_{z1}^4}
+ \frac{1}{2}\Big(- f_2n_2^3 + 3f_3n_2^2n_3 + 3f_2n_2n_3^2 - f_3n_3^3\Big)\overline{a_{z2}^4} \\& 
+ \frac{3}{2}\Big(- f_2n_1^2n_2 + f_3n_1^2n_3 - f_1n_1n_2^2 + f_1n_1n_3^2 + f_3n_2^2n_3 + f_2n_2n_3^2 -\frac{2}{3} f_3n_3^3\Big)\overline{a_{z1}^2a_{z2}^2} \\ & 
+ \frac{1}{2}\Big(f_1n_1^3 - 4f_3n_1^2n_3 + f_1n_1n_2^2 - 2f_1n_1n_3^2 - f_3n_2^2n_3 + f_3n_3^3\Big)\overline{a_{z1}^2} \\ & 
+ \frac{1}{2}\Big(f_2n_1^2n_2 - f_3n_1^2n_3 + f_2n_2^3 - 4f_3n_2^2n_3 - 2f_2n_2n_3^2 + f_3n_3^3\Big)\overline{a_{z2}^2} 
+ \frac{1}{2}\Big(f_3n_3n_1^2 + f_3n_3n_2^2\Big)
\end{split}
\label{eq:s_fznx2nz}
\end{equation}
\begin{equation}
\begin{split}
\overline{f_zn^3_z} = & + \Big(f_1n_1^3 - 3f_3n_1^2n_3 - 3f_1n_1n_3^2 + f_3n_3^3\Big)\overline{a_{z1}^4}
+ 3\Big(f_3n_1^2n_3 + f_1n_1n_3^2 - \frac{2}{3}f_3n_3^3\Big)\overline{a_{z1}^2} \\ & 
+ 3\Big(f_2n_1^2n_2 - f_3n_1^2n_3 + f_1n_1n_2^2 - f_1n_1n_3^2 - f_3n_2^2n_3 - f_2n_2n_3^2 + \frac{2}{3}f_3n_3^3\Big)\overline{a_{z1}^2a_{z2}^2} \\ & 
+ \Big(f_2n_2^3 - 3f_3n_2^2n_3 - 3f_2n_2n_3^2 + f_3n_3^3\Big)\overline{a_{z2}^4} 
+ 3\Big(f_3n_2^2n_3 + f_2n_2n_3^2 - \frac{2}{3}f_3n_3^3\Big)\overline{a_{z2}^2}
+ f_3n_3^3
\end{split}
\label{eq:s_fznz3}
\end{equation}
\begin{equation}
\begin{split}
\overline{ga_{z1}\delta_1}= & \frac{2}{3\pi}\Big(6n_1n_3r_2u_x^2u_z - 4n_1n_3r_2u_z^3\Big)\overline{a_{z1}^4}
+ \frac{1}{4}\Big(n_1r_3u_x^2 - 2n_1r_3u_z^2\Big)\overline{a_{z1}^2a_{z2}} \\ &
+ \frac{2}{3\pi}\Big(6n_1n_2r_3u_x^2u_z - 4n_1n_3r_2u_z^3 - 4n_1n_2r_3u_z^3 + 6n_1n_3r_2u_x^2u_z\Big)\overline{a_{z1}^2a_{z2}^2} \\ &
+ \frac{4}{3\pi}\Big(2n_1n_3r_2u_z^3 - n_1n_2r_3u_x^2u_z - 2n_1n_3r_2u_x^2u_z\Big)\overline{a_{z1}^2}
\end{split}
\end{equation}
\begin{equation}
\begin{split}
\overline{ga_{z2}\delta_2}= & \frac{2}{3\pi}\Big(4n_1n_2r_3u_z^3 + 4n_2n_3r_1u_z^3 - 6n_1n_2r_3u_x^2u_z - 6n_2n_3r_1u_x^2u_z\Big)\overline{a_{z1}^2a_{z2}^2} \\ &
- \frac{3}{12}\Big(n_1r_3u_x^2 + n_3r_1u_x^2 - 2n_1r_3u_z^2 - 2n_3r_1u_z^2\Big)\overline{a_{z1}^2a_{z2}} \\ &
+ \frac{2}{3\pi}\Big(4n_2n_3r_1u_z^3 - 6n_2n_3r_1u_x^2u_z\Big)\overline{a_{z2}^4}
- \frac{1}{4}\Big(n_3r_1u_x^2 - 2n_3r_1u_z^2\Big)\overline{a_{z2}^3} \\ &
+ \frac{4}{3\pi}\Big(n_1n_2r_3u_x^2u_z - 2n_2n_3r_1u_z^3 + 2n_2n_3r_1u_x^2u_z\Big)\overline{a_{z2}^2} 
+ \frac{1}{4}\Big(n_1r_3u_x^2 - 2n_3r_1u_z^2\Big)\overline{a_{z2}}
\end{split}
\end{equation}
\begin{equation}
\begin{split}
\overline{ga_{z3}\delta_3}= & \frac{2}{3\pi}\Big(6n_1n_3r_2u_x^2u_z - 4n_1n_3r_2u_z^3\Big)\overline{a_{z1}^2a_{z3}^2}
+ \frac{2}{3\pi}\Big(4n_2n_3r_1u_z^3 - 6n_2n_3r_1u_x^2u_z\Big)\overline{a_{z2}^2a_{z3}^2} \\ &
 - \frac{1}{4}\Big(n_3r_1u_x^2 - 2n_3r_1u_z^2\Big)\overline{a_{z2}a_{z3}^2}
 + \frac{4}{3\pi}\Big(n_2n_3r_1u_x^2u_z - n_1n_3r_2u_x^2u_z\Big)\overline{a_{z3}^2}
\end{split}
\end{equation}

\section*{\MakeUppercase{Acknowledgements}}
\normalsize
This work was supported by a NASA Space Technology Research Fellowship through grant NNX16AM53H. DJS acknowledges support from AFOSR through grant FA9550-18-1-0313.

\bibliography{references}

\end{document}